\documentclass[useAMS,usenatbib]{mn2e}

\usepackage{times}
\usepackage{graphicx}
\usepackage{subfigure}

\newcommand{\gsim}{\mathrel{\hbox{\rlap{\lower.55ex \hbox {$\sim$}}
                   \kern-.3em \raise.4ex \hbox{$>$}}}}
\newcommand{\lsim}{\mathrel{\hbox{\rlap{\lower.55ex \hbox {$\sim$}}
                   \kern-.3em \raise.4ex \hbox{$<$}}}}

\title[The dependence of stellar properties on metallicity: opacity]{The statistical properties of stars and their dependence on metallicity: the effects of opacity}
\author[M.R. Bate]{Matthew R. Bate\thanks{E-mail:
mbate@astro.ex.ac.uk}\\ School of Physics and Astronomy, University of Exeter, Stocker
Road, Exeter EX4 4QL}

\date{\today}
\begin{document}
\maketitle
\begin{abstract}
We report the statistical properties of stars and brown dwarfs obtained from four radiation hydrodynamical simulations of star cluster formation that resolve masses down to the opacity limit for fragmentation.   The calculations are identical except for their dust and gas opacities.  Assuming dust opacity is proportional to metallicity, the calculations span a range of metallicities from 1/100 to 3 times solar, although we emphasise that changing the metallicity has other thermodynamic effects that the calculations do not capture (e.g.  on the thermal coupling between gas and dust).

All four calculations produce stellar populations whose statistical properties are difficult to distinguish from observed stellar systems, and we find no significant dependence of stellar properties on opacity.  The mass functions and properties of multiple stellar systems are consistent with each other.  However, we find protostellar mergers are more common with lower opacities.  Combining the results from the three calculations with the highest opacities, we obtain a stellar population consisting of more than 500 stars and brown dwarfs.  Many of the statistical properties of this population are in good agreement with those observed in our Galaxy, implying that gravity, hydrodynamics, and radiative feedback may be the primary ingredients for determining the statistical properties of low-mass stars.  However, we do find indications that the calculations may be slightly too dissipative. Although further calculations will be required to understand all of the effects of metallicity on stellar properties, we conclude that stellar properties are surprisingly resilient to variations of the dust and gas opacities.
\end{abstract}

\section{Introduction}
\label{introduction}

Understanding the origin of the statistical properties of stellar systems is the fundamental goal of a complete theory of star formation.  Much attention has been paid to the origin of the stellar initial mass function (IMF), and there are many models that have been proposed for its origin \citep*[see the review of][]{BonLarZin2007}.  However, a complete model must be able to explain the origin of all the statistical properties of stellar systems (e.g., the star formation rate and efficiency, and the abundance and properties of multiple stellar systems) and how these depend on variations in environment and initial conditions.  While simplified analytic or semi-analytic models are useful for understanding the role that different processes play in the origin of some stellar properties, numerical simulations are almost certainly necessary to help us understand the full complexity of the star formation process.

Numerical simulations first became powerful enough to begin tackling the question of the origin of the statistical properties of stars in the late 1990s and early 2000s, with calculations that could follow the formation of groups of stars \citep[e.g.][]{Bonnelletal1997, KleBurBat1998, KleBur2000, BatBonBro2003}.  However, until recently, the stellar populations produced by such calculations have always differed significantly from the properties of observed stellar systems.  For example, some calculations used large sink particles to model the protostars and could not resolve most brown dwarfs, thus producing incomplete mass functions \citep[e.g.][]{Bonnelletal2001b, BonBatVin2003}.  Hydrodynamical simulations that resolved the opacity limit for fragmentation (and hence the lowest mass brown dwarfs) but used barotropic equations of state tended to overproduce brown dwarfs \citep{BatBonBro2003, BatBon2005, Bate2009a,Offneretal2009}, particularly when the molecular cloud was modelled with decaying rather than driven turbulence \citep{OffKleMcK2008}.  Moreover, using a barotropic equation of state results in a characteristic stellar mass that depends primarily on the initial mean thermal Jeans mass in the cloud \citep{BatBon2005, Jappsenetal2005, BonClaBat2006}.  This seems to be the case even when  the barotropic equation of state is modified as might be expected for a different metallicity \citep{Bate2005}.  Including radiative transfer and a more realistic equation of state is found to dramatically decrease the amount of fragmentation, increase the characteristic stellar mass, decrease the proportion of brown dwarfs \citep{Bate2009b,Offneretal2009,UrbMarEva2010}, and weaken the dependence of the characteristic mass of the IMF on the initial Jeans mass \citep{Bate2009b}.  The latter effect may help to explain why the IMF is not observed to be strongly dependent on initial conditions, at least in our Galaxy \citep*{BasCovMey2010}. However, the introduction of radiative transfer can also lead to problems in reproducing the observed IMF as `overheating' of the gas due to protostellar radiative feedback can produce a top-heavy IMF in calculations of massive dense molecular cloud cores \citep{KruKleMcK2011}.

The most successful numerical simulation of star formation published to date, in terms of reproducing a wide variety of the observed statistical properties of stellar systems, is that of \cite{Bate2012}.  This radiation hydrodynamical calculation produced more than 180 stars and brown dwarfs, including 40 multiple systems, whose properties were difficult to distinguish from observed stellar systems.  The mass function of the stellar population was in good agreement with the observed Galactic IMF, the multiplicity of the stellar systems was found to increase with primary mass with values in agreement with the results from various field star surveys, and the properties of these multiple systems (e.g. their mass ratios and separation distributions) also reproduced many of the observed characteristics.  Although the earlier barotropic calculation of \cite{Bate2009a} was able to reproduce many of the observed characteristics of multiple stellar systems, it greatly overproduced brown dwarfs due to the absence of radiative feedback.  More recently, \cite{KruKleMcK2012} have also presented results from a radiation hydrodynamical calculation whose stellar mass distribution is in statistical agreement with the observed IMF and with a stellar multiplicity that increases with primary mass.  They find that including protostellar outflows and large-scale turbulent driving are important for avoiding the `overheating' problem  \citep{KruKleMcK2011}.  However, this calculation underproduces low-mass multiple systems and only produces two dozen multiple systems in total which limits further comparison with observed systems.

Now that we are able to produce simulations that create stellar populations whose statistical properties are in close agreement with observed stellar systems in our Galaxy, we can begin to use further calculations to reveal how the statistical properties of stellar systems may depend on initial conditions and environment.  In this paper, we report the results from three new simulations which are identical to the calculation of \cite{Bate2012} except that they employ different opacities.  Although each of the calculations is started from the same initial conditions, the calculations soon differ because the different opacities affect the thermodynamics.  As the calculations diverge, they produce stellar systems whose dynamics are, in general, chaotic.  Thus, particularly on small-scales, the calculations each produce a different set of stellar systems.

Our aim is to {\em begin} investigating the extent to which stellar properties may depend on the metallicity of a star-forming region.  However, it is important to realise a change in the metallicity does much more to a star-forming cloud than simply change the opacity of the matter, particularly if the metallicity is reduced.  There have been many studies that have investigated the thermodynamics of molecular gas with different metallicities (e.g. \citealt{Omukai2000, Omukaietal2005, GloJap2007, Jappsenetal2007, Jappsenetal2009a}a, \citeyear{Jappsenetal2009b}b; \citealt{HocSpa2010, OmuHosYos2010, SchOmu2010, Dopckeetal2011, Walchetal2011, GloCla2012b, GloCla2012a, GloCla2012c, Omukai2012, Schneideretal2012, Dopckeetal2013}).  The thermal behaviour of collapsing molecular gas is found to be almost independent of its metallicity once it becomes opaque to long-wavelength radiation, but at lower densities the gas temperature is complicated and depends on a large number of heating and cooling processes \citep{Omukai2000}.  These studies show that changing the metallicity can change the thermal evolution of the low-density gas in several different ways.  First, while Galactic star formation calculations often assume that the gas and dust temperatures are well coupled, due to collisions, this depends on both on the gas density and the density (and properties) of dust grains.  The gas and dust are well coupled at molecular densities $\gsim 10^5$~cm$^{-3}$ and with solar metallicities, but they become poorly coupled as either the density or metallicity are reduced \citep[e.g][]{TsuOmu2006, Dopckeetal2011, NozKozNom2012, ChiNozYos2013, Dopckeetal2013}. This has a huge impact on the gas temperatures.  When the gas and dust are thermally well coupled, both are typically cold and nearly isothermal ($\approx 10$~K) because thermal dust emission is the primary coolant and the dust cooling rate is a strong function of temperature \cite[typically scaling as $\sim T_{\rm d}^6$, e.g. ][]{Goldsmith2001}.  However, when they are decoupled, the dust remains cold, but the gas tends to be much hotter.  An added complication is that the dust properties themselves are likely to change as the metallicity varies \citep[e.g.][]{RemyRuyer2014}.  Second, in the absence of dust cooling, the gas cools directly via atomic and molecular line emission (e.g. from C$^+$ and CO).  Clearly, as the metallicity is reduced, so is the effectiveness of these coolants.  However, the gas cooling rate also increases much more slowly with increasing temperature than the dust, so that when the gas and dust become decoupled, an even higher gas temperature is required to make up for the lost dust cooling.  Third, a star-forming cloud is heated by external radiation and cosmic rays.  At the very least it will receive cosmic background radiation, but typically it is also irradiated by other stars in its galaxy and perhaps by the radiation from an active galactic nucleus.  The reduced dust opacities associated with a reduced metallicity will mean that this radiation penetrates further into the cloud, again tending to increase the temperature of the cloud.  Overall, this means that the typical temperature of low-density gas with 1/100 solar metallicity tends to be more like $\sim 100$~K rather than $\sim 10$~K \citep[e.g.][]{GloCla2012c}.  Therefore, it is essential to recognise that the low-opacity calculations in this paper in particular, are only a first step in the direction of helping us to understanding how star formation may vary with metallicity.

Only one other study has used radiation hydrodynamical simulations of star cluster formation to begin to address the  question of how star formation depends on metallicity \citep{Myersetal2011}.  They also changed only the opacity of the matter, and they found no significant variation of the IMF.  However, they only explored opacities ranging over a factor of 20 (from solar metallicity to 1/20 of the solar value), their calculations were unable to resolve the low-mass end of the IMF, and each calculation produced only a few dozen stars, limiting their sensitivity to variations.  They also used relatively large sink particles (radii of 28~AU) so they could not explore the effects of opacity on the properties of multiple stellar systems.  In contrast, we explore an opacity range of 300 (from three times solar to one one-hundredth of solar metallicity), each calculation produces 170 or more protostars (including low-mass brown dwarfs), and we employ sink particles with radii of only 0.5~AU, allowing us to follow the formation of most multiple systems in some detail.

\section{Computational method}
\label{sec:method}

The calculations presented here were performed 
using a three-dimensional smoothed particle
hydrodynamics (SPH) code based on the original 
version of \citeauthor{Benz1990} 
(\citeyear{Benz1990}; \citealt{Benzetal1990}), but substantially
modified as described in \citet*{BatBonPri1995},
\citet*{WhiBatMon2005}, \citet{WhiBat2006},
\cite{PriBat2007}, and 
parallelised using both OpenMP and MPI.

Gravitational forces between particles and a particle's 
nearest neighbours are calculated using a binary tree.  
The smoothing lengths of particles are variable in 
time and space, set iteratively such that the smoothing
length of each particle 
$h = 1.2 (m/\rho)^{1/3}$ where $m$ and $\rho$ are the 
SPH particle's mass and density, respectively
\cite[see][for further details]{PriMon2007}.  The SPH equations are 
integrated using a second-order Runge-Kutta-Fehlberg 
integrator with individual time steps for each particle
\citep{BatBonPri1995}.
To reduce numerical shear viscosity, we use the
\cite{MorMon1997} artificial viscosity
with $\alpha_{\rm_v}$ varying between 0.1 and 1 while $\beta_{\rm v}=2 \alpha_{\rm v}$
\citep[see also][]{PriMon2005}.

\subsection{Equation of state and radiative transfer}

As in \cite{Bate2012}, the calculations presented in this paper were performed using radiation hydrodynamics
with an ideal gas equation of state for the gas pressure
$p= \rho T_{\rm g} \cal{R}/\mu$, where 
$T_{\rm g}$ is the gas temperature, $\mu$ is the mean molecular weight of the gas,
and $\cal{R}$ is the gas constant.  
The thermal evolution takes into account the translational,
rotational, and vibrational degrees of freedom of molecular hydrogen 
(assuming a 3:1 mix of ortho- and para-hydrogen; see
\citealt{Boleyetal2007}).  It also includes molecular
hydrogen dissociation, and the ionisations of hydrogen and helium.  
The hydrogen and helium mass fractions are $X=0.70$ and 
$Y=0.28$, respectively.  For this composition, the mean molecular weight of the gas is initially
$\mu = 2.38$.
The contribution of metals to the equation of state is neglected.

Two temperature (gas and radiation) radiative transfer in the flux-limited
diffusion approximation is implemented as described by \citet{WhiBatMon2005}
and \citet{WhiBat2006}, except that the standard explicit SPH contributions 
to the gas energy equation due to the work and artificial viscosity are used 
when solving the (semi-)implicit energy equations to provide better 
energy conservation.  Energy is generated when work is done on the gas or 
radiation fields, radiation is transported via flux-limited diffusion and energy 
is transferred between the gas and radiation fields depending on their 
relative temperatures, and the gas density and opacity.  The gas and dust
temperatures are assumed to be the same.  

The clouds have free boundaries.  To provide a boundary condition for the
radiative transfer we use the same method as \cite{Bate2009b} and \cite{Bate2012}.  All particles
with densities less than $10^{-21}$~g~cm$^{-3}$ have their gas and radiation 
temperatures set to the initial values of 10.3 K. This gas is two orders of magnitude 
less dense that the initial cloud (see Section \ref{initialconditions}) and, thus, these boundary particles 
surround the region of interest in which the star cluster forms.

\begin{figure}
\centering \vspace{-0.5cm}
    \includegraphics[width=17.0cm]{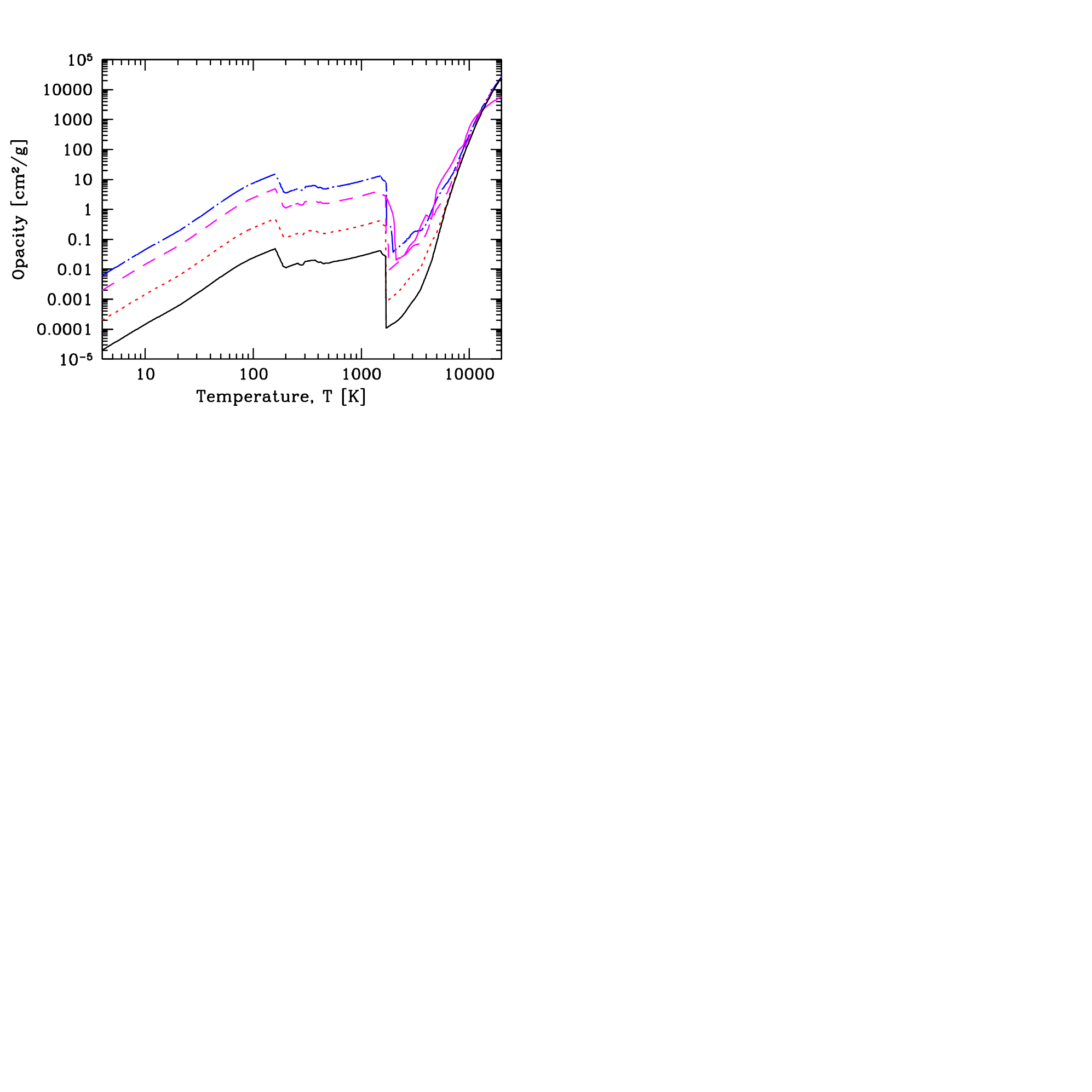}\vspace{-10.5cm}
\caption{Examples of the Rosseland mean opacities  for different metallicities: 1/100 Z$_\odot$ (solid black line), 1/10 Z$_\odot$ (short-dashed red line), Z$_\odot$ (long-dashed magneta line), and 3~Z$_\odot$ (dot-dashed blue line).  We also give the opacity curve as used by Bate (2012) above 1500~K (solid magenta line) which differs slightly from the Z$_\odot$ case because this calculation used the gas opacities of Alexander (1975) rather than Ferguson et al.~(2005).
The opacities are functions of both temperature and density.  For this graph, we plot the opacity as a function of temperature in which the density at each temperature satisfies the equation $(T/10~K)=(\rho/10^{-13}$~{\rm g~cm}$^{-3}$)$^{0.3}$ which (very roughly) approximates the typical densities and temperatures found during the collapse of a molecular cloud core.
}
\label{opacities}
\end{figure}

\subsection{Opacities and metallicity}

\cite{Bate2012} assumed solar metallicity gas, with the opacity set to be the maximum of 
the interstellar dust grain opacity tables of \citet{PolMcKChr1985} and, at higher 
temperatures when the dust is destroyed, the gas opacity
tables of \citet{Alexander1975} (the IVa King model)  \citep[see][for further details]{WhiBat2006}.

In this paper, we wish to investigate how variation of the opacity, due to the molecular gas having
a different metallicity, may affect 
the star formation process.  Therefore, we use a range of metallicities.  For the dust opacities, we
use scaled versions of the \citet{PolMcKChr1985} opacities, in which we assume that the
dust opacity scales linearly with the metallicity.  This assumes that the dust properties are
independent of metallicity and that their number density is simply proportional to the overall
metallicity.  Observations of the gas to dust ratios in other galaxies show that this may be
a reasonable assumption for metallicities $\gsim 1/10$ of the solar value, but at lower metallicities
the gas-to-dust ratio appears to be substantially greater than that given by a strict linear relation
\citep[][and reference therein]{RemyRuyer2014}.

For the gas opacities, we use the newer models of \cite{Fergusonetal2005} with $X=0.70$.
They provide opacities for heavy element abundances from $Z=0$ to $Z=0.1$.  We take the
solar abundance to be ${\rm Z}_\odot=0.02$.  The tables of \citeauthor{Fergusonetal2005} provide 
the logarithm of the Rosseland mean opacities as functions of the logarithms of temperature 
and density.  We use bilinear interpolation from the tables to provide opacities our desired
heavy element abundance.  As in \cite{Bate2012}, the total opacity is set to be the maximum 
of the dust and gas opacities (in the regions of parameter space where the tables overlap).
Typical opacities and their dependence on metallicity are illustrated in Fig.~\ref{opacities}.

In this paper, we compare the results of new calculations with the solar-metallicity calculation of \cite{Bate2012}.
At the same metallicity, the gas opacities of \citet{Alexander1975} are somewhat different than those
of \cite{Fergusonetal2005} (see Fig.~\ref{opacities}).  However,  these opacity differences are only relevant at high temperatures 
(beyond the dust sublimation temperature) and, thus, only affect the regions very close to the protostars, 
not larger scales.  The differences in the opacities are also very small compared with the range of a factor of 300 that we explore in this paper.

\subsection{Sink particles}
\label{sinks}

As in \cite{Bate2012}, using the above realistic equation of state and radiation hydrodynamics
means that as the gas collapses, each of the phases of
protostar formation can be captured \citep{Larson1969}, including the formation
of a first hydrostatic core and its second collapse due to the dissociation of molecular hydrogen.
We could follow the collapse to the formation of a stellar core \citep[e.g.][]{Bate2010,Bate2011},
however, as the collapse proceeds, the timesteps required to obey the
Courant-Friedrichs-Levy criterion become smaller and smaller. 
Because we wish to evolve the large scales over timescales of
$\sim 10^5$ years, we cannot afford to follow the small scales 
(e.g.\ the stellar cores themselves).
Instead, we follow the evolution of each protostar through the first core
phase and into the second collapse (which begins at densities of
$\sim 10^{-7}$~g~cm$^{-3}$), but we insert a sink particle 
\citep{BatBonPri1995} when the density exceeds
$10^{-6}$~g~cm$^{-3}$, approximately three orders of magnitude before the
stellar core would begin to form (density $\sim 10^{-3}$~g~cm$^{-3}$).

As in \cite{Bate2012}, a sink particle is formed by 
replacing the SPH gas particles contained within $r_{\rm acc}=0.5$ AU 
of the densest gas particle in region undergoing second collapse 
by a point mass with the same mass and momentum.  Any gas that 
later falls within this radius is accreted by the point mass 
if it is bound and its specific angular momentum is less than 
that required to form a circular orbit at radius $r_{\rm acc}$ 
from the sink particle.  Thus, gaseous discs around sink 
particles can only be resolved if they have radii $\gsim 1$ AU.
Sink particles interact with the gas only via gravity and accretion.
There is no gravitational softening between sink particles.
The angular momentum accreted by a sink particle is recorded
but plays no further role in the calculation.

Since all sink particles are created within pressure-supported 
fragments, their initial masses are several Jupiter-masses (M$_{\rm J}$), 
as given by the opacity limit for fragmentation \citep{LowLyn1976,Rees1976}.  
Subsequently, they may accrete large amounts of material 
to become higher-mass brown dwarfs ($\lsim 75$ M$_{\rm J}$) or 
stars ($\gsim 75$ M$_{\rm J}$), but {\it all} the stars and brown
dwarfs begin as these low-mass pressure-supported fragments.

In \cite{Bate2012}, sink particles were permitted to merge in the calculation if they
passed within 0.01 AU of each other (i.e., $\approx 2$~R$_\odot$) but no mergers occurred.
In the new calculations performed for this paper, this was increased 
slightly to 0.015~AU (i.e., $\approx 3$~R$_\odot$), since it is likely that 
young protostars accreting at high rates are somewhat larger than the Sun \citep{HosOmu2009}.
Some mergers occurred during the calculations, as will be discussed below.

The benefits and potential problems associated with introducing 
sink particles into radiation hydrodynamical star formation calculations 
have been discussed in detail by \cite{Bate2012} and will not be repeated
here.  The interested reader is referred to this earlier paper for a detailed discussion.
Briefly, we use sink particles from which there is no radiative feedback from inside the sink 
particle radius, but we use as small an accretion radius as is computationally feasible to
minimise missing luminosity.  
\cite{Bate2012} showed empirically that the main effects of radiative feedback
on the fragmentation of a collapsing molecular cloud is captured when using sink particles with 
$r_{\rm acc}=0.5$~AU or smaller.

\begin{table*}
\begin{tabular}{lccccccccccc}\hline
Calculation & Initial Gas & Metallicity & No. Stars & No. Brown  & Mass of Stars \&  & Mean  & Mean & Median & Stellar\\
& Mass  &   & Formed & Dwarfs Formed & Brown Dwarfs & Mass & Log-Mass & Mass & Mergers \\
 & M$_\odot$ &  Z$_\odot$ & & & M$_\odot$ & M$_\odot$ &M$_\odot$ & M$_\odot$ \\ \hline
Metallicity 1/100 & 500 &  0.01 & $\geq 134$& $\leq 64$& 78.3 & $0.40\pm0.04$ & $0.16\pm0.02$ & 0.18 & 21 \\   
Metallicity 1/10 & 500 &  0.1 & $\geq 136$& $\leq 34$& 84.0 & $0.49\pm0.05$ & $0.24\pm0.02$ & 0.25 & 7\\   
B2012 & 500 &  1.0 & 
$\geq 147$& $\leq 36$& 88.2 & $0.48\pm0.05$ & $0.22\pm0.02$ & 0.21 & 0\\  
Metallicity 3 & 500 & 3.0 & $\geq 143$& $\leq 39$& 86.1 & $0.47\pm0.05$ & $0.21\pm0.02$ & 0.19 & 2 \\  \hline 
\end{tabular}
\caption{\label{table1} The parameters and overall statistical results for the calculation of Bate (2012) using solar metallicities and the three new calculations presented here.  The initial conditions for all of the calculations were taken as the state of the Bate (2009a) calculation at 0.70~$t_{\rm ff}$ (initial cloud free-fall times), and all calculations were run to 1.20~$t_{\rm ff}$.  All calculations employ sink particles with $r_{\rm acc}=0.5$~AU and no gravitational softening.  Brown dwarfs are defined as having final masses less than 0.075 M$_\odot$.  The numbers of stars (brown dwarfs) are lower (upper) limits because some brown dwarfs were still accreting when the calculations were stopped.  Changing the opacities results in no significant difference in the statistical quantities for opacities corresponding to metallicities $\ge 1/10$~Z$_\odot$, except for the numbers of stellar mergers.  However, the lowest opacity calculation converts gas into stars at a slower rate and produces objects slightly more quickly which results in a slightly higher fraction of low-mass objects than the other calculations. }
\end{table*}

\subsection{Initial conditions}
\label{initialconditions}

The initial conditions for the calculations presented in this paper are taken from the calculation of 
\cite{Bate2009a} and are identical to those of \cite{Bate2012}.  We followed the collapse 
of an initially uniform-density molecular cloud containing 500 M$_\odot$ of molecular gas.
The cloud's radius was 0.404 pc (83300 AU) giving an initial density of 
$1.2\times 10^{-19}$~g~cm$^{-3}$.  At the initial temperature of 10.3 K, the mean 
thermal Jeans mass was 1 M$_\odot$ (i.e., the cloud contained 500 thermal Jeans masses).  
Although the cloud was uniform in density, we imposed an initial 
supersonic `turbulent' velocity field in the same manner
as \citet*{OstStoGam2001} and \cite{BatBonBro2003}.  
We generated a divergence-free random Gaussian velocity field with 
a power spectrum $P(k) \propto k^{-4}$, where $k$ is the wavenumber.  
In three dimensions, this results in a
velocity dispersion that varies with distance, $\lambda$, 
as $\sigma(\lambda) \propto \lambda^{1/2}$ in agreement with the 
observed Larson scaling relations for molecular clouds 
\citep{Larson1981}.
The velocity field was generated on a $128^3$ uniform grid and the
velocities of the particles were interpolated from the grid.  
The velocity field was normalised so that the kinetic energy 
of the turbulence was equal to the magnitude of the gravitational potential 
energy of the cloud.
Thus, the initial root-mean-square (rms) Mach number of the turbulence 
was ${\cal M}=13.7$.
The initial free-fall time of the cloud was $t_{\rm ff}=6.0\times 10^{12}$~s or 
$1.90\times 10^5$ years.

Molecular clumps of this mass, radius, and internal velocity dispersion
are not found in nearby star-forming regions, but these initial conditions are very similar 
to the clumps found in many infrared dark clouds 
(e.g. \citealt*{RatJacSim2006}; \citealt{Battersbyetal2010,Raganetal2012a}a, \citeyear{Raganetal2012b}b).

As for the calculation performed for \cite{Bate2012}, since the initial conditions for the 
calculation are identical to those of
\cite{Bate2009a} and including radiative transfer does not alter the 
temperature of the gas significantly until just before the first protostar forms,
the early evolution of the cloud was not repeated for any of the calculations presented
in this paper.  Instead, all of the radiation
hydrodynamical calculations were begun from a dump file taken from the 
original \cite{Bate2009a}  calculation at $t=0.70~t_{\rm ff}$, just before the maximum density 
exceeded $10^{-15}$~g~cm$^{-3}$.

Three new calculations were performed for this paper, with opacities relevant for gas with metallicities of 1/100, 1/10, and 3 times solar 
(i.e. $Z=2\times 10^{-4}$, 0.002, and 0.06), assuming a linear dependence of the dust opacity on metallicity as discussed above.  When combined with the calculation presented by \cite{Bate2012},
this gives four calculations whose metallicities and opacities range over a factor of 300.  
We restrict the highest metallicity to three times the solar value 
for two reasons.  First, there are not many stars known with higher metallicities.  Second, the contribution of metals to the equation of state of the gas is neglected.  While this is standard practice for solar-metallicity star formation calculations, the approximation will break down for sufficiently high metallicities.  

At the other end of the metallicity range, we do not study metallicities less than 1/100 solar because our method ignores the many other effects that a decreased the metallicity would have on the thermal behaviour of the gas that we listed in Section \ref{introduction}.  These effects become more and more important as the metallicity is decreased.  In particular, our method assumes that the gas and dust are thermally well coupled and, therefore, that the gas cools primarily via thermal dust emission.  These are standard assumptions for star formation calculations that begin with dense molecular gas with solar metallicity, but they quickly break down at low metallicites and/or low densities.  In fact, even though our initial cloud has a relatively high density, the $Z=1/100$~Z$_\odot$ calculation is almost certainly very unrealistic and even the $Z=1/10$~Z$_\odot$ calculation may only be realistic in extreme circumstances.  This can be easily seen by taking a simple model  \citep[e.g.][]{Goldsmith2001} for the gas temperature of our initial conditions.  We can estimate the gas temperature deep inside a molecular cloud (i.e. where the interstellar radiation field is sufficiently attenuated by dust absorption), by balancing the cosmic ray heating of the gas by its cooling due to line emission and collisions with dust grains.  Following \cite{Goldsmith2001}, we take the cosmic ray heating rate (measured in  ${\rm erg~cm}^{-3}~{\rm s}^{-1}$) to be
\begin{equation}
\label{eq:cr} 
\Gamma_{\rm gas, cr} = 10^{-27} n_{\rm H2},
\end{equation}
and the gas-dust cooling rate to be
\begin{equation}
\label{eq:gd} 
\Lambda_{\rm gd} = 2 \times 10^{-33}n_{\rm H2}^2 \left( \frac{Z}{{\rm Z}_\odot} \right) \left(\frac{T_{\rm g}}{10~K}\right)^{1/2}\left( T_{\rm g} - T_{\rm d} \right),
\end{equation}
where $n_{\rm H2}$ is the number density of molecular hydrogen, $T_{\rm d}$ is the dust temperature, and we have assumed that the gas-dust cooling rate scales linearly with metallicity.  \cite{Goldsmith2001} parameterises the line cooling as $\Lambda_{\rm gas, line} = \alpha (T_{\rm g}/10~{\rm K})^\beta$, where $\alpha$ and $\beta$ are tabulated as functions of $n_{\rm H2}$.  For our initial conditions ($n_{\rm H2} \approx 3\times 10^4$~cm$^{-3}$), we can estimate  $\alpha \approx 1\times 10^{-23}$ and $\beta \approx 2.85$ at solar metallicities from \citeauthor{Goldsmith2001}'s values, and again we assume that the line cooling scales linearly with metallicity.  These equations allow us to estimate the equilibrium gas temperature for our initial cloud by solving
\begin{equation}
\label{eq:solve}
\Gamma_{\rm gas, cr}  - \Lambda_{\rm gas, line} -  \Lambda_{\rm gd} = 0,
\end{equation}
but we need to estimate the dust temperature to evaluate the last term. Assuming the thermal coupling between the gas and the dust is weak (which is the regime we are interested in), the dust temperature can be estimated by balancing the absorption of radiation from the interstellar radiation field (attenuated by extinction) and thermal dust emission.  Because the latter scales as $\sim T_{\rm d}^6$, the dust temperature is only weakly dependent on the local level of the interstellar radiation field and will lie within a factor of two of  $10$~K for a wide range of assumptions \citep{Goldsmith2001}.  Taking $T_{\rm d} =10$~K and solving equation \ref{eq:solve}, we find that for our initial cloud conditions $T_{\rm g}\approx 30$ for $Z=0.1~{\rm Z}_\odot$, and $T_{\rm g}\approx 70$ for $Z=0.01~{\rm Z}_\odot$.  Clearly these temperatures are much higher than the initial temperatures assumed by \cite{Bate2009a,Bate2012}.  We note that for $Z={\rm Z}_\odot$, we obtain $T_{\rm g}\approx 13$ (i.e. the gas and dust are well coupled and the initial conditions used by the earlier papers are reasonable).  To obtain the same equilibrium temperature for the $Z=0.1~{\rm Z}_\odot$ case as the $Z={\rm Z}_\odot$ case, we need to assume that the cosmic ray heating rate is a factor of ten lower than that given by equation \ref{eq:cr}.  Therefore, the $Z=0.1~{\rm Z}_\odot$ calculation can only be considered to be reasonable in star-forming regions where the cosmic ray heating rate is much lower than that estimated locally.  The $Z=0.01~{\rm Z}_\odot$ calculation is unlikely to be realistic in any situation -- even if the cosmic ray flux was 100 times lower, the fact that the gas and dust are thermally decoupled will mean that even compressional heating of the gas due to the turbulent initial conditions is likely to result in much higher temperatures than those assumed by \cite{Bate2009a, Bate2012}.  Because of this fact, we only discuss the results of the $Z=0.01~{\rm Z}_\odot$ calculation in Sections \ref{sec:clouds} and \ref{sec:imf}, simply to illustrate the extreme case of using very low opacities.  It should not be taken as a realistic calculation, and even the results from the $Z=0.1~{\rm Z}_\odot$ calculation should be treated with care.

\subsection{Resolution}
\label{resolution}

The local Jeans mass must be resolved throughout the calculation 
to model fragmentation correctly (\citealt{BatBur1997, Trueloveetal1997, Whitworth1998, Bossetal2000}; \citealt*{HubGooWhi2006}).  
The original barotropic calculation of \cite{Bate2009a} used $3.5 \times 10^7$ particles to model the 
500-M$_\odot$ cloud and resolve the Jeans mass down to its minimum 
value of $\approx 0.0011$ M$_\odot$ (1.1 M$_{\rm J}$) at the maximum 
density during the isothermal phase of the collapse, 
$\rho_{\rm crit} = 10^{-13}$ g~cm$^{-3}$.  Using radiation hydrodynamics results in
temperatures at a given density no less than those given by the original 
barotropic equation of state \citep[e.g.][]{WhiBat2006} and, thus, the Jeans mass is also resolved in
the calculations presented here.

The calculations were performed on the University of Exeter 
Supercomputer, an SGI Altix ICE 8200 that was upgraded in 2011 to dual 2.80~GHz Intel Westmere nodes.  
Using 96 compute cores, each of the new calculations required between 0.7 and 1 million CPU hours (i.e. 10--13 months of wall time).



\begin{figure*}
\centering \vspace{-0.8cm} \hspace{-0.4cm}
    \includegraphics[width=18cm]{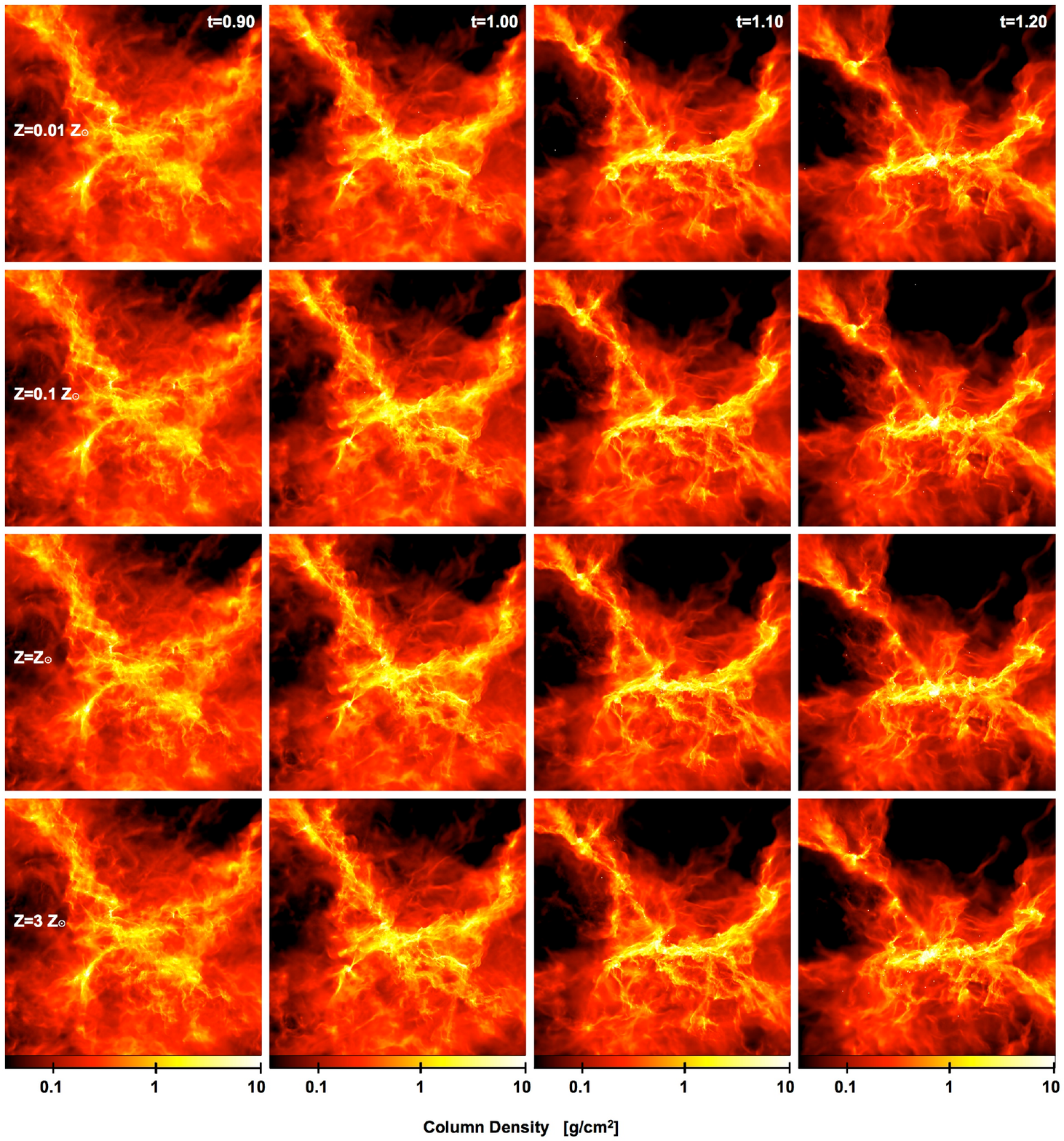} \vspace{-7cm}
\caption{The global evolution of column density for each of the radiation hydrodynamical calculations from time $t=0.90$ to 1.20$t_{\rm ff}$.  From top to bottom, the rows show the evolution of the calculations with opacities corresponding to metallicities of 1/100, 1/10, 1, and 3 times solar, respectively.  Shocks lead to the dissipation of the turbulent energy that initially supports the cloud, allowing parts of the cloud to collapse.  By $t=1.10t_{\rm ff}$ each calculation has produced several main sub-clusters. Each panel is 0.4 pc (82,500 AU) across.  Time is given in units of the initial free-fall time, $t_{\rm ff}=1.90\times 10^5$ yr.  The panels show the logarithm of column density, $N$, through the cloud, with the scale covering $-1.4<\log N<1.0$ with $N$ measured in g~cm$^{-2}$. White dots represent the stars and brown dwarfs.}
\label{global_density}
\end{figure*}

\begin{figure*}
\centering \vspace{-0.8cm} \hspace{-0.4cm}
     \includegraphics[width=18cm]{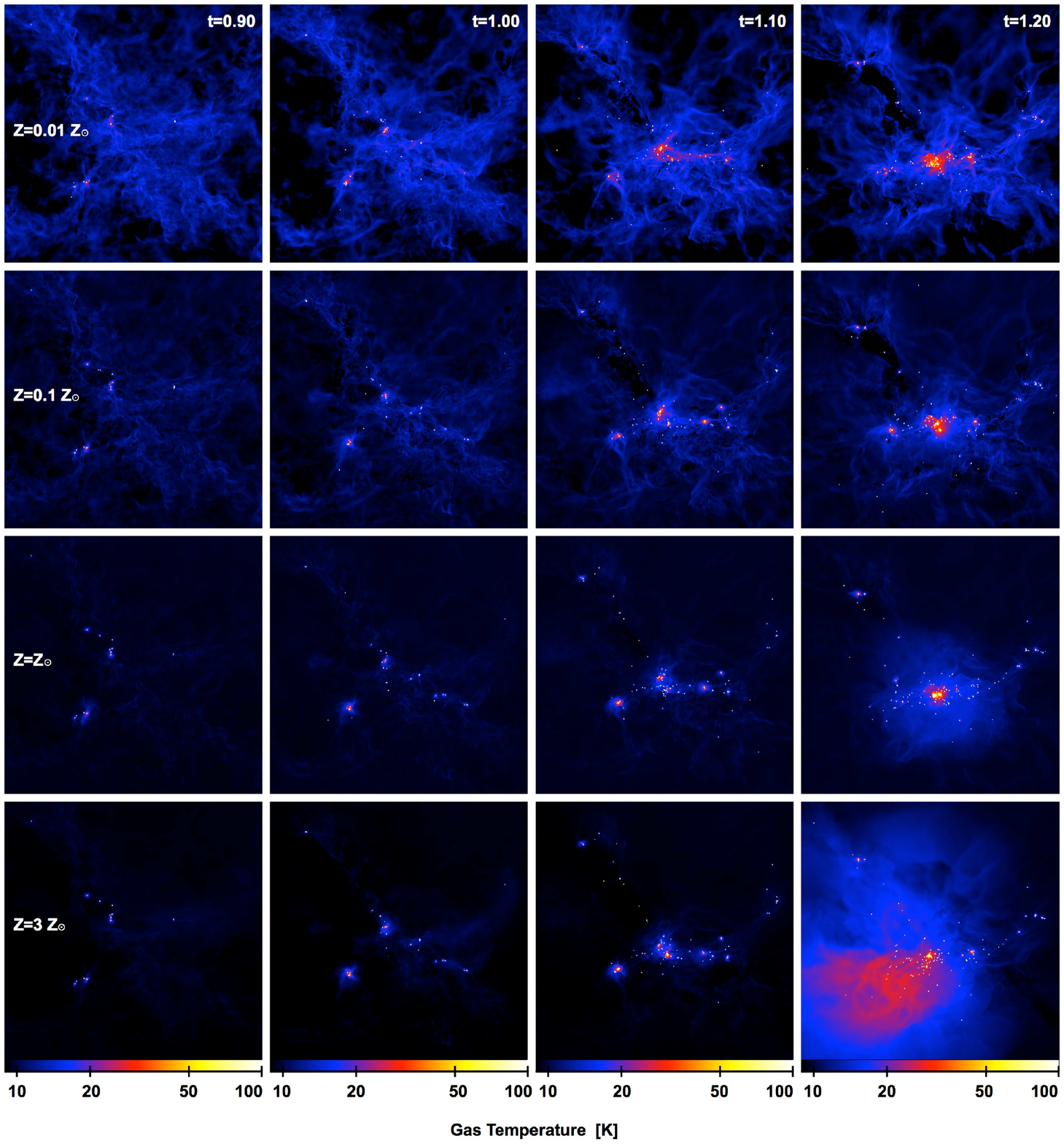} \vspace{-7cm}
\caption{The global evolution of gas temperature for each of the radiation hydrodynamical calculations from time $t=0.90$ to 1.20$t_{\rm ff}$.  From top to bottom, the rows show the evolution of the calculations with opacities corresponding to metallicities of 1/100, 1/10, 1, and 3 times solar, respectively.  At early times, the gas in the shocks is hotter with lower opacities as the dust cooling is inefficient.  At later times, the higher opacity, more optically-thick clouds are heated more strongly by the thermal feedback from the protostars. 
Each panel is 0.4 pc (82,500 AU) across.  Time is given in units of the initial free-fall time, $t_{\rm ff}=1.90\times 10^5$ yr.  The panels show the logarithm of mass weighted gas temperature, $T_{\rm g}$, through the cloud, with the scale covering $9-100$~K.  White dots represent the stars and brown dwarfs.}
\label{global_temp}
\end{figure*}

\section{Results}
\label{sec:results}

We present results from four radiation hydrodynamical calculations that are essentially identical, except for their opacities.  Assuming a linear dependence of the dust opacity on metallicity, the opacities correspond to metallicities $Z = 0.01, 0.1, 1,$ and $3~{\rm Z}_\odot$.  See Table \ref{table1} for a summary of the statistics from each of the calculations, including the numbers of stars and brown dwarfs produced, the total mass that was converted to stars and brown dwarfs, and the mean and median stellar masses. We use the same analysis methods as those used by \cite{Bate2009a} and \cite{Bate2012}, but we discuss fewer properties.  We consider the mass functions for each calculation individually.  However, we only discuss other statistical properties for the three calculations with the highest opacities ($Z \geq 0.1~{\rm Z}_\odot$) because, as discussed in Section \ref{initialconditions}, we consider the lowest opacity calculation to be too unrealistic.  We discuss the multiplicities, and the separations and mass ratios of the multiple systems for individual calculations.  In Section \ref{sec:combined}, we construct a combined sample consisting of the 535 stars produced by the three calculations with $Z \geq 0.1~{\rm Z}_\odot$.  In addition to presenting the mass function, multiplicity, separations and mass ratios of the systems in this combined sample, we also consider the eccentricity distributions of multiple systems and the orientations of orbits in triple systems or stars and discs in binary systems.  We do not consider the accretion histories or kinematics of the stars or the distributions of closest encounters at all.  These omissions are made for the purpose of brevity, but we note that we find no evidence that these properties vary with opacity.

\begin{figure*}
\centering
    \includegraphics[width=5.8cm]{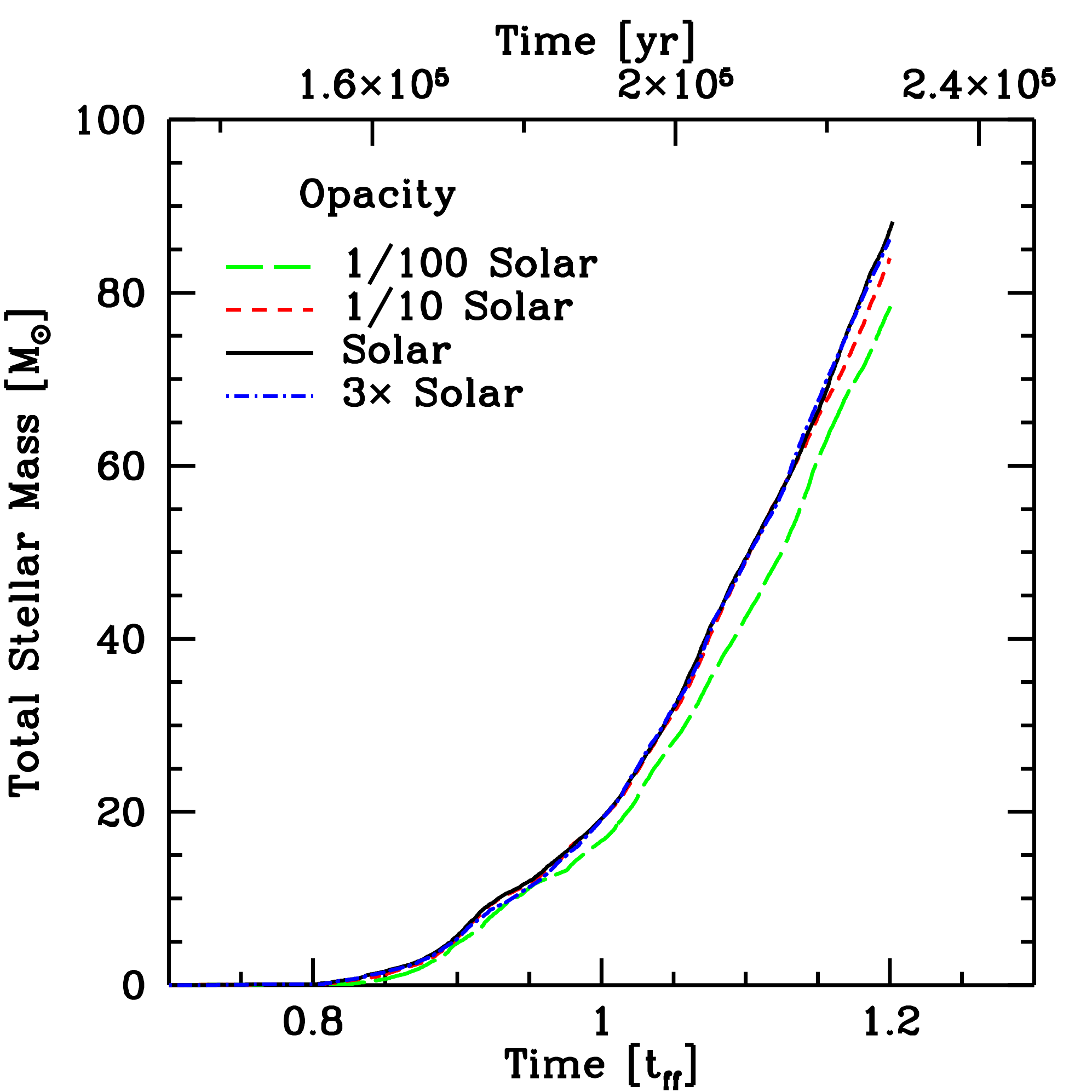}
    \includegraphics[width=5.8cm]{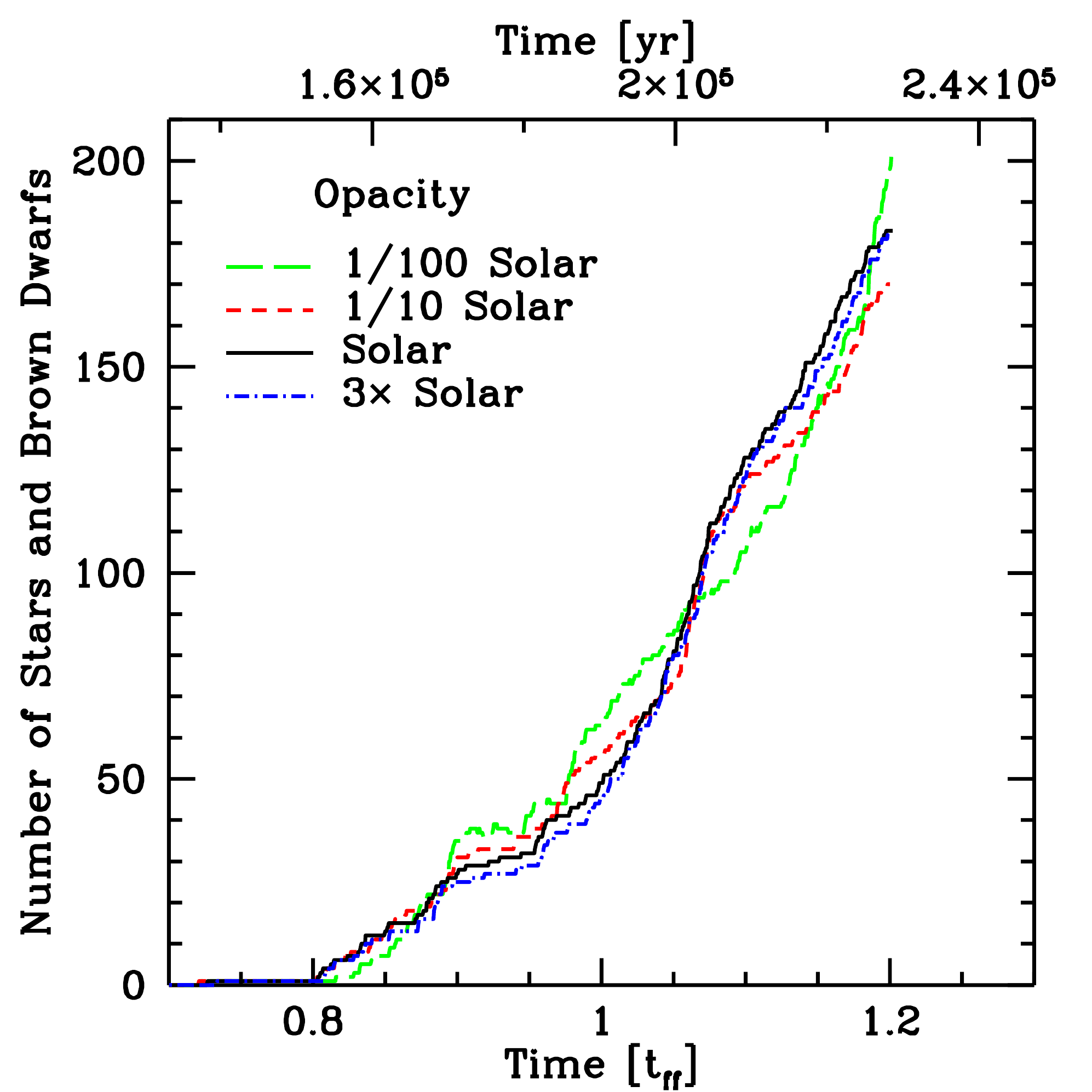}
    \includegraphics[width=5.8cm]{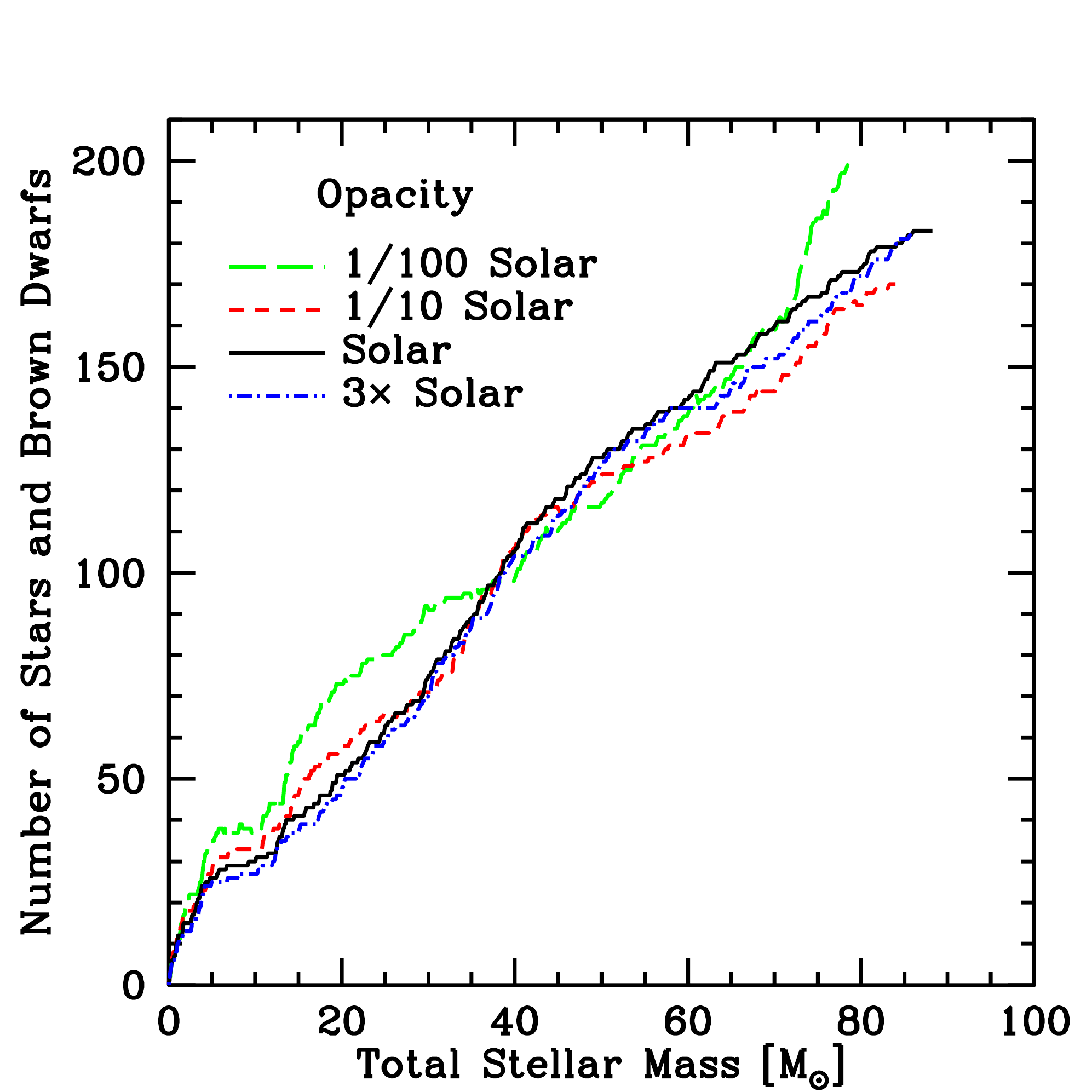}
\caption{The star formation rates obtained for each of the four radiation hydrodynamical calculations.  We plot: the total stellar mass (i.e. the mass contained in sink particles) versus time (left panel), the number of stars and brown dwarfs (i.e. the number of sink particles) versus time (centre panel), and the number of stars and brown dwarfs versus the total stellar mass (right panel).  The different line types are for opacities corresponding to metallicities of 1/100 (long dashed), 1/10 (dotted), 1 (solid), and 3 (dot-dashed) times solar. Time is measured from the beginning of the calculation in terms of the free-fall time of the initial cloud (bottom) or years (top). The rate at which mass is converted into stars is almost independent of the opacity, but for the lowest opacity the rate appears to be slightly lower and the rate at which new stars and brown dwarfs form seems to be more variable.  }
\label{massnumber}
\end{figure*}

\subsection{The evolution of the star-forming clouds}
\label{sec:clouds}

As mentioned in Section \ref{initialconditions}, all the calculations were begun from a dump file at $t=0.70~t_{\rm ff}$ from the original calculation of Bate (2009a), before the maximum density exceeded $10^{-15}$ g~cm$^{-3}$.  Before this point the initial `turbulent' velocity field had generated density structure in the gas, some of which was collected into dense cores which had begun to collapse.  Those readers interested in this early phase should refer to \cite{Bate2009a} for figures and further details.

In the solar-metallicity calculation, the first sink particle was inserted at $t=0.727~t_{\rm ff}$.  In the low-opacity calculations, the first sink particles were inserted slightly earlier at $t=0.722~t_{\rm ff}$ for both the $Z=0.01~{\rm Z}_\odot$ and $Z=0.1~{\rm Z}_\odot$ calculations.  In the highest opacity calculation, the first sink particle was inserted at $t=0.733~t_{\rm ff}$.  The general increase in the formation time of the first object with increasing opacity occurs because cooling at high densities becomes slower with the increased optical depth and the protostars spend longer in the first hydrostatic core phase of evolution before undergoing the second collapse.

In the panels of Figs.~\ref{global_density} and \ref{global_temp}, we present snapshots during the evolution of each calculation from $t=0.90-1.20~t_{\rm ff}$ (we omit earlier times because they show little of interest).  Fig.~\ref{global_density} displays the column density (using a red-yellow-white colour scale), while Fig.~\ref{global_temp} displays the mass-weighted temperature in the cloud (using a blue-red-yellow-white colour scale).  Animations of each of the calculations can be downloaded from http://www.astro.ex.ac.uk/people/mbate/~.  As in the calculations of \cite{BonBatVin2003}, \cite{Bate2009a}, and \cite{Bate2012}, the star formation in the clouds occurs in small groups, embedded within larger-scale filaments that are formed by the turbulent initial conditions.  Initially, each group contains only a few low-mass objects and the heating of the surrounding gas is limited to their immediate vicinity (a few thousand AU).  However, as the stellar groups grow in number and some of the stars grow to greater masses, the heating can be seen to extend to larger and larger scales, particularly in the higher opacity calculations.

Despite the very different evolution of the temperature distributions in the four calculations (Fig.~\ref{global_temp}), the evolution of the column density is very similar on large scales ($\gsim 0.01$~pc).  This is because the gravitational and turbulent energies are dominant over the thermal energy on large scales.  Differences in the thermal energy only have large effects on scales of thousands of AU where the fragmentation of discs and filaments may be inhibited from occurring \citep[c.f.][]{Bate2009b, Offneretal2009}.  There are two main effects that produce different temperature distributions with the different opacities.  The first is visible even at early times in the left panels of Fig.~\ref{global_temp}.  Much of this material is optically-thin in all of the calculations, but in the lower opacity calculations the matter (gas and dust) is less well coupled to the radiation field and so the matter does not cool as effectively.  Thus, the temperatures of 15-20~K occupying much of the volume in the $Z=0.01$ and 0.1~Z$_\odot$ calculations are due to inefficient cooling of the shocks formed by the supersonic motions in the clouds.  We emphasise, however, that the gas temperatures in the low-opacity calculations are still lower than would be expected in more realistic calculations that take account of the thermal decoupling of the gas and dust.  In the $Z=0.01$~Z$_\odot$ in particular, although the dust temperature would be expected to be $\approx 10$~K, the gas would be essentially uncoupled from the dust except at very high densities, and is expected to have temperatures of $\sim 100$~K \citep[e.g.][]{GloCla2012c}.  The second main difference is visible at late times (the right-most panels in Fig.~\ref{global_temp}).  At $t\approx 1.15~t_{\rm ff}$ two subclusters of protostars merge near the centre of the cloud.  The merger of the dense clumps and dynamical interactions between protostars lead to increased protostellar accretion rates and a burst of radiation which heats the central region of the cloud.  However, with higher opacities, the cloud is more optically thick and the radiation trapped by the cloud heats the matter significantly.  This results in heating of the cloud to distances of $\approx 0.1$~pc from the centre in the $Z=1~{\rm Z}_\odot$ calculation \citep[reported by][]{Bate2012}, and even more dramatic heating to distances of $\approx 0.3$~pc in the $Z=3~{\rm Z}_\odot$ calculation.

\begin{figure*}
\centering  \vspace{-0.5cm}
    \includegraphics[width=8cm]{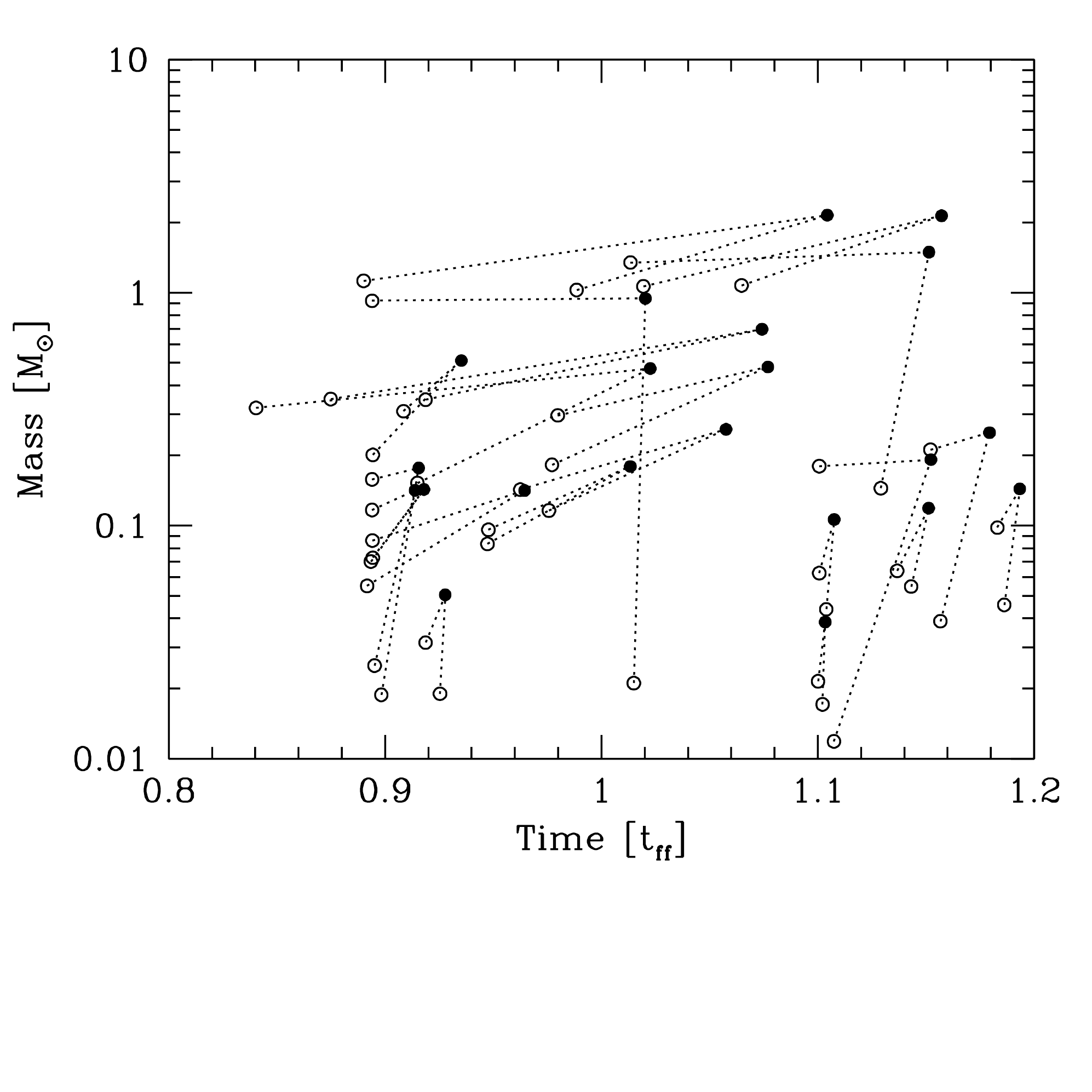}
    \includegraphics[width=8cm]{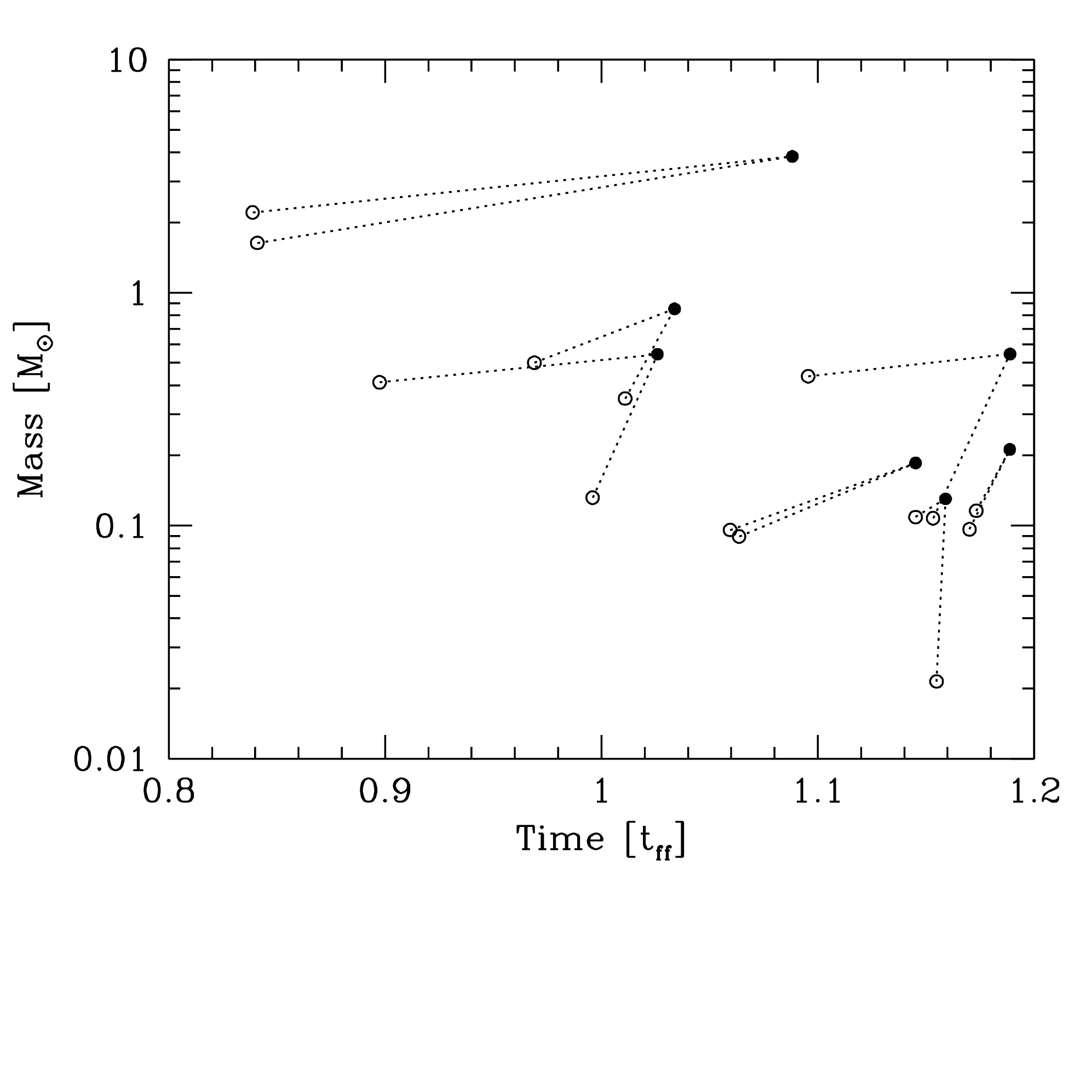} \vspace{-1.5cm}
\caption{A summary of the protostellar mergers that occurred in the low opacity $Z= 0.01~{\rm Z}_\odot$ (left) and $Z= 0.1~{\rm Z}_\odot$ (right) calculations.  For each merger, we plot the masses of each of the two progenitors at the time they were formed as open circles, and each of these is linked by a dotted line to a filled circle which is plotted at the time of the merger and gives the mass of the merged object.  It can be seen that brown dwarfs, low-mass stars, and super-solar stars are all involved in protostellar mergers.  There is no plot for the $Z=3~{\rm Z}_\odot$ calculation because only two mergers occur, involving objects of 0.15 and 0.075 M$_\odot$ and 1.5 and 0.8~M$_\odot$, repectively.  Both of these occurred at $t \approx 1.14~t_{\rm ff}$.   }
\label{mergers}
\end{figure*}

We follow the calculations to $1.20~t_{\rm ff}$ (228\,280 yr) which is $9.0 \times 10^4$ years after the star formation began.  At this stage $78 - 88$ M$_\odot$ of gas (16--18\%) has been converted into 170--198 stars and brown dwarfs, depending on the calculation (Table \ref{table1}).
Despite the huge variation in opacity (a factor of 300), the amount of gas converted into stars and brown dwarfs, the numbers of objects, and their mean and median masses shows little variation between the four calculations (Table \ref{table1}, columns 7--9).  The median mass varies by 42\% at most (from 0.18 to 0.25 M$_\odot$), while the mean mass varies by 25\% at most (from 0.40 to 0.49 M$_\odot$) and the values are within the $2\sigma$ formal statistical uncertainties of each other.  We also provide the mean values of the logarithm of the masses.  It is interesting to note that the two calculations with the most different characteristic masses are the two calculations with the lowest opacities.  The $Z=0.1~{\rm Z}_\odot$ has the highest mean and median stellar masses, while the $Z=0.01~{\rm Z}_\odot$ calculation has the lowest.  The $Z=0.1~{\rm Z}_\odot$ calculation also produces the most massive star (4.56~M$_\odot$), while the other calculations produce stars with masses up to 2.92 M$_\odot$ ($Z=0.01~{\rm Z}_\odot$), 3.84 M$_\odot$ ($Z=1~{\rm Z}_\odot$), and 3.71 M$_\odot$ ($Z=3~{\rm Z}_\odot$).

We investigate the significance of these variations in the mass distributions in the next section.  Before that, we examine the star formation rates in terms of mass and the numbers of stars and brown dwarfs (Fig.~\ref{massnumber}).  In the left panel, we plot the total stellar mass as a function of time for each of the calculations.  It can be seen that in terms of stellar mass, there is a slow star formation rate of $\approx 5\times 10^{-4}$~M$_\odot$~yr$^{-1}$ from $\approx 0.8-1.0~t_{\rm ff}$ followed by an increase to $\approx 2\times 10^{-3}$~M$_\odot$~yr$^{-1}$ after $\approx 1.0~t_{\rm ff}$.  The star formation rate is quite constant after this transition.  Gas is converted into stars slightly more slowly in the lowest opacity calculation, presumably due to the slightly higher temperatures in the low-density gas due to the less effective cooling, but the other three calculations are indistinguishable.  In terms of the number of stars and brown dwarfs versus time (centre panel), there is no obvious difference between the calculations.  This is also true of the number of objects versus the total stellar mass (right panel), except that the lowest opacity calculation seems to have two `bursts' where it forms a lot of objects at $t \approx 1 t_{\rm ff}$ and again near the end of the calculation.  The latter burst is partially responsible for the lower median and mean stellar masses --- at $t=1.18~t_{\rm ff}$, the median and mean masses for the $Z=0.01~{\rm Z}_\odot$ calculation are 0.20 and 0.44 M$_\odot$, respectively.

Finally, we note that in the each of the new calculations some stellar mergers occurred.  The $Z=0.01$~Z$_\odot$ calculation had 21 stellar mergers (i.e. $\approx 10$\% of the stars), the $Z=0.1$~Z$_\odot$ calculation had 7 stellar mergers, and the $Z=3$~Z$_\odot$ calculation had only 2 stellar mergers.  No mergers occurred in the solar metallicity calculation, but this calculation had a slightly smaller merger radius (2~R$_\odot$ rather than 3~R$_\odot$).  Examining the records of the sink particle trajectories from the solar metallicity calculation, if the larger merger radius was used one merger would have occurred.   Thus, we find that stellar mergers occur more frequently with decreasing opacity.  The reason for the opacity dependence of the numbers of mergers will be discussed in Section \ref{discussion}. In Fig.~\ref{mergers} we plot the masses and times involved in the stellar mergers.  There is no apparent dependence of the frequency of mergers on stellar mass --- sink particles with masses ranging from 12 Jupiter masses to 2.2~M$_\odot$ were involved in collisions with 18 of the 30 mergers involving stars with masses in the $0.1-1$~M$_\odot$ range.  As expected, most of the brown dwarfs involved in stellar mergers are involved shortly after they first form, since the reason they have low masses is that they haven't had long to accrete to higher masses \citep{BatBon2005}.  \cite*{BonBatZin1998} proposed that protostellar collisions may be an important ingredient in the formation of massive stars ($M \gsim 10$~M$_\odot$) in a cluster environment.  Here we find that protostars of all masses may undergo collisions, but we also note that by the end of the $Z=0.01~{\rm Z}_\odot$ calculation two of its four most massive stars have suffered collisions, and that in the $Z=0.1~{\rm Z}_\odot$ calculation the most massive star was also formed in through a collision.




\subsection{The initial mass function}
\label{sec:imf}

\begin{figure*}
\centering \vspace{-0.5cm}
\mbox {    \includegraphics[width=8.cm]{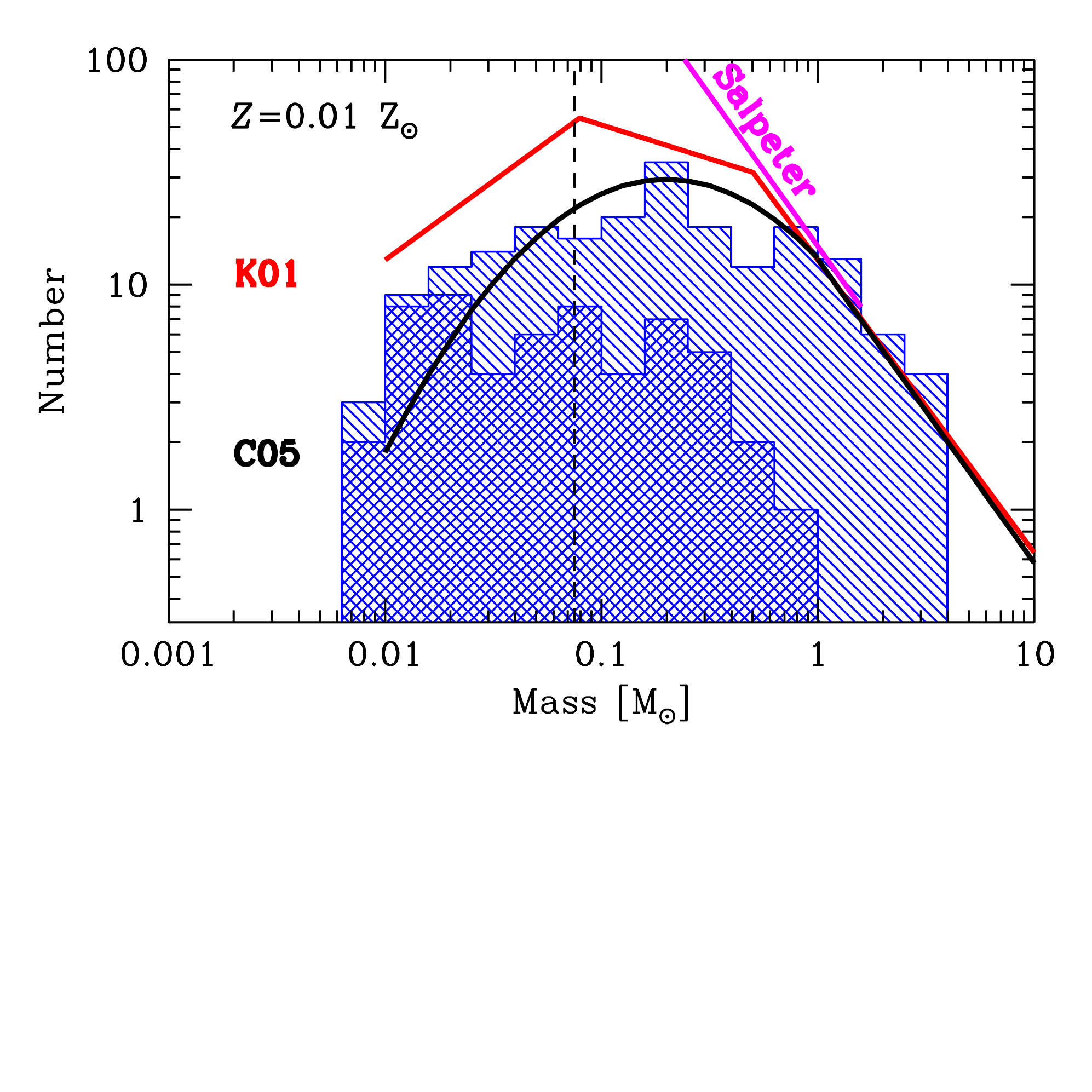}
    \includegraphics[width=8.cm]{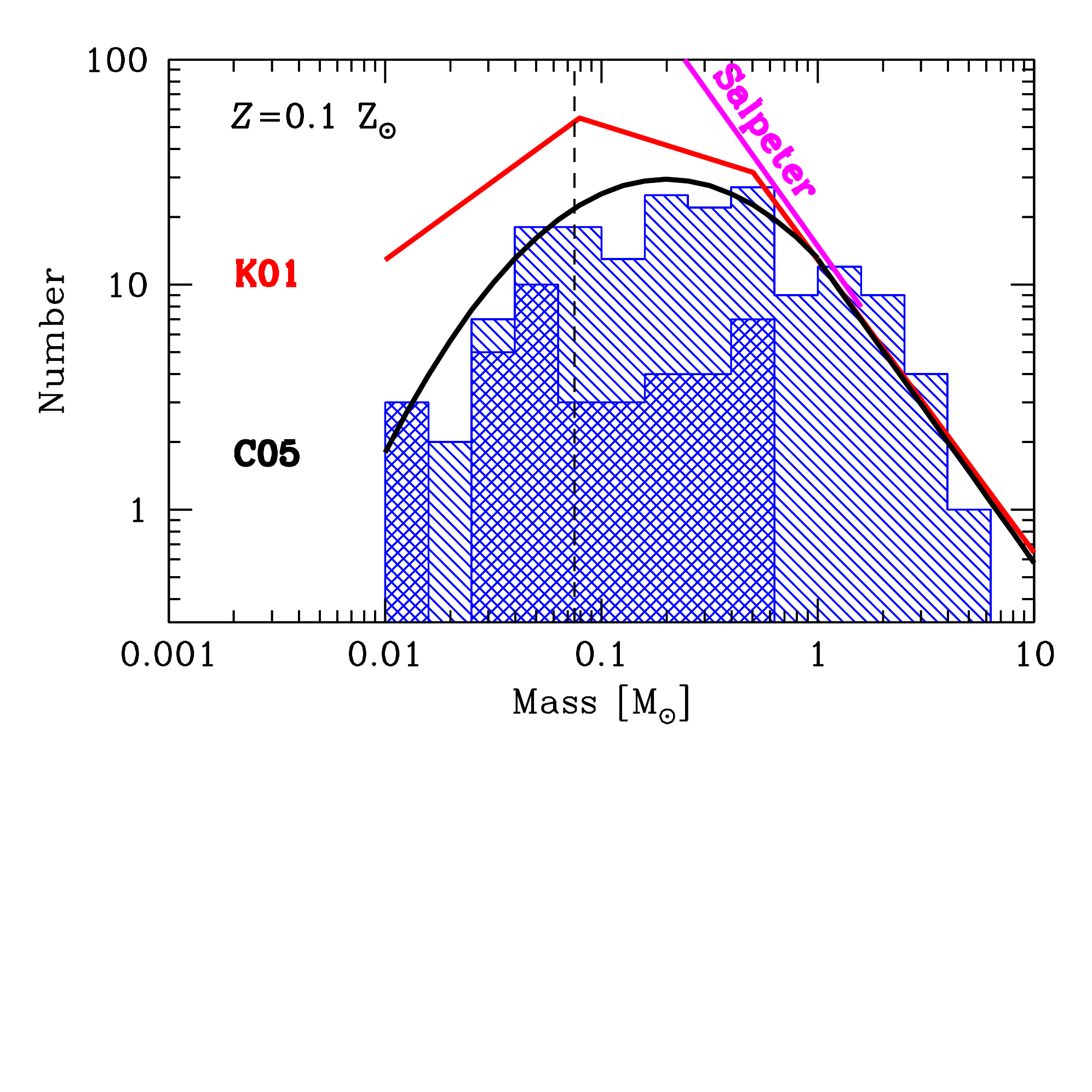} }\vspace{-2.8cm}
    \includegraphics[width=8.cm]{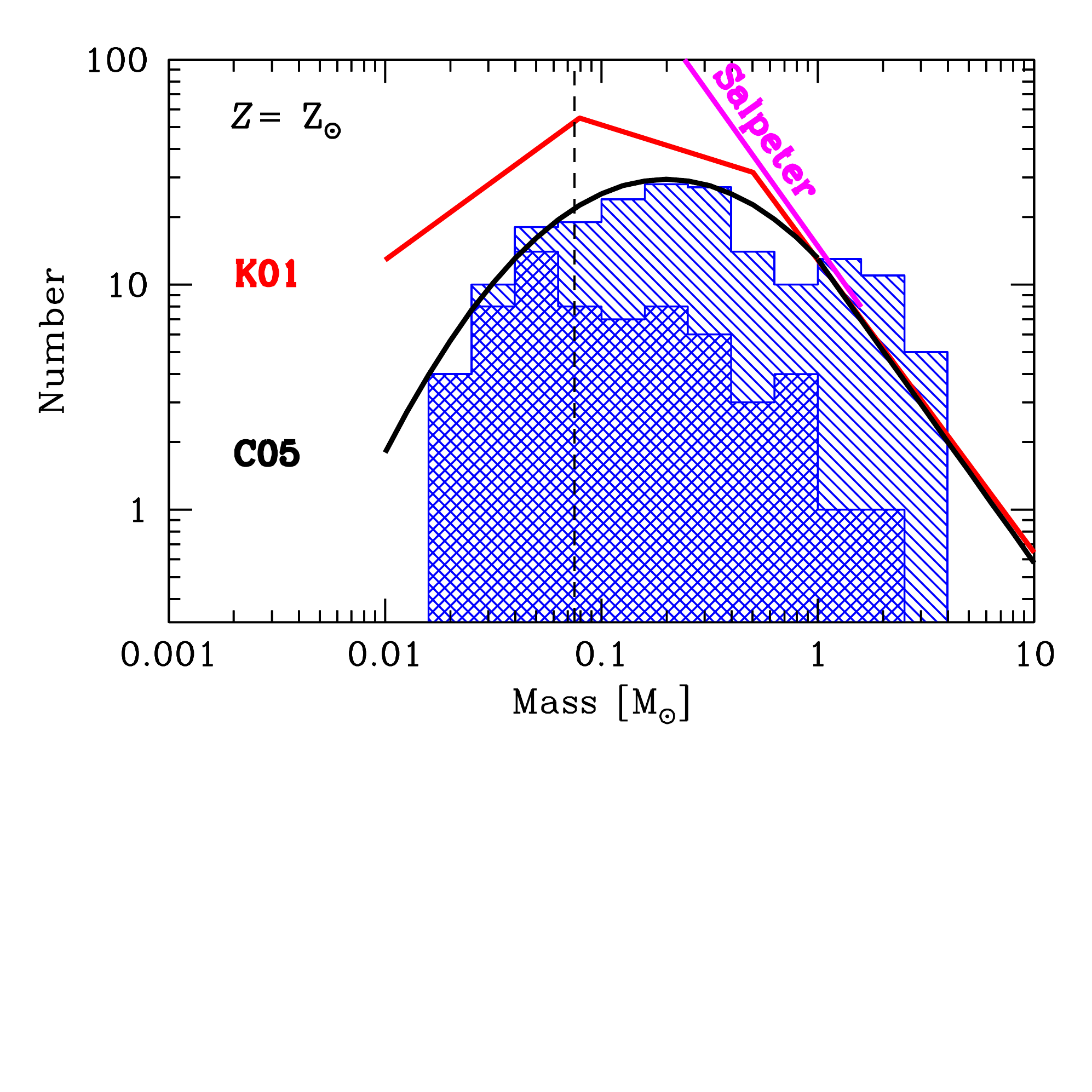}
    \includegraphics[width=8.cm]{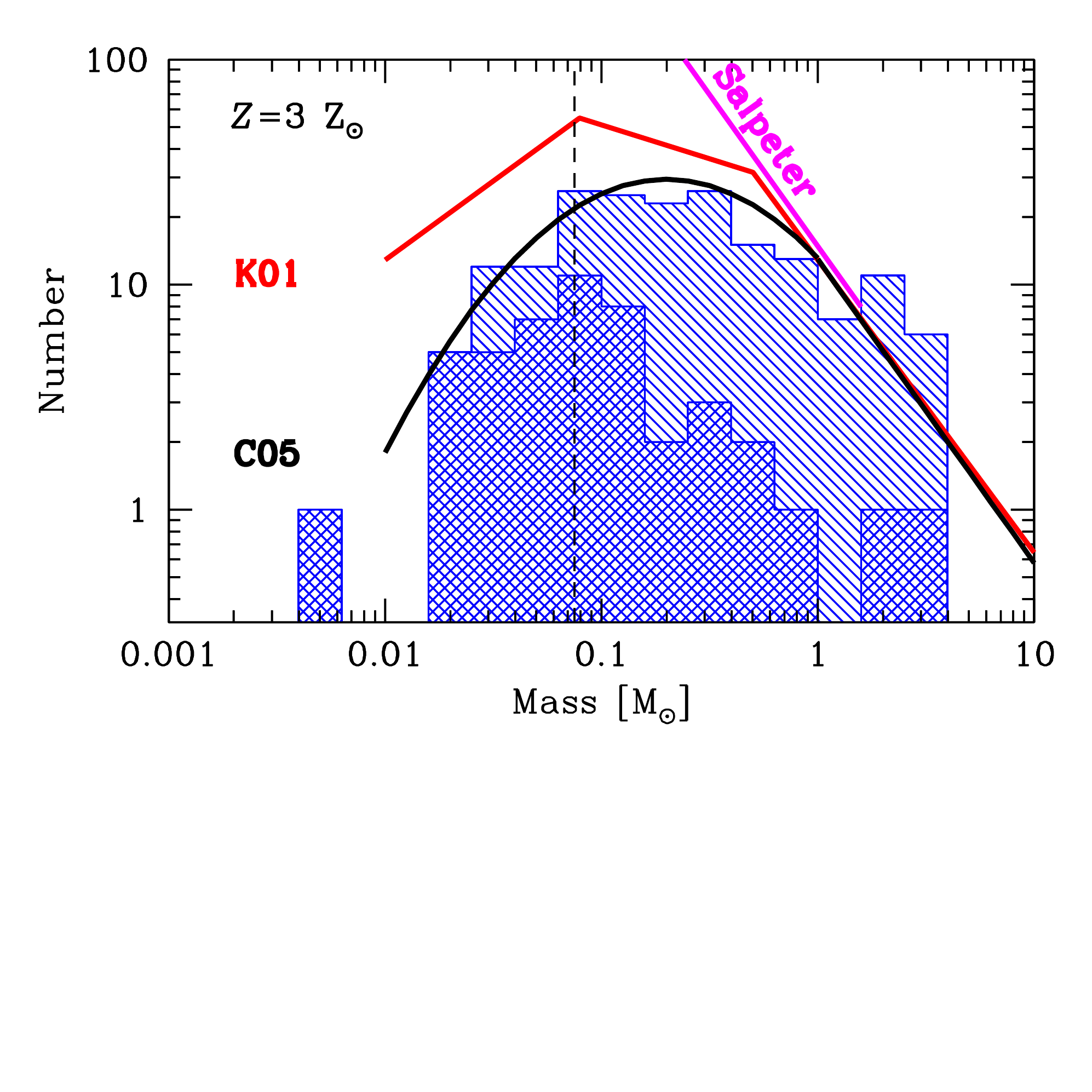}
    \vspace{-2.8cm}
\caption{Histograms giving the initial mass functions of the stars and brown dwarfs produce by the four radiation hydrodynamical calculations, each at $t=1.20t_{\rm ff}$.  The double hatched histograms are used to denote those objects that have stopped accreting (defined as accreting at a rate of less than $10^{-7}$~M$_\odot$~yr$^{-1}$), while those objects that are still accreting are plotted using single hatching.  Each of the mass functions are in good agreement with the Chabrier (2005) fit to the observed IMF for individual objects.  Two other parameterisations of the IMF are also plotted: Salpeter (1955) and Kroupa (2001). Despite the opacity varying by a factor of up to 300 between the calculations, the IMFs are indistinguishable, though we note that there is a potential excess of brown dwarfs for the calculation with the lowest opacity ($Z=0.01~{\rm Z}_\odot$). }
\label{imfcomp}
\end{figure*}

\begin{figure}
\centering
    \includegraphics[width=8.0cm]{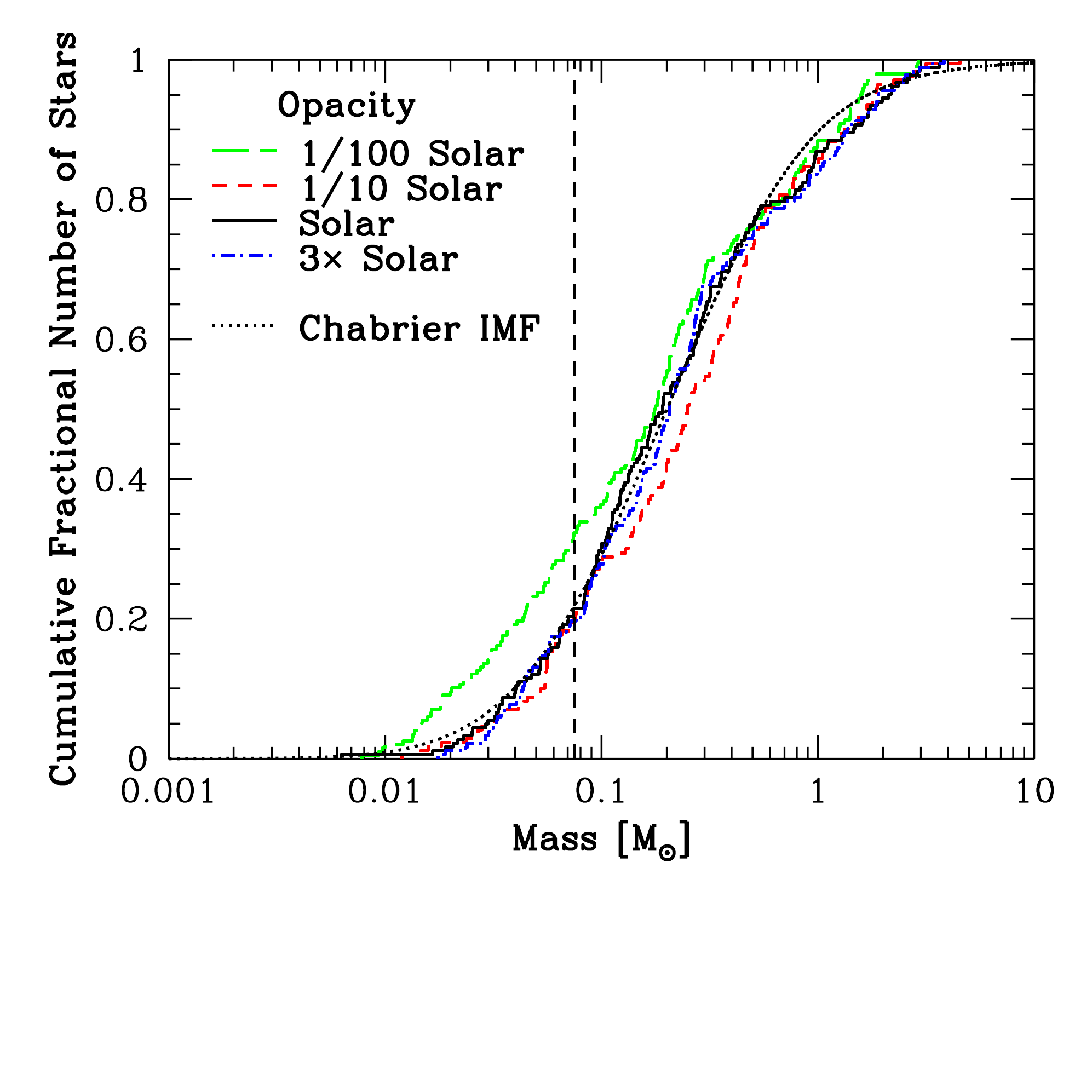}\vspace{-1.8cm}
\caption{The cumulative stellar mass distributions produced by the four radiation hydrodynamical calculations with different opacities, corresponding to metallicities of $Z=0.01~{\rm Z}_\odot$ (green long-dashed line), $Z=0.1~{\rm Z}_\odot$ (red dashed line), $Z={\rm Z}_\odot$ (black solid line), and $Z=3~{\rm Z}_\odot$ (blue dot-dashed line).  We also plot the Chabrier (2005) IMF (black dotted line).  The vertical dashed line marks the stellar/brown dwarf boundary.  The form of the stellar mass distribution does not vary significantly with different opacities: Kolmogorov-Smirnov tests show that even the two most different distributions ($Z=0.01~{\rm Z}_\odot$ and $Z=0.1~{\rm Z}_\odot$) have a 1.2\% probability of being drawn from the same underlying distribution (equivalent to a $\approx 2.5 \sigma$ difference).  However, in the lowest opacity case there does seem to be a slight excess of brown dwarfs.}
\label{cumimf_comp}
\end{figure}

\begin{table}
\begin{tabular}{lccccc}\hline
Mass Range ~ [M$_\odot$]& Single & Binary  & Triple & Quadruple  \\ \hline
\multicolumn{5}{c}{Metallicity $Z=0.1~{\rm Z}_\odot$}  \\ \hline
\hspace{0.83cm}$M<0.03$       &      7     &     0     &      0      &     0    \\
$0.03\leq M<0.07$      &    17     &   1    &       0     &      0   \\
$0.07\leq M<0.10$      &      11      &    0      &     0     &      0   \\
$0.10\leq M<0.20$      &      10     &     2    &       1     &      0   \\
$0.20\leq M<0.50$      &      25     &     9    &       0     &     4   \\
$0.50\leq M<0.80$      &     6      &     2      &     1     &      2   \\
$0.80\leq M<1.2$        &       0      &     2      &     0     &      1   \\
\hspace{0.83cm}$M>1.2$        &       0       &    4      &     4     &      2   \\ \hline
\multicolumn{5}{c}{Metallicity $Z={\rm Z}_\odot$}  \\ \hline
\hspace{0.83cm}$M<0.03$       &      7     &     0     &      0      &     0    \\
$0.03\leq M<0.07$      &    20     &   0    &       0     &      0   \\
$0.07\leq M<0.10$      &      8      &    3      &     0     &      0   \\
$0.10\leq M<0.20$      &      17     &     7    &       1     &      0   \\
$0.20\leq M<0.50$      &      21     &     9    &       2     &     2   \\
$0.50\leq M<0.80$      &     5      &     2      &     0     &      1   \\
$0.80\leq M<1.2$        &       2      &     1      &     1     &      0   \\
\hspace{0.83cm}$M>1.2$        &       4       &    6      &     1     &      4   \\ \hline
\multicolumn{5}{c}{Metallicity $Z=3~{\rm Z}_\odot$}  \\ \hline
\hspace{0.83cm}$M<0.03$       &      8     &     0     &      0      &     0    \\
$0.03\leq M<0.07$      &    24     &   0    &       0     &      0   \\
$0.07\leq M<0.10$      &      13      &    1      &     0     &      0   \\
$0.10\leq M<0.20$      &      18    &     5    &       2     &      0   \\
$0.20\leq M<0.50$      &      18     &     5    &       3    &     2   \\
$0.50\leq M<0.80$      &     4      &     2      &     0     &      2   \\
$0.80\leq M<1.2$        &       3      &     3      &     0     &      1   \\
\hspace{0.83cm}$M>1.2$        &       4       &    1      &     3     &      3   \\ \hline
All masses, 3 calculations                    &   252   &     65     &     19   &      24           \\ \hline
\end{tabular}
\caption{\label{tablemult} The numbers of single and multiple systems for different primary mass ranges at the end of the three radiation hydrodynamical calculations with the highest opacities ($Z \ge 0.1~{\rm Z}_\odot$). }
\end{table}

In Fig.~\ref{imfcomp}, we compare the differential IMFs at the end of the four radiation hydrodynamical calculations with different opacities.  Each is compared with the parameterisations of the observed IMF given by \citet{Chabrier2005}, \citet{Kroupa2001}, and \citet{Salpeter1955}.  There is no obvious difference between the mass functions, indicating that the IMFs produced by the calculations do not depend strongly on opacity.  We do note, however, that the calculation with the lowest opacity seems to produce somewhat more brown dwarfs than the other calculations.  

The cumulative IMFs from the four calculations are compared with each other in Fig.~\ref{cumimf_comp}.  Also plotted is the parameterisation of the observed IMF given by \citet{Chabrier2005}.  The apparent excess of brown dwarfs in the lowest opacity calculation can be seen, with around 30\% of the objects being brown dwarfs in the $Z=0.01~{\rm Z}_\odot$ calculation, while only 20\% of the objects are brown dwarfs in the other three calculations.  However, apart from this difference, there is little to distinguish between the four IMFs.  This conclusion is born out by running Kolmogorov-Smirnov tests on each pair of distributions.  Formally, they are all indistinguishable from each other.  The two most different mass functions are those from the $Z=0.01~{\rm Z}_\odot$ and $Z=0.1~{\rm Z}_\odot$ calculations, but even these have a 1.2\% probability of being drawn from the same underlying distribution (i.e. they only differ at the level of approximately $2.5 \sigma$).  Each of the four mass functions are also indistinguishable from the \cite{Chabrier2005} IMF.

Note that, in fact, the calculations produce protostellar mass functions (PMFs) rather than IMFs \citep{FleSta1994a, FleSta1994b,McKOff2010} because some of the objects are still accreting when the calculation is stopped.  In this paper, we refer to each mass function as an `IMF' because we compare it to the observed IMF since the PMF cannot yet be determined observationally.  However, it should be noted that how a PMF transforms into the IMF depends on the accretion history of the protostars and how the star formation process is terminated.  \cite{Bate2012} found that the distribution of stellar masses in the solar-metallicity calculation evolved such that, no matter when the distribution was examined, it was always consistent with being drawn from a constant underlying mass function.  Within the statistical uncertainties, the median stellar mass and the overall shape of the distribution did not change with time.  The same is also true of the three new calculations presented here.  Therefore, in stopping the calculations at $t=1.20~t_{\rm ff}$ we do not seem to have stopped at a special point in the evolution of the clusters.  Rather, at any given time, the IMFs are always `fully-formed', even though the number of stars and the maximum stellar mass both increase with time.

\begin{figure}
\centering \vspace{-0.2cm}
    \includegraphics[width=8.cm]{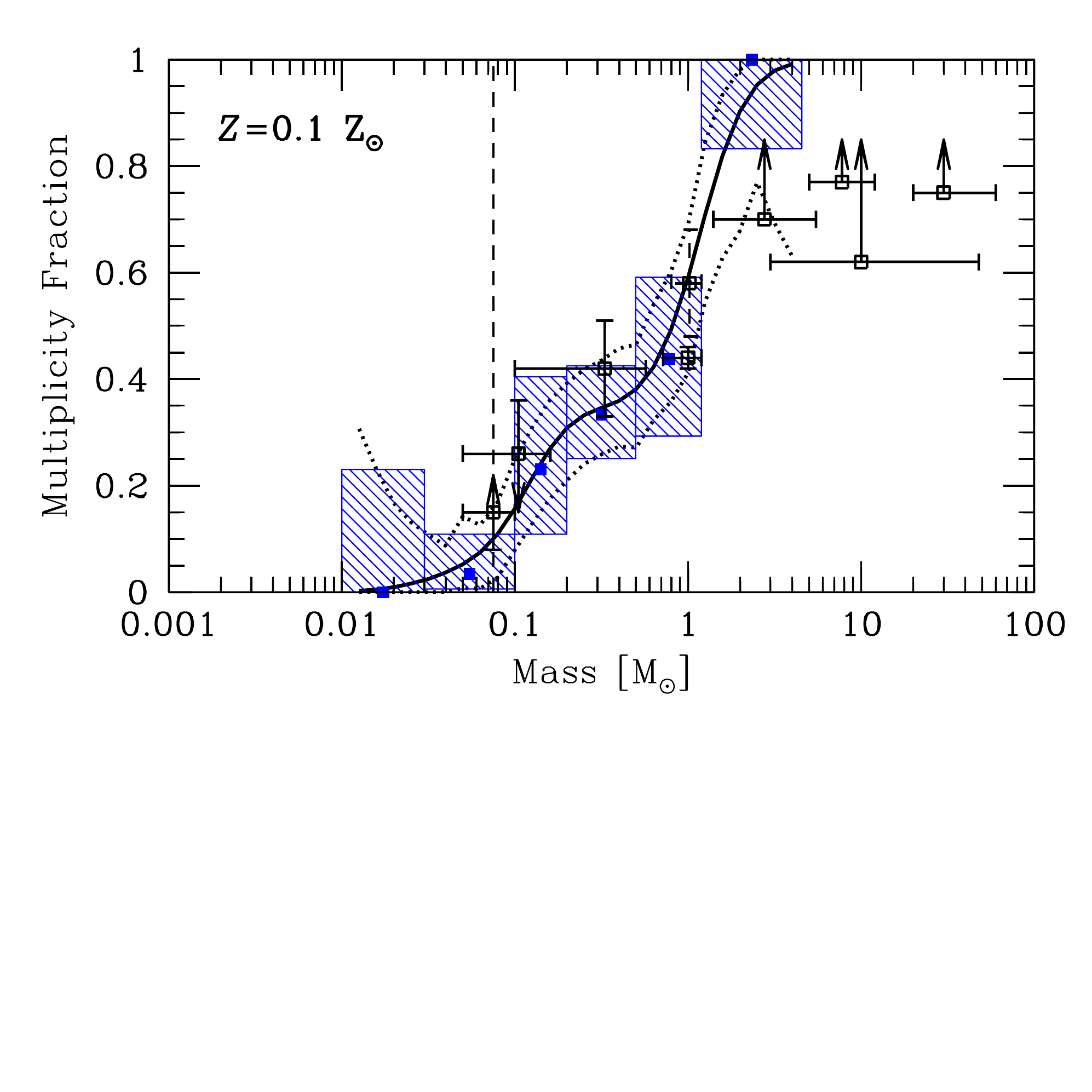}\vspace{-3cm}
    \includegraphics[width=8.cm]{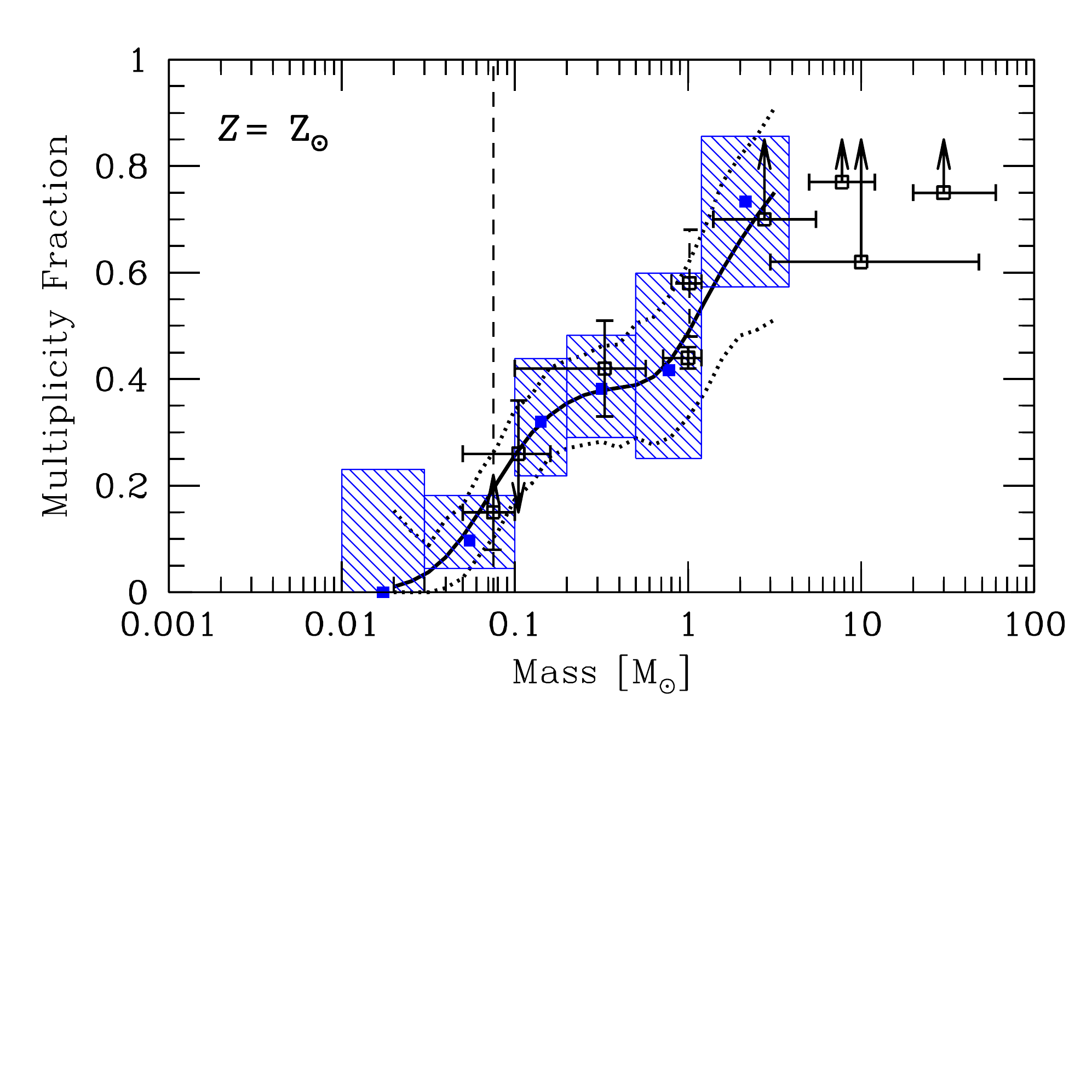}\vspace{-3cm}
    \includegraphics[width=8.cm]{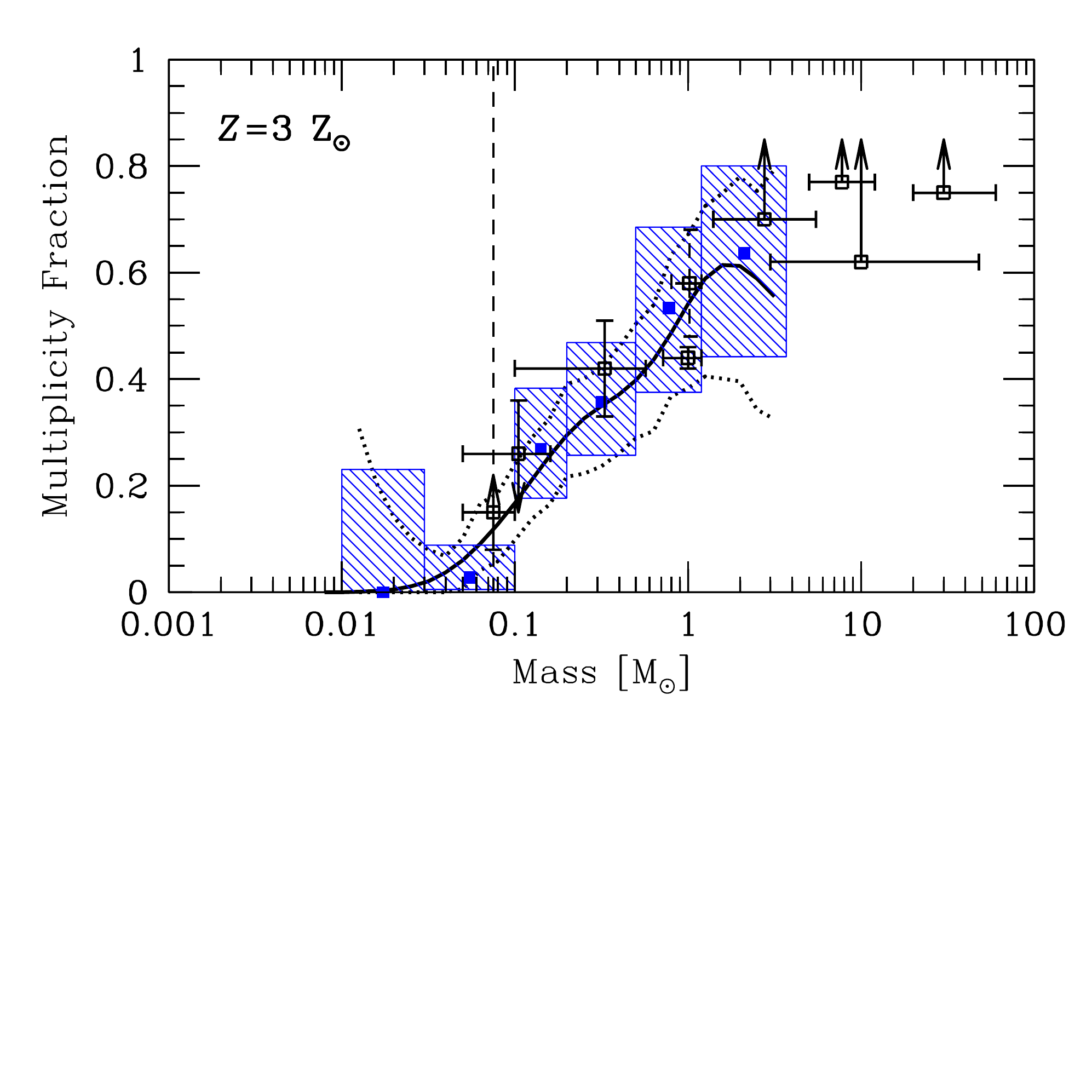}\vspace{-3cm}
\caption{Multiplicity fraction as a function of primary mass at the end of each of the three radiation hydrodynamical calculations with the highest opacities.  The blue filled squares surrounded by shaded regions give the results from the calculations with their statistical uncertainties.  The thick solid lines give the continuous multiplicity fractions from the calculations computed using a sliding log-normal average and the dotted lines give the approximate $1\sigma$ confidence intervals around the solid line. The open black squares with error bars and/or upper/lower limits give the observed multiplicity fractions from the surveys of Close et al. (2003), Basri \& Reiners (2006), Fisher \& Marcy (1992), Raghavan et al. (2010), Duquennoy \& Mayor (1991), Kouwenhoven et al. (2007), Rizzuto et al. (2013), Preibisch et al. (1999) and Mason et al. (1998), from left to right.  Note that the error bars of the Duquennoy \& Mayor (1991) results have been plotted using dashed lines since this survey has been superseded by Raghavan et al. (2010).  The observed trend of increasing multiplicity with primary mass is well reproduced by all calculations.  Note that because the multiplicity is a steep function of primary mass it is important to ensure that similar mass ranges are used when comparing the simulations with observations. }
\label{multiplicity}
\end{figure}

\subsection{Multiplicity as a function of primary mass}
\label{multiplicity_sec}

The formation of multiple systems in a radiation hydrodynamical calculation and the evolution of their properties (e.g.\@ separations) during their formation was discussed in some detail by \cite{Bate2012} and will not be repeated here.  As mentioned above, in this paper our primary purpose is to investigate the dependence of the resulting statistical properties of stars and brown dwarfs on opacity.

As in \cite{Bate2009a} and \cite{Bate2012}, to quantify the fraction of stars and brown dwarfs that are in multiple systems we use the multiplicity fraction, $mf$, defined as a function of stellar mass.  We define this as
\begin{equation}
mf = \frac{B+T+Q}{S+B+T+Q},
\end{equation}
where $S$ is the number of single stars within a given mass range and, $B$, $T$, and $Q$ are the numbers of binary, triple, and quadruple systems, respectively, for which the primary has a mass in the same mass range.  This measure of multiplicity is relatively insensitive to both observational incompleteness (e.g.\@ if a binary is found to be a triple it is unchanged) and further dynamical evolution (e.g.\@ if an unstable quadruple system decays the numerator only changes if it decays into two binaries)  \citep{HubWhi2005, Bate2009a}.

\begin{table}
\begin{center}
\begin{tabular}{lccccl}\hline
Object Number & Mass & $t_{\rm form}$ & Accretion Rate\\
& [M$_\odot$] & [$t_{\rm ff}$] & [M$_\odot$~yr$^{-1}$] \\ \hline
 1 & 1.3749 & 0.7266 & $3.18\times10^{-5}$ \\
  2 & 1.8626 & 0.8034 & $2.3\times10^{-6}$ \\
  3 & 2.2732 & 0.8046 & 0 \\
  4 & 1.3284 & 0.8066 & $3.0\times10^{-6}$ \\
  5 & 2.5311 & 0.8120 & $4.3\times10^{-6}$ \\
\hline
\end{tabular}
\end{center}
\caption{\label{tablestars} For each of the four calculations, we provide online tables of the stars and brown dwarfs that were formed, numbered by their order of formation, listing the mass of the object at the end of the calculation, the time (in units of the initial cloud free-fall time) at which it began to form (i.e. when a sink particle was inserted), and the accretion rate of the object at the end of the calculation (precision  $\approx 10^{-7}$~M$_\odot$~yr$^{-1}$).  The first five lines of the table for the solar metallicity calculation are provided above.}
\end{table}

\begin{table*}
\begin{tabular}{lcccccccccccccl}\hline
Object Numbers & No. of &  No. in & $M_{\rm max}$ & $M_{\rm min}$  & $M_1$ & $M_2$  & $q$ & $a$  & P & $e$  & Relative Spin  & Spin$_1$ & Spin$_2$ \\
& Objects & System & & & & & & & & & or Orbit  & -Orbit & -Orbit \\
&  &  & & & & & & & & & Angle & Angle & Angle\\
     & &    & [M$_\odot$] & [M$_\odot$] & [M$_\odot$] & [M$_\odot$] &  & [AU] & [yr] & & [deg] & [deg] & [deg] \\ \hline
25,             26            & 2 & 4 & 1.807 & 1.233 & 1.807 & 1.233 & 0.682 &        0.91 & 0.50 & 0.610 &    7 &   31 &   27 \\
64,             79            & 2 & 3& 0.837 & 0.103  & 0.837 & 0.103 & 0.123 &        1.95 & 2.82 & 0.242 &   48 &   28 &   57 \\
44,             82            & 2 & 4 &  1.028 & 0.908 & 1.028 & 0.908 & 0.883 &       14.28 & 38.79 & 0.008 &    8 &   35 &   31 \\
4,             84            & 2 & 4 &   1.328 & 1.062 & 1.328 & 1.062 & 0.800 &       19.29 & 54.80 & 0.018 &   12 &   40 &   41 \\
\\
(25,  26),        37            & 3 & 4 &  1.807 & 1.233 & 3.041 & 1.684 & 0.554 &        5.53 & 5.98 & 0.188 &  34 &   -- &    -- \\
(64,  79),        55            & 3 & 3 &  0.859 & 0.103 & 0.939 & 0.859 & 0.914 &       18.08 & 57.31 & 0.104 &    4 &    -- &    -- \\
\\
(4,  84),        (44,  82)       & 4 & 4 &   1.328 & 0.908 & 2.391 & 1.935 & 0.810 &      138.88 & 786.64 & 0.033 &      -- &    -- &    -- \\
((25,  26),  37),   40            & 4 & 4 &  3.379 & 1.233 & 4.725 & 3.379 & 0.715 &      176.55 & 823.79 & 0.308 &     -- &    -- &    -- \\
\hline
\end{tabular}
\caption{\label{tablemultprop} For each of the three calculations with the highest opacities  ($Z \ge 0.1~{\rm Z}_\odot$), we provide online  tables of the properties of the multiple systems at the end of each calculation.  The structure of each system is described using a binary hierarchy.  For each `binary' we give the masses of the most massive star $M_{\rm max}$ in the system, the least massive star $M_{\rm min}$ in the system, the masses of the primary $M_1$ and secondary $M_2$, the mass ratio $q=M_2/M_1$, the semi-major axis $a$, the period $P$, the eccentricity $e$.  For binaries, we also give the relative spin angle, and the angles between orbit and each of the primary's and secondary's spins.  For triples, we give the relative angle between the inner and outer orbital planes. For binaries, $M_{\rm max}=M_1$ and $M_{\rm min}=M_2$.  However, for higher-order systems $M_1$ gives the combined mass of the most massive sub-system (which may be a star, binary, or a triple) and $M_2$ gives the combined mass of the least massive sub-system (which also may be a star, a binary, or a triple).  Multiple systems of the same order are listed in order of increasing semi-major axis.  As examples, we provide selected lines from the table from the solar metallicity calculation.}
\end{table*}

The method we use for identifying multiple systems is the same as that used by \cite{Bate2009a} and \cite{Bate2012}, and a full description of the algorithm is given in the former paper.
When analysing the simulations, some subtleties arise.  For example, many `binaries' are in fact members of triple or quadruple systems and some `triple' systems are components of quadruple or higher-order systems.  From this point on, unless otherwise stated, we define the numbers of multiple systems as follows.  The number of binaries excludes those that are components of triples or quadruples.  The number of triples excludes those that are members of quadruples.  However, higher order systems are ignored (e.g. a quintuple system may consist of a triple and a binary in orbit around each other, but this would be counted as one binary and one triple).  We choose quadruple systems as a convenient point to stop as it is likely that most higher order systems will not be stable in the long-term and would decay if the cluster was evolved for many millions of years.  

The numbers of single and multiple stars produced by each of the three calculations with the highest opacities are given in Table \ref{tablemult} following these definitions.  We do not provide the statistics of the multiple systems for the lowest opacity calculation because the thermal behaviour of the gas is so unrealistic in this calculation that we do not believe that it is worthwhile discussing the calculation any further.  We do note, however, that, as with the mass function, we find no statistically significant difference between the lowest opacity calculation and the other calculations in terms of the multiple systems that are produced.

\cite{Bate2012} provided a table of the properties of each of the multiple systems produced by the solar metallicity calculation.  However, in total the three calculations with the highest opacities produce 108 multiple systems.  Therefore, rather than include them with the paper, this information is provided electronically in ASCII tables.  For all  four calculations, we provide tables that list the masses, formation times, and final accretion rates of the stars and brown dwarfs (see Table \ref{tablestars} for an example).  These tables are given file names such as {\tt Table3\_Stars\_Metal01.txt} for the $Z=0.1~{\rm Z}_\odot$ calculation.  For each calculation with $Z \geq 0.1~{\rm Z}_\odot$, we also provide tables than list the properties of each multiple system (see Table \ref{tablemultprop} for an example).  These tables are given file names such as {\tt Table4\_Multiples\_Metal3.txt} for the $Z=3~{\rm Z}_\odot$ calculation.

The overall multiplicities for all stars and brown dwarfs from each of the three remaining calculations are 32\%, 32\%, and 26\%, each with $1\sigma$ uncertainties of $\pm 5$\% for opacities corresponding to metallicities of 1/10, 1, and 3 times solar, respectively.  Therefore, there is no significant overall dependence of the multiplicity on opacity.

However, observationally, it is clear that the fraction of stars or brown dwarfs that are in multiple systems increases with stellar mass \citep[e.g.][]{KraHil2012,DucKra2013}, with different surveys examining primaries with different masses: massive stars \citep{Masonetal1998, Preibischetal1999, ShaTok2002, KobFry2007, Masonetal2009, Chinietal2012, Peteretal2012, Sanaetal2012, Rizzutoetal2013, Sotaetal2014}, intermediate-mass stars: \citep{Patienceetal2002, Kouwenhovenetal2007, Chinietal2012, FuhChi2012, DeRosaetal2014}, solar-type stars \citep{DuqMay1991,Raghavanetal2010}, M-dwarfs \citep{FisMar1992, ReiGiz1997, Jansonetal2012}, and very-low-mass stars and brown dwarfs (\citealt{Burgasseretal2003,Closeetal2003, Siegleretal2005,Burgasseretal2006,BasRei2006,Reidetal2006,Reidetal2008,Allen2007,Ducheneetal2013}; \citealt*{PopMarTut2013}). 

\begin{figure*}
\centering
    \includegraphics[width=5cm]{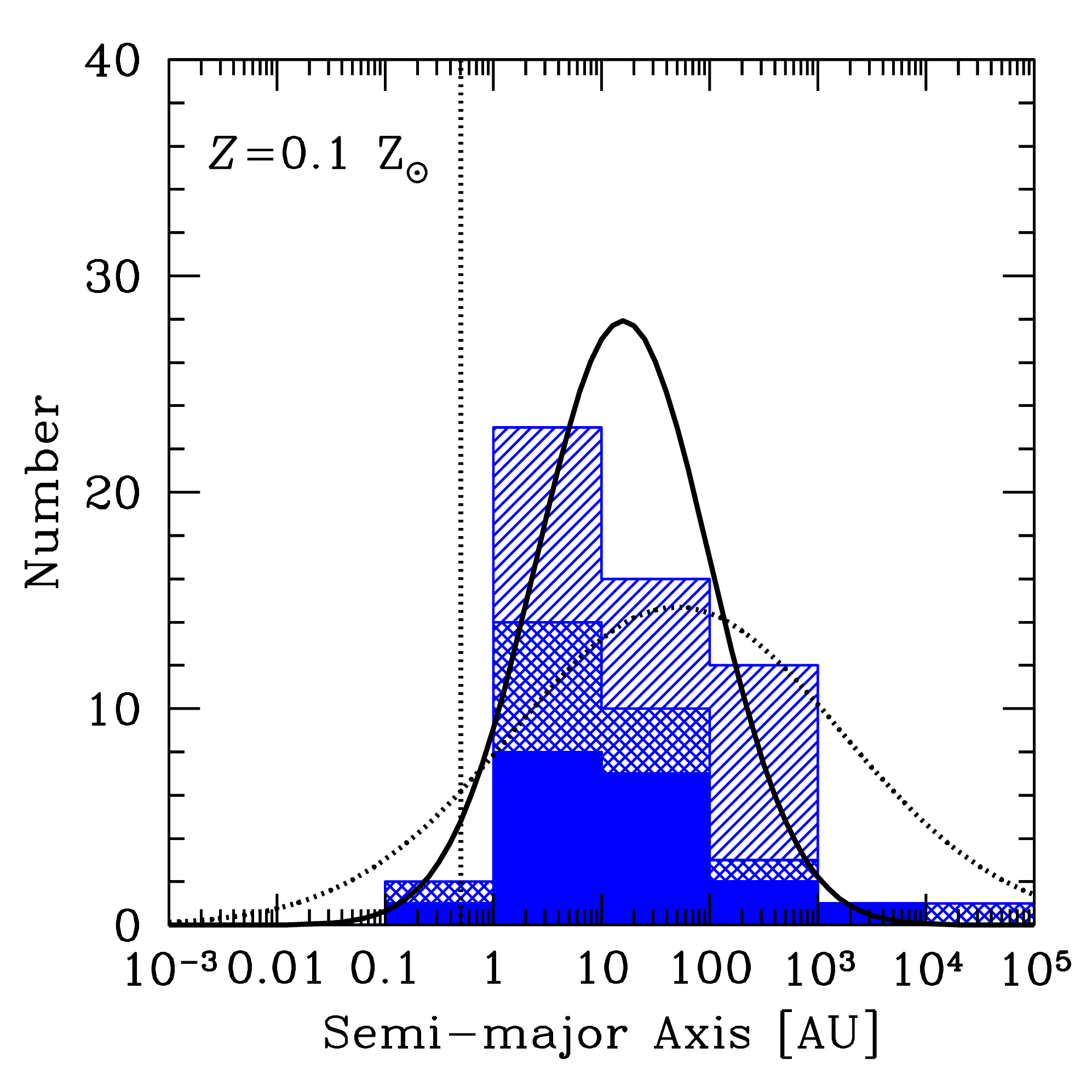} 
    \includegraphics[width=5cm]{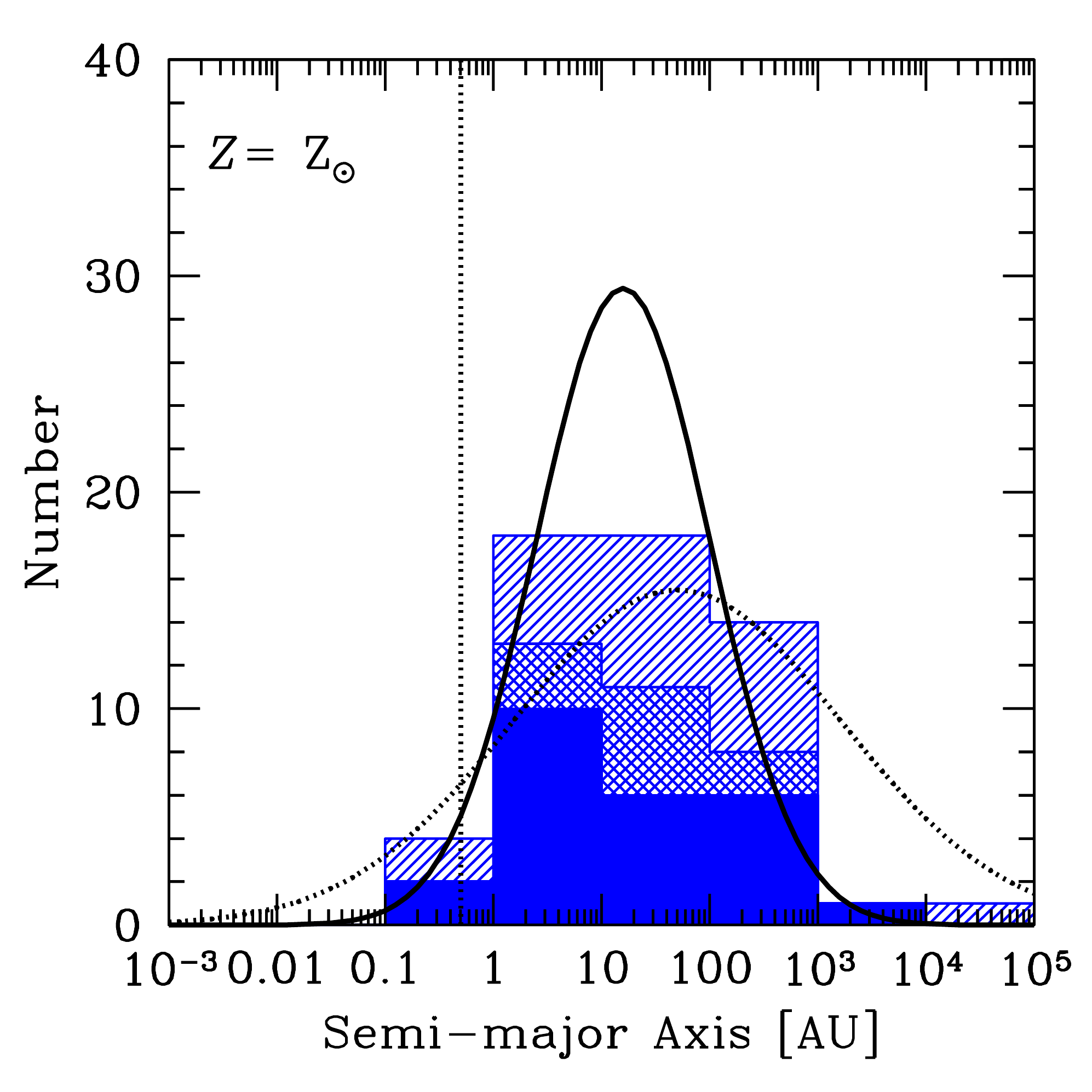} 
    \includegraphics[width=5cm]{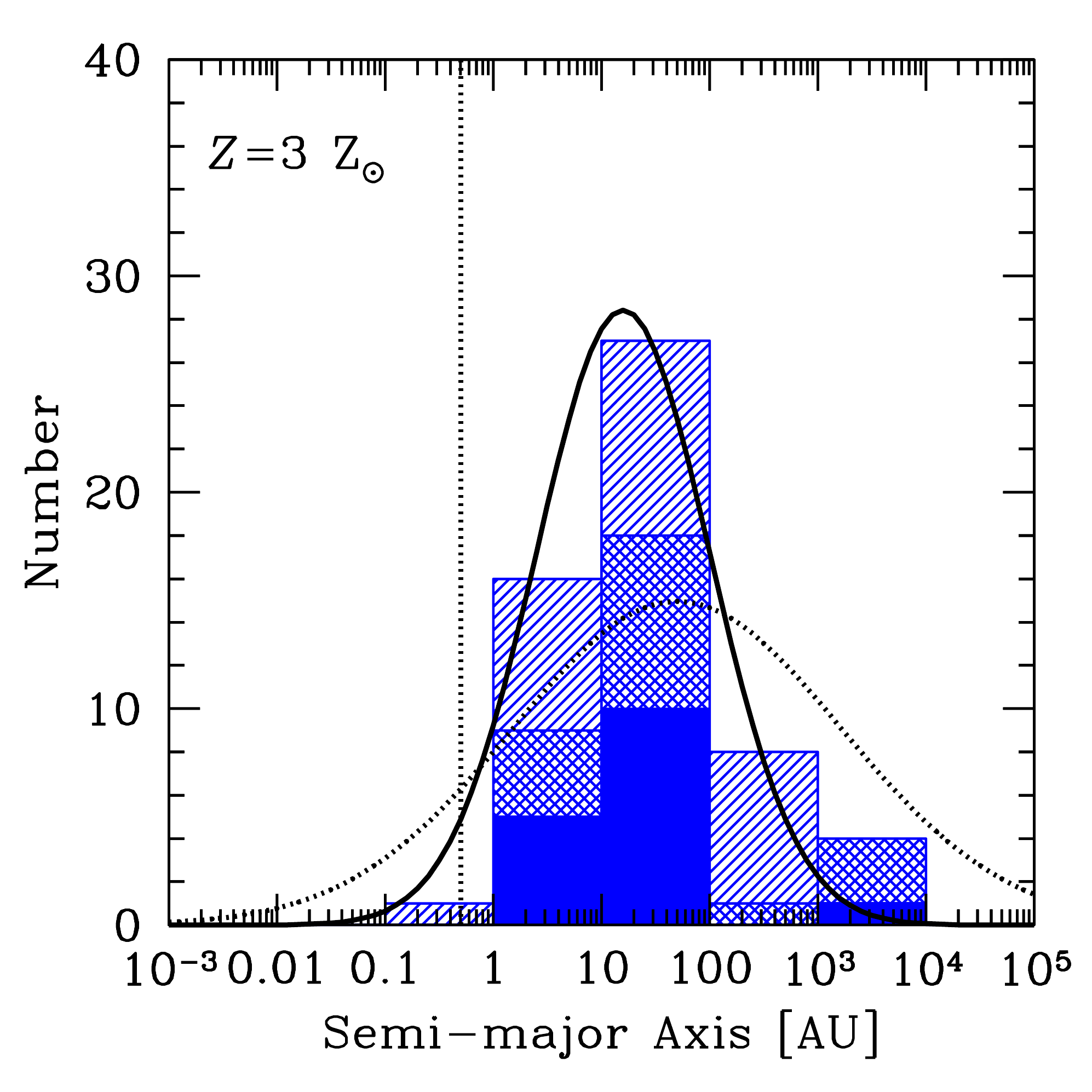}
\caption{The distributions of separations (semi-major axes) of multiple systems with stellar primaries ($M_*>0.1$~M$_\odot$) produced by the three radiation hydrodynamical calculations with the highest opacities.  The solid, double hatched, and single hatched histograms give the orbital separations of binaries, triples, and quadruples, respectively (each triple contributes two separations, each quadruple contributes three separations).  The solid curve gives the M-dwarf separation distribution (scaled to match the area) from the M-dwarf survey of Janson et al.\ (2012), and the dotted curve gives the separation distribution for solar-type primaries of Raghavan et al.\ (2010). 
Note that since most of the simulated systems are low-mass, the distributions are expected to match the Janson et al. distribution better than that of Raghavan et al. (see also Fig. 14).  The vertical dotted line gives the resolution limit of the calculations as determined by the accretion radii of the sink particles (0.5 AU).}
\label{separation_dist}
\end{figure*}

In Fig.~\ref{multiplicity}, for each of the three radiation hydrodynamical calculations with the highest opacities we compare the multiplicity fraction of the stars and brown dwarfs as functions of stellar mass with the values obtained from observational surveys.  The results from a variety of observational surveys (see the figure caption) are plotted using black open boxes with associated error bars and/or upper/lower limits.  The data point from the survey of \cite{DuqMay1991} is plotted using dashed lines for the error bars since this survey has been superseded by the lower value of \cite{Raghavanetal2010}.  The results from the simulations have been plotted in two ways.  First, using the numbers given in Table \ref{tablemult} we compute the multiplicities in six mass ranges: low-mass brown dwarfs (masses $<0.03$~M$_\odot$), very-low-mass (VLM) objects excluding the low-mass brown dwarfs (masses $0.03-0.10$ M$_\odot$), low-mass M-dwarfs (masses $0.10-0.20$ M$_\odot$), high-mass M-dwarfs (masses $0.20-0.50$ M$_\odot$), K-dwarfs and solar-type stars (masses $0.50-1.20$ M$_\odot$), and intermediate mass stars (masses $>1.2$ M$_\odot$). The filled blue squares give the multiplicity fractions in these mass ranges, while the surrounding blue hatched regions give the range in stellar masses over which the fraction is calculated and the $1\sigma$ (68\%) uncertainty on the multiplicity fraction calculated using Poisson statistics.  Note that the uncertainties in the equivalent figure presented by \cite{Bate2012} for the solar metallicity calculation were a factor of two too small by mistake.  In addition to the blue squares, a thick solid line gives the continuous multiplicity fraction computed using a sliding log-normal-weighted average from the results from each simulation.  The width of the log-normal average is half decade a in stellar mass.  The dotted lines give the approximate $1\sigma$ (68\%) uncertainty on the sliding log-normal average.  

All three calculations clearly produce multiplicity fractions that strongly increase with increasing primary mass.  Furthermore, the values in each mass range are in agreement with observations of field stars.  There is no significant dependence of the multiplicity on opacity.  For primary masses up to 1.2~M$_\odot$ the values are in close agreement between all the calculations.  The $Z=0.1$~Z$_\odot$ calculation has a higher multiplicity for intermediate-mass stars ($M_1>1.2$~M$_\odot$) than the other two calculations, but the result is not statistically significant.

\begin{figure}
\centering
    \includegraphics[width=8.4cm]{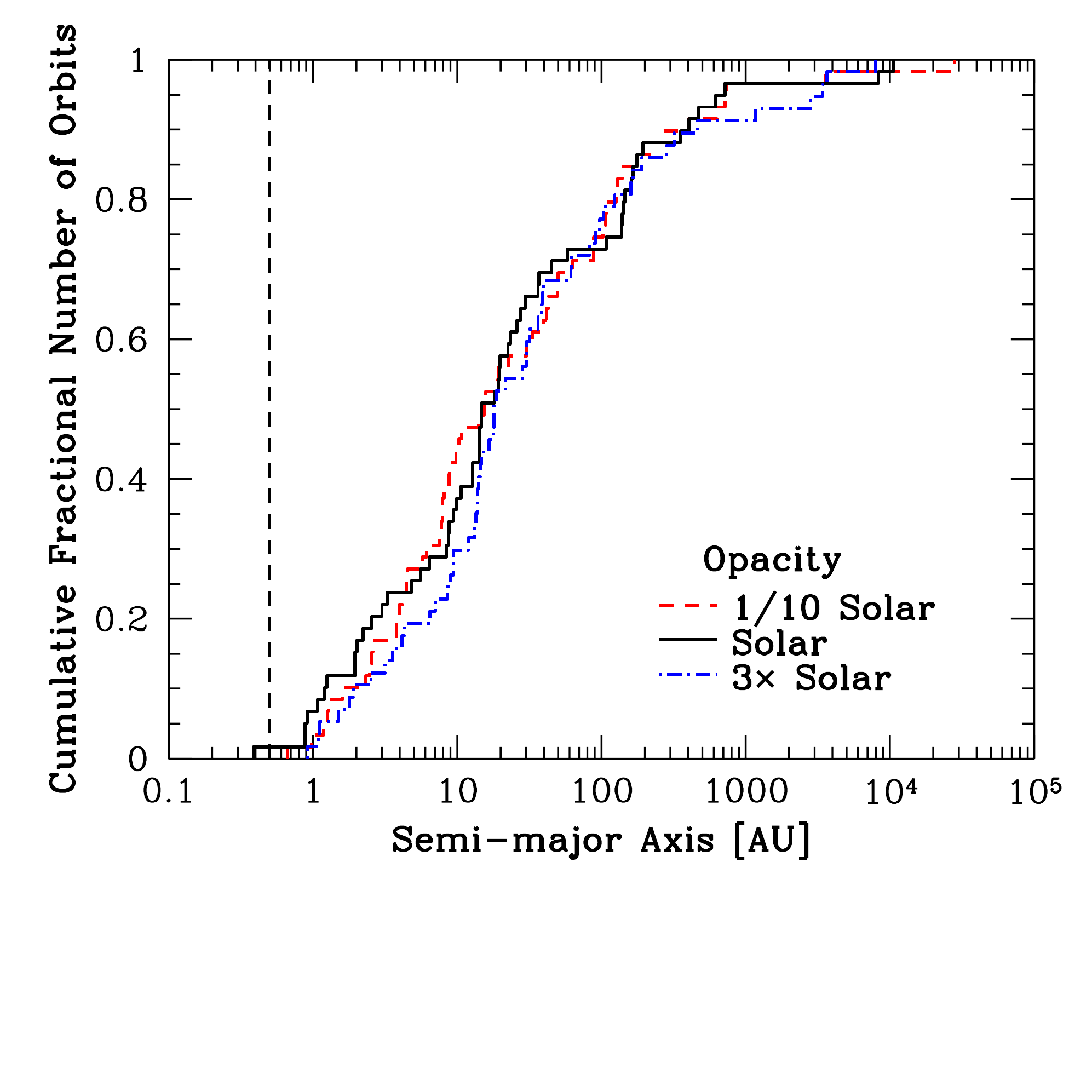}\vspace{-1.5cm}
\caption{The cumulative semi-major axis (separation) distributions of the multiple systems with stellar primaries ($M_*>0.1$~M$_\odot$) produced by the three radiation hydrodynamical calculations with the highest opacities.  All orbits are included in the plot (i.e. two separations for triple systems, and three separations for quadruple systems).  The opacities correspond to metallicities of $Z=0.1~{\rm Z}_\odot$ (red dashed line), $Z={\rm Z}_\odot$ (black solid line), and $Z=3~{\rm Z}_\odot$ (blue dot-dashed line).  The vertical dashed line marks the resolution limit of the calculations as determined by the accretion radii of the sink particles.  Performing Kolmogorov-Smirnov tests on the distributions shows that they are statistically indistinguishable.}
\label{cumsep_comp}
\end{figure}

\begin{figure*}
\centering
    \includegraphics[width=5cm]{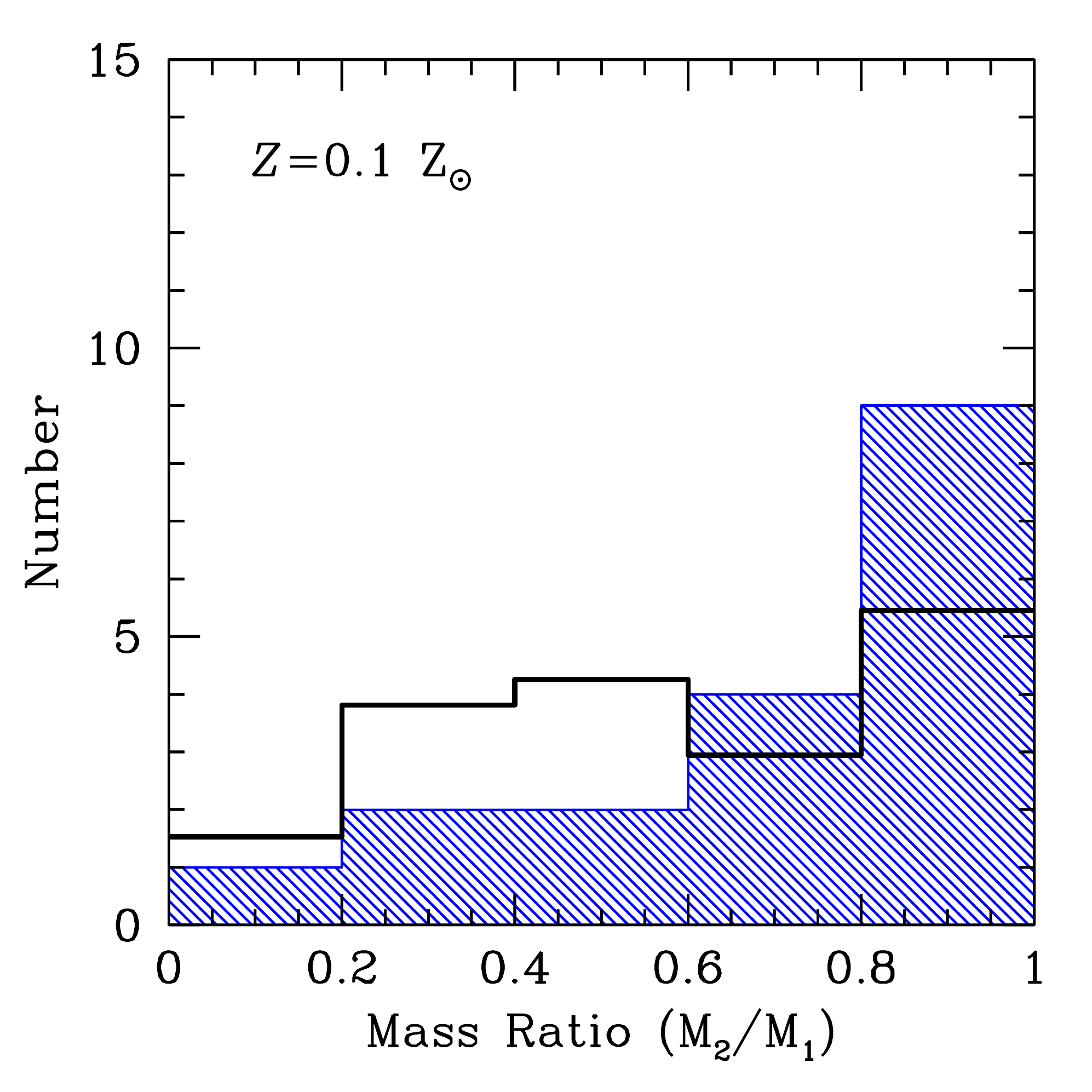}
    \includegraphics[width=5cm]{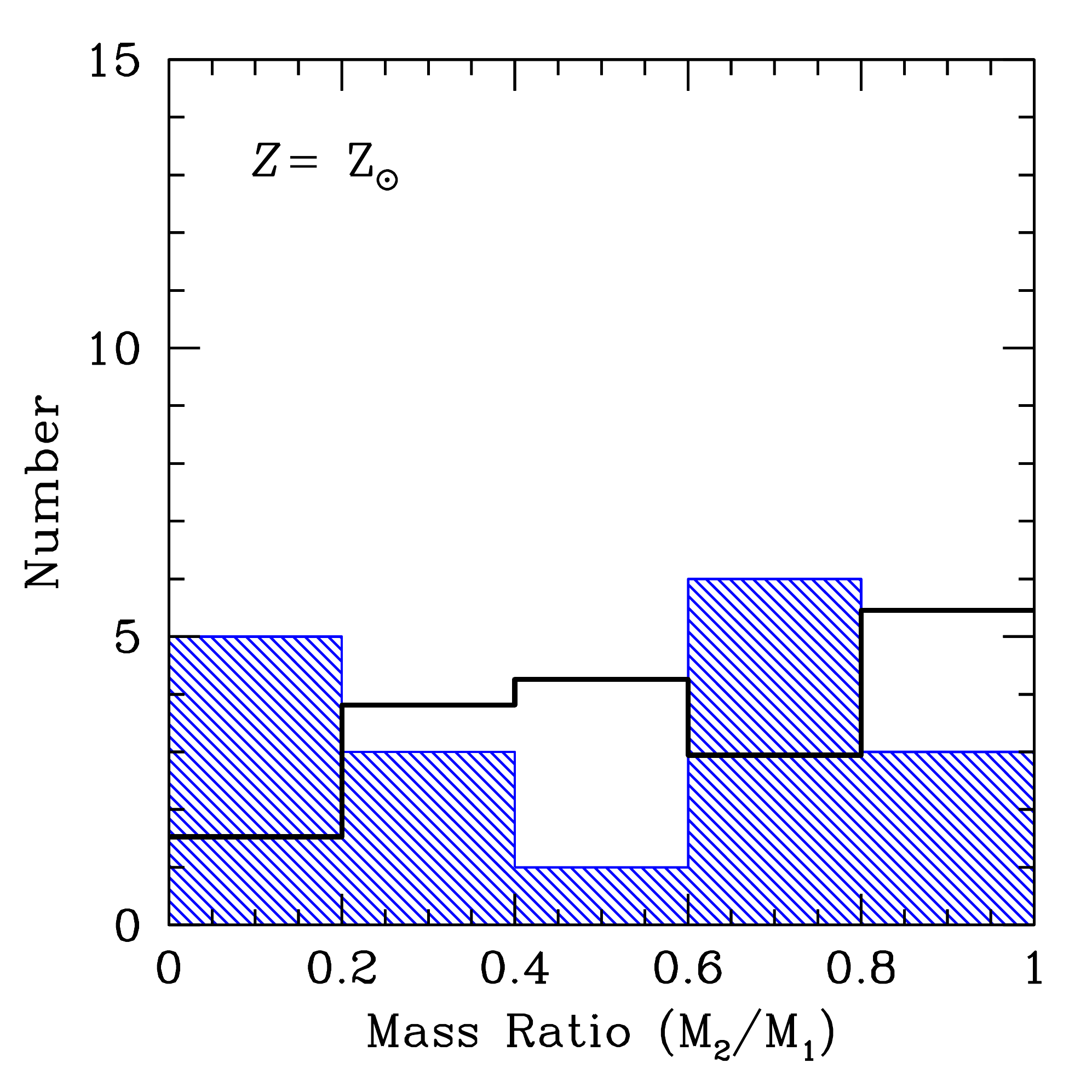}
    \includegraphics[width=5cm]{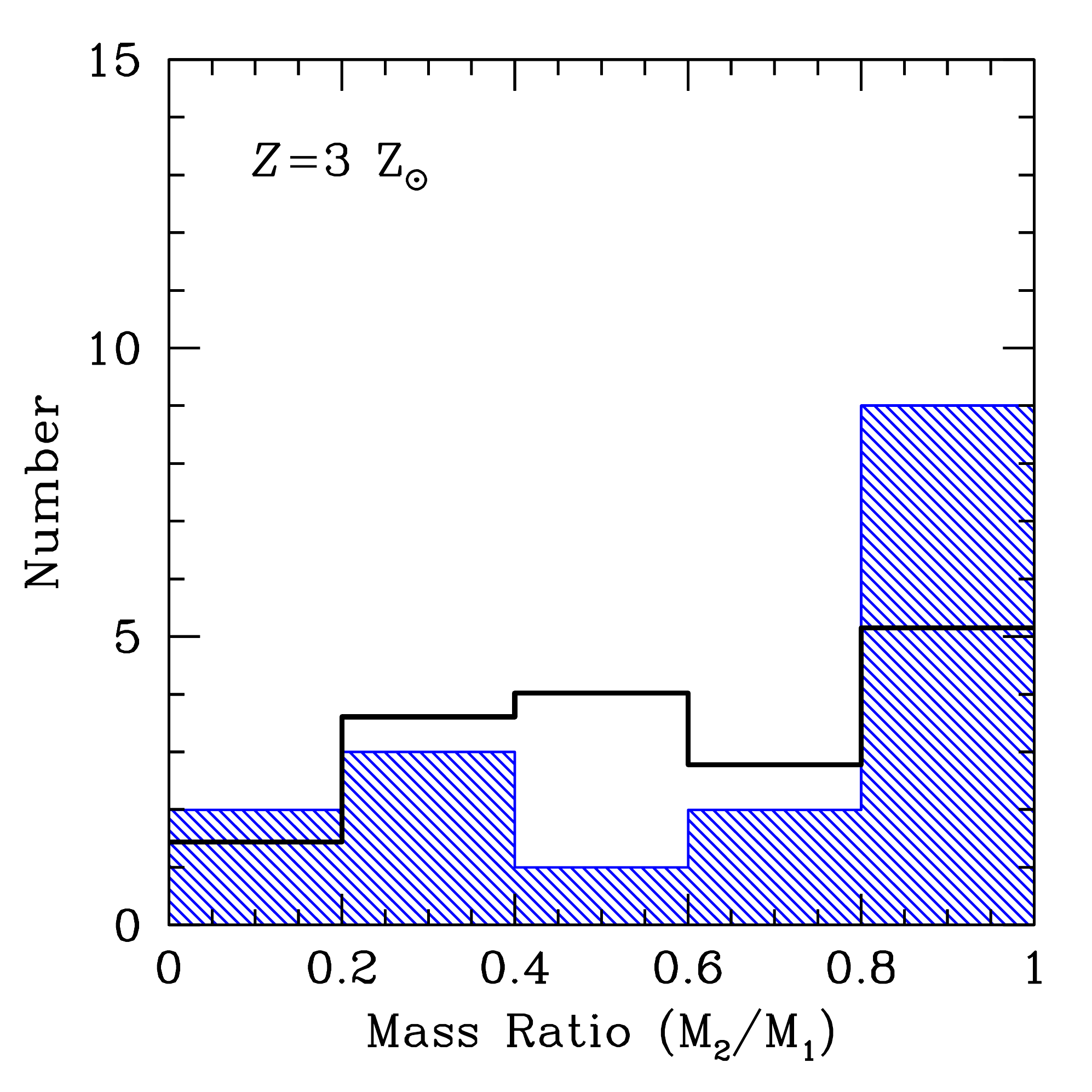}
    \includegraphics[width=5cm]{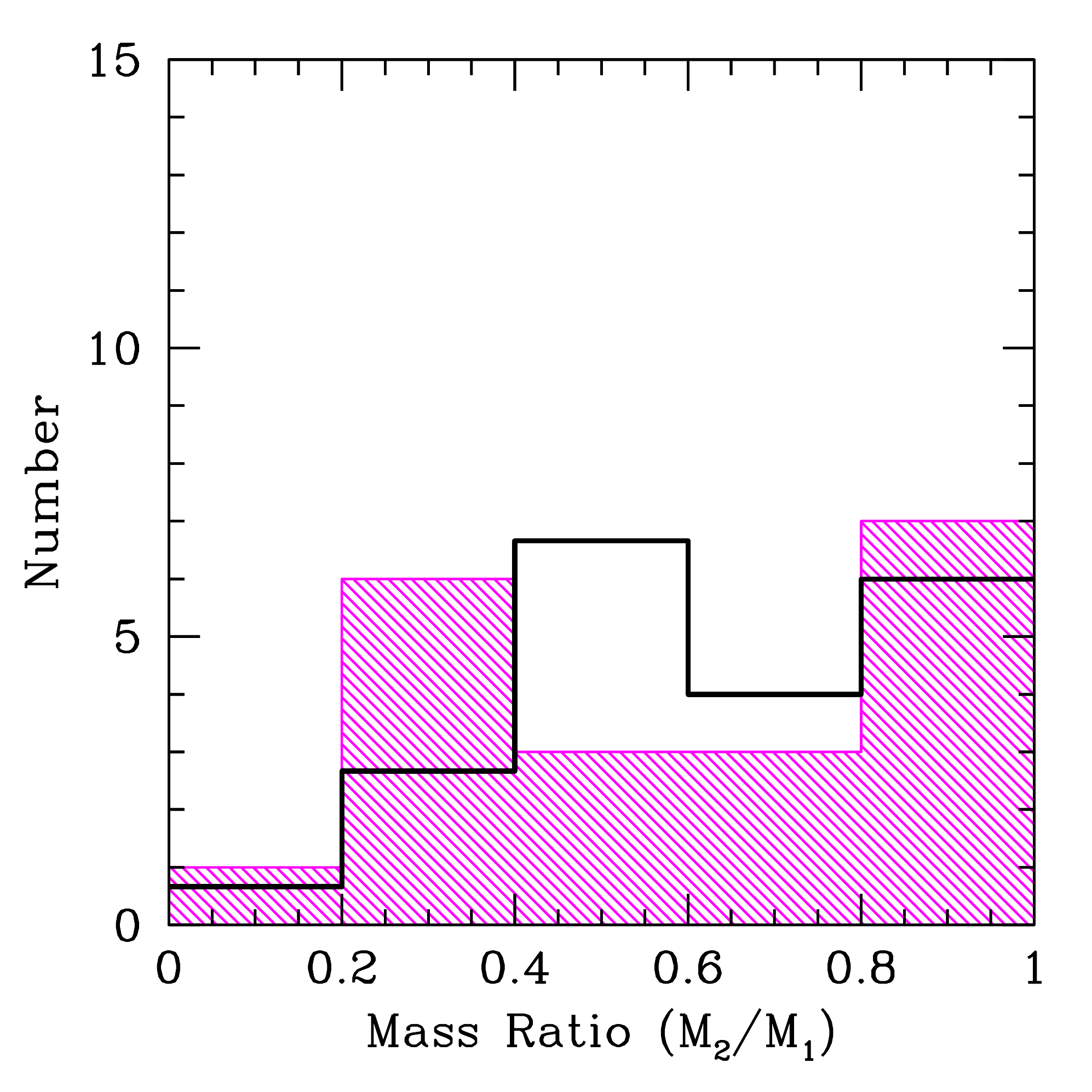}
    \includegraphics[width=5cm]{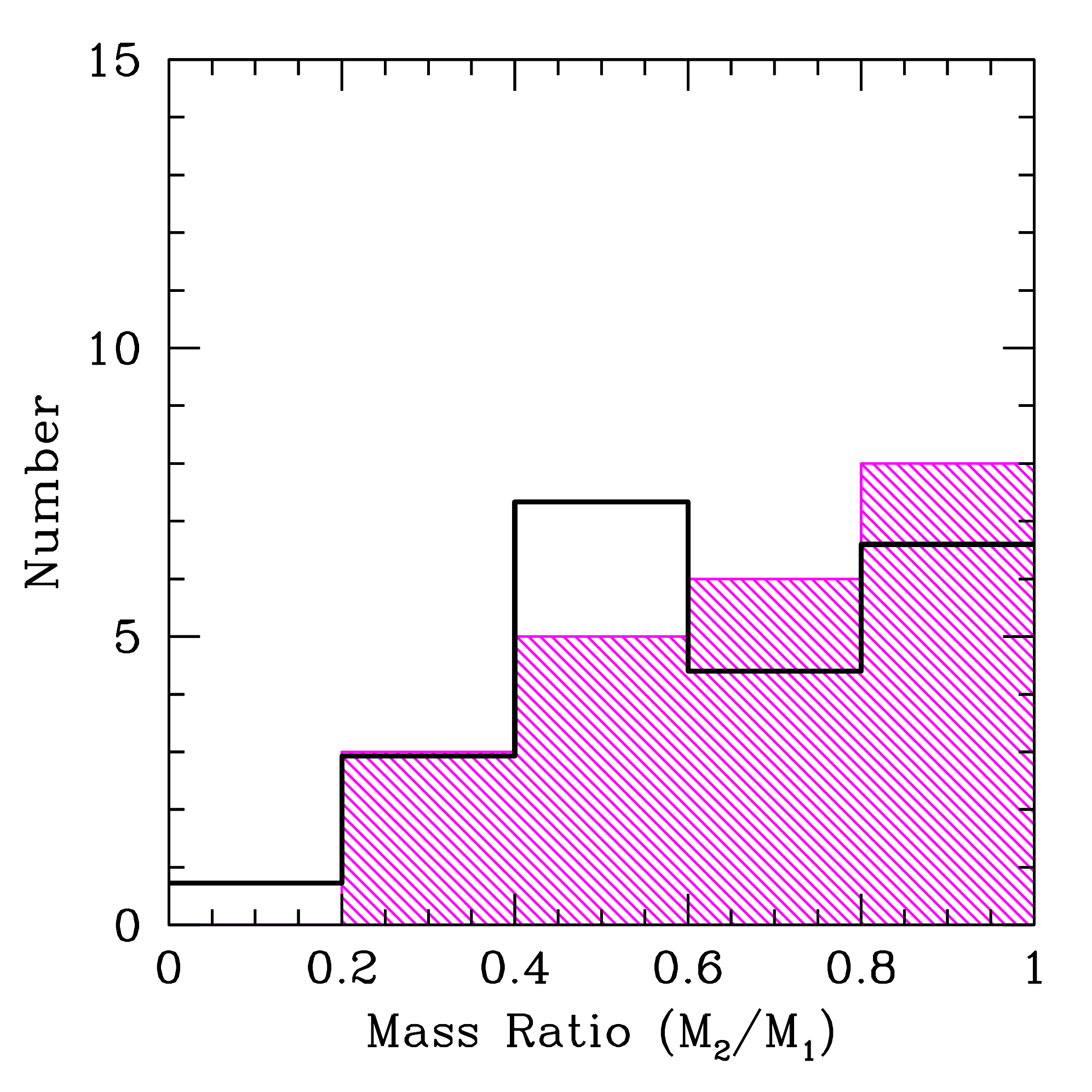}
    \includegraphics[width=5cm]{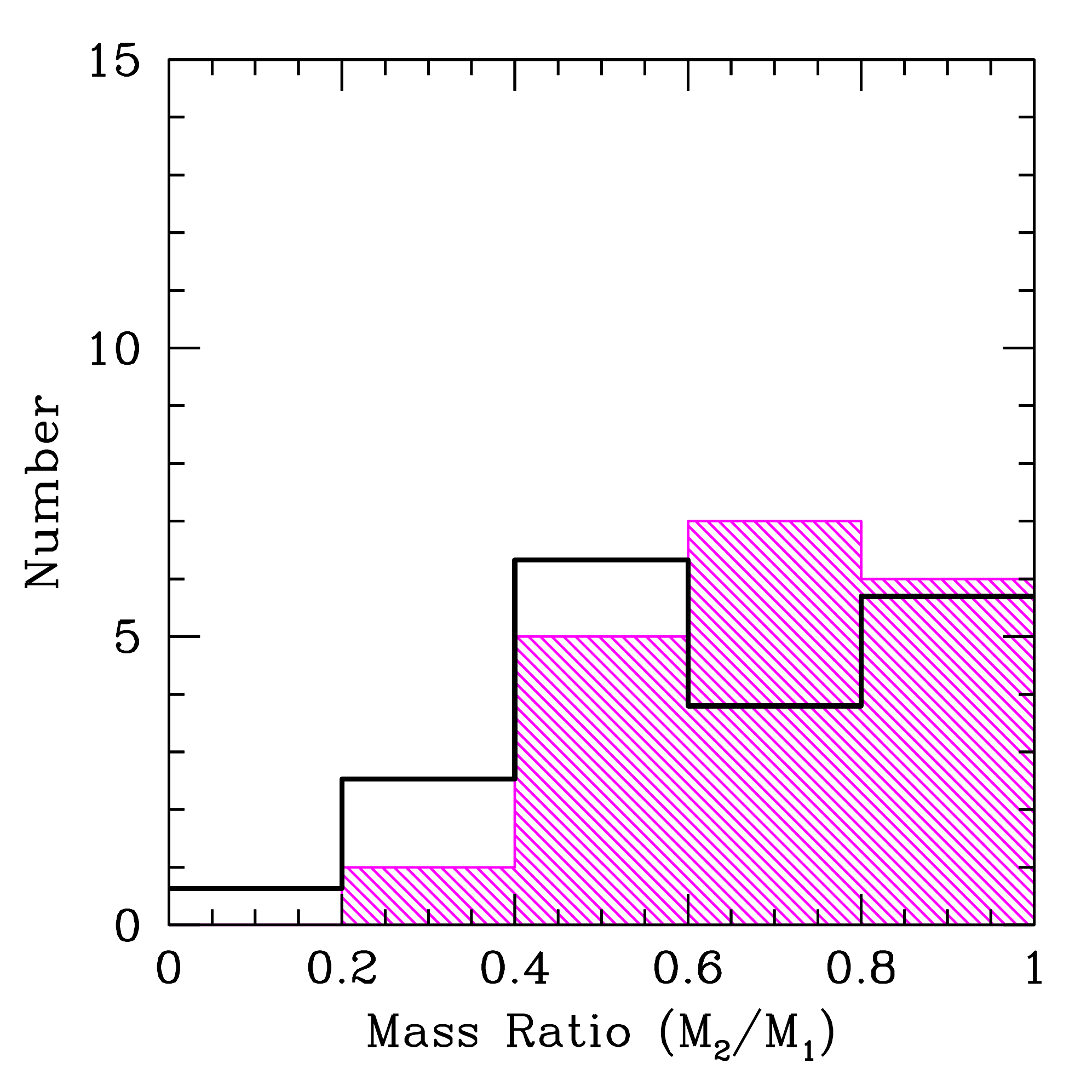}
    \caption{The mass ratio distributions of binary systems with stellar primaries in the mass ranges $M_1>0.5$ M$_\odot$ (top row) and $M_1=0.1-0.5$ M$_\odot$ (bottom row) produced by the three radiation hydrodynamical calculations with the highest opacities (left to right).  The solid black lines give the observed mass ratio distributions of \citet{Raghavanetal2010} for binaries with solar-type primaries (top row) and \citet{Jansonetal2012} for M-dwarfs (bottom row).  The observed mass ratio distributions have been scaled so that the areas under the distributions match those from the simulation results.  There is no obvious dependence of the mass ratio distributions on opacity.}
\label{massratios}
\end{figure*}

\subsection{Separation distributions of multiples}
\label{sec:separations}

Observationally, the mean and median separations of binaries are found to depend on primary mass \citep[see the review of][]{DucKra2013}.  \citet{DuqMay1991} found that the mean separation (in the logarithm of separation) for solar-type binaries was $\approx 30$ AU.  In the recent larger survey of solar-type stars, \cite{Raghavanetal2010} found $\approx 40$~AU.   \citet{FisMar1992} and \cite{Jansonetal2012} found indications of smaller mean separations for M-dwarf binaries of $\approx 10$ and $16$ AU, respectively.  Finally, VLM binaries (those with primary masses $<0.1$~M$_\odot$) are found to have a mean separation $\lsim 4$ AU \citep{Closeetal2003, Closeetal2007, Siegleretal2005}, with few VLM binaries found to have separations greater than 20 AU, particularly in the field \citep{Allenetal2007}.  A list of VLM multiple systems can be found at http://vlmbinaries.org/.  

Although we are able to follow binaries as close as 0.015~AU before they are assumed to merge in the radiation hydrodynamical calculations carried out for this paper, the sink particle accretion radii are 0.5~AU.  Thus, dissipative interactions between stars and gas are omitted on these scales which likely affects the formation of very close systems \citep{BatBonBro2002a}.  

In Fig.~\ref{separation_dist}, we present the separation (semi-major axis) distributions of the stellar (primary masses greater than 0.10 M$_\odot$) multiples.  The distributions are compared with the log-normal distributions from the surveys of M-dwarfs by \cite{Jansonetal2012} (solid lines) and solar-type stars by \cite{Raghavanetal2010} (dotted lines).  In each case the filled histogram gives the separations of binary systems, while the double hatched region adds the separations from triple systems (two separations for each triple, determined by decomposing a triple into a binary with a wider companion), and the single hatched region includes the separations of quadruple systems (three separations for each quadruple which may be comprised of two binary components or a triple with a wider companion).

In Fig.~\ref{separation_dist}, it appears that there may be a weak trend whereby the peak of the distribution moves from the $1-10$~AU bin with low opacities to the $10-100$~AU bin at the highest opacity.  To investigate this further, in Figure \ref{cumsep_comp} we provide the cumulative separation distributions for each of the calculations.  This demonstrates that the apparent trend is an artefact of the binning.  The median separations are 15.5~AU ($Z=0.1$~Z$_\odot$), 14.8~AU ($Z=1$~Z$_\odot$), and 16.7~AU ($Z=3$~Z$_\odot$), and performing Kolmogorov-Smirnov tests between the distributions shows that they are statistically indistinguishable.

The $Z=0.1$~Z$_\odot$ and $Z=3$~Z$_\odot$ calculations each produced only one very-low-mass (VLM) binary (primary masses $M_1<0.1$~M$_\odot$), and the solar-metallicity calculation only produced three.  Because of these small numbers, we defer discussion of the VLM binaries until Section \ref{sec:combined} where we discuss the statistical properties of the combined sample.

\begin{figure*}
\centering
    \includegraphics[width=9.2cm]{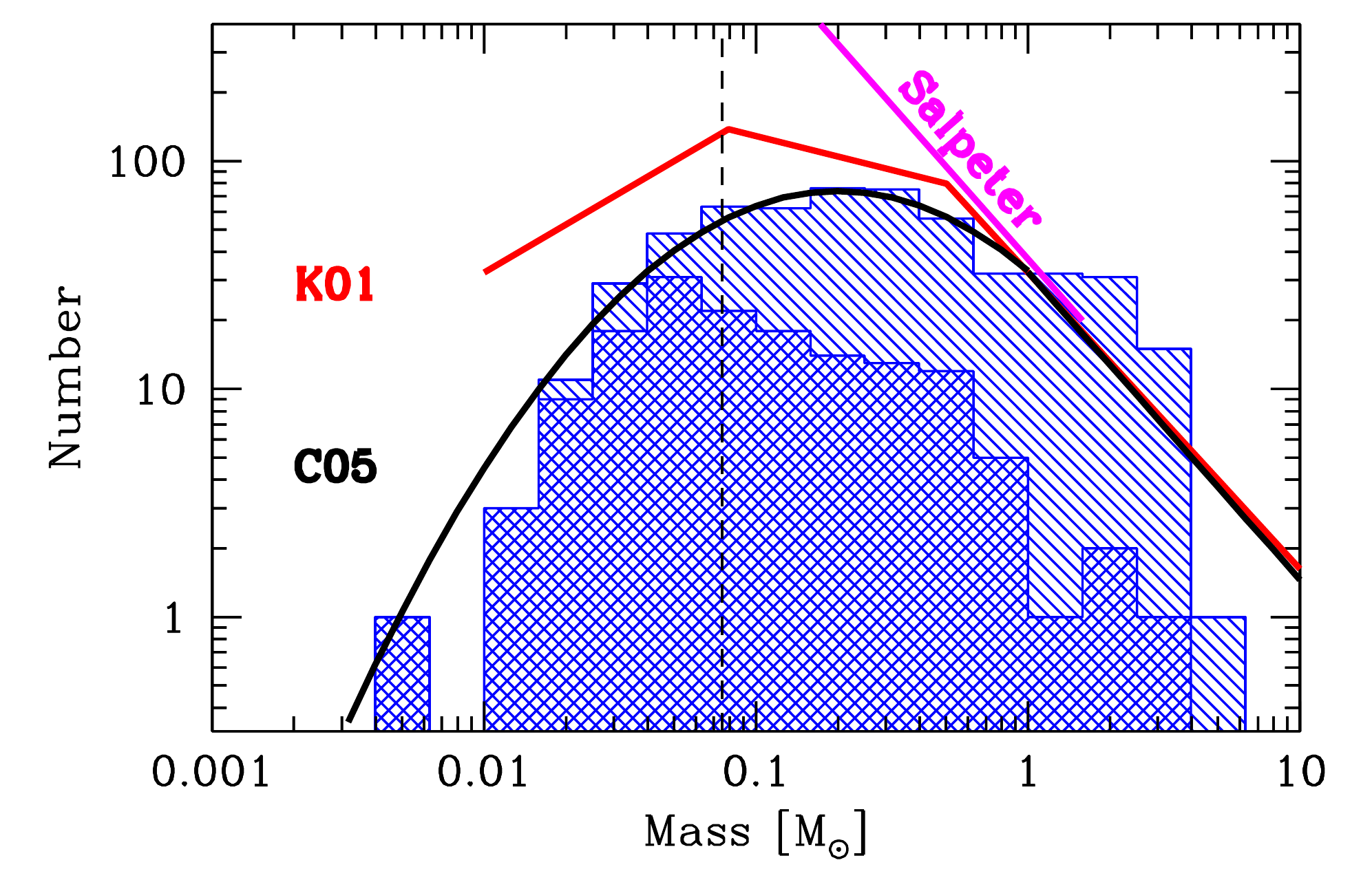} \hspace{0.5cm}
    \includegraphics[width=7.45cm]{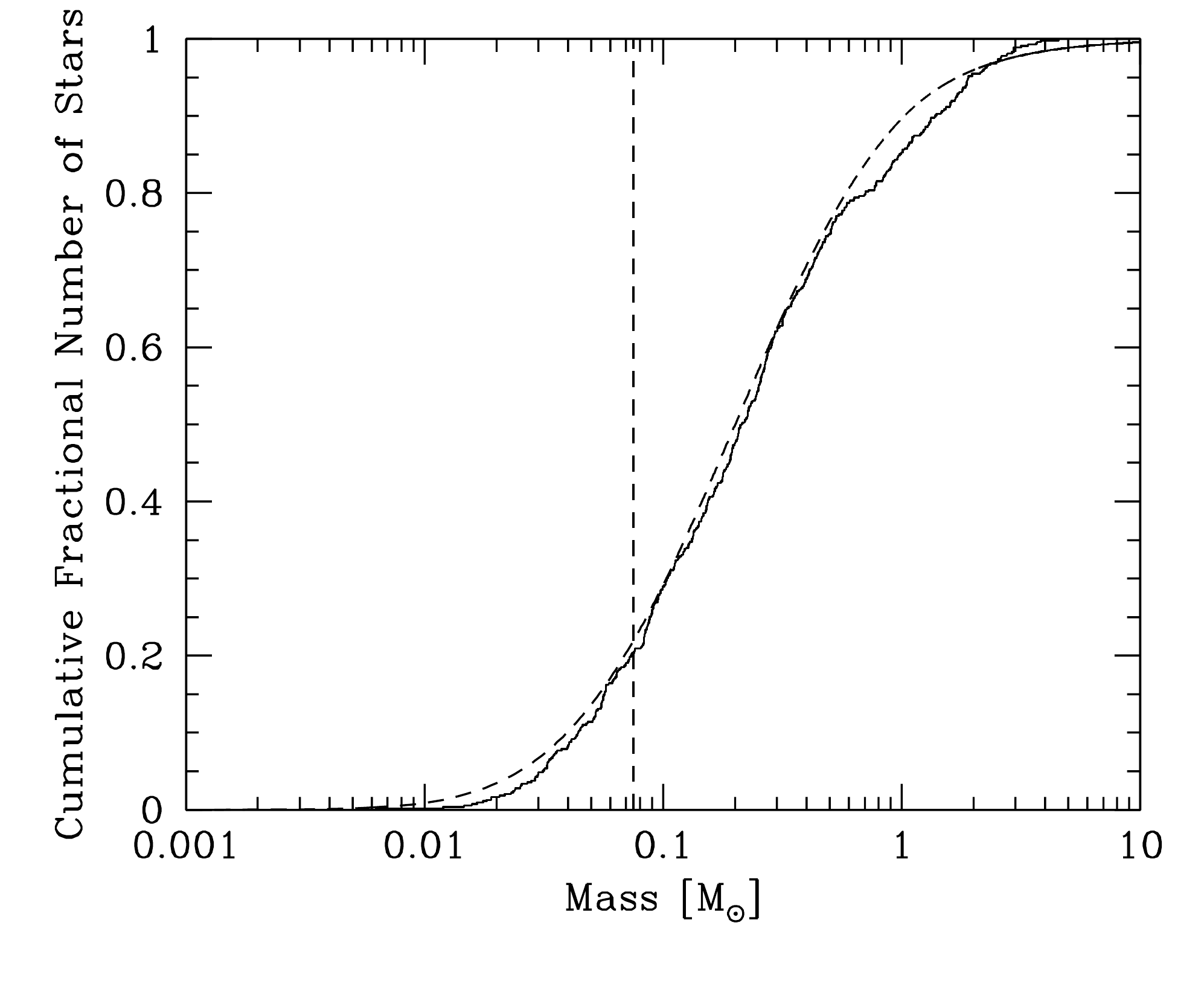}\vspace{-0.0cm}
\caption{The differential (histogram; left panel) and cumulative (solid line; right panel) IMFs produced by combining the stars and brown dwarfs from the three calculations with the highest opacities.  In total, there are 535 stars and brown dwarfs.  The double hatched histogram denotes those objects that have stopped accreting (defined as accreting at a rate of less than $10^{-7}$~M$_\odot$~yr$^{-1}$), while those objects that are still accreting when the calculations were stopped are plotted using single hatching.  The differential IMF is compared with the parameterisations of the observed IMF by Salpeter (1955), Kroupa (2001), and Chabrier (2005).  The cumulative IMF is compared with the cumulative IMF  (dashed line; right panel) from the parameterisation of the observed IMF by Chabrier (2005).  The vertical dashed line marks the stellar/brown dwarf boundary.  A Kolmogorov-Smirnov test shows the numerical IMF to be indistinguishable from Chabrier's parameterisation of the IMF (there is a 9\% probability that the numerical IMF could have been drawn from Chabrier's fit to the observed IMF).}
\label{imf_combined}
\end{figure*}

\subsection{Mass ratios distributions of multiples}
\label{sec:massratios}

In addition to the separation distributions of the multiple systems, we can investigate their mass ratio distributions.  We only consider binaries, but we include binaries that are inner components of triple and quadruple systems.  A triple system composed of a binary with a wider companion contributes the mass ratio from the closest pair, as does a quadruple composed of a triple with a wider companion.  A quadruple composed of two pairs orbiting each other contributes two mass ratios --- one from each of the pairs.

In Fig.~\ref{massratios}, we present the mass ratio distributions of the binaries with primary masses $\geq 0.5$ M$_\odot$ (top panels) and M-dwarfs with masses $0.1~{\rm M}_\odot \leq M_1<0.5$ M$_\odot$ (bottom panels) for each of the three calculations the highest opacities (left to right).  We compare the M-dwarf mass ratio distribution to that of \citet{Jansonetal2012}, and the higher mass stars to the mass ratio distribution of binaries with solar-type primaries obtained by \citet{Raghavanetal2010}.  As for the separations of the VLM binaries, we defer discussion of the mass ratios of the five VLM binaries until Section \ref{sec:combined}, but we note that all have mass ratios $M_2/M_1>0.6$.

Examining the distributions, there is no firm evidence of a dependence of the mass ratio distributions on opacity.  It could be noted that the lowest opacity calculation produces a higher proportion of low-mass ratio M-dwarf systems ($M_2/M_1<0.4$) than the two higher opacity calculations.  However, the reverse is true of the solar-type ($M_1>0.5$~M$_\odot$) binaries (the two higher opacity calculations give the greatest proportions of low-mass ratio systems).  Interestingly, both of the new calculations produce large numbers of solar-type `twins' -- binaries with mass ratios $M_2/M_1>0.8$.  The solar-metallicity calculation of \citet{Bate2012} produced an essentially flat mass ratio distribution for solar-type stars, in good agreement with \cite{Raghavanetal2010}.  But in each of the new calculations approximately half of the binaries have mass ratios $M_2/M_1>0.8$.  This will be discussed further below when we discuss the combined sample.

\section{The combined statistics}
\label{sec:combined}

As discussed above, there is little evidence that the stellar properties produced by the calculations depend on opacity, and formally we cannot distinguish between the four calculations whose opacities range over a factor of 300.  Therefore, in this section we amalgamate the stellar systems from three of the radiation hydrodynamical calculations with the highest opacities ($Z\ge 0.1~{\rm Z}_\odot$) and discuss the combined sample of 535 stars and brown dwarfs.  We exclude the lowest opacity calculation from this combined sample because, as discussed in Section \ref{initialconditions}, the thermal behaviour of the low-density gas in this calculation cannot be considered to be realistic.  However, we note that our conclusions in this section are the same, whether this calculation is included or not.  This combined sample provides lower statistical uncertainties than each of the calculations individually, allowing us to better compare the statistical properties with the results of observational surveys.  We also note that the comparisons discussed below are usually with field populations, which anyway contain stars with a range of metallicities.

\subsection{The combined initial mass function}

The combined IMF is given in Figure \ref{imf_combined} in both differential and cumulative forms.  The lowest mass object has a mass of just 6.3 M$_{\rm J}$, while the most massive star has a mass of 4.56~M$_\odot$ (and is still accreting when the calculations are stopped).  The close agreement with the parameterisation of the observed IMF given by \cite{Chabrier2005} is astonishing, and a Kolmogorov-Smirnov test shows the two IMFs to be indistinguishable (there is a 9\% probability that the numerical IMF could have been drawn randomly from Chabrier's IMF).  Observationally, the form of the IMF in the sub-stellar regime is still quite uncertain.  Rather than fit a function to the IMF in the low-mass regime, another method is to compare the number of brown dwarfs to the number of stars with masses less than that of the Sun. \citet{Andersenetal2008} analyse a large number of young clusters and find that the ratio of stars with masses $0.08-1.0$ M$_\odot$ to brown dwarfs with masses $0.03-0.08$ M$_\odot$ is $N(0.08-1.0)/N(0.03-0.08)\approx 5\pm 2$.  The recent analysis of the mass functions in NGC1333 and IC348 by \cite{Scholzetal2013} finds ratios of 1.9--2.4 and 2.9--4.0 in these two clusters, respectively.  For the combined sample this ratio is 344/87 = 3.95, broadly consistent with observations.  The ratio of all stars to brown dwarfs ($M_*<0.08$~M$_\odot$) is 423:112 = 3.8.  

Although the IMF of the combined sample and Chabrier's IMF are formally indistinguishable, we do note that there appears to be a deficit of stars with masses greater than 4~M$_\odot$.  The simulations produce one, but the Chabrier IMF would predict that there should be 7 or 8.  This may be related to the age old question of whether the IMF obtained by adding together the stars from several low-mass clouds is the same as the IMF produced by a single large cloud of the same total mass.  In other words, does the mass of the most massive star depend on the mass of the stellar group it is contained within, or is it consistent with being drawn randomly from a universal mass function \citep{Larson1982, Larson2003, Elmegreen1983, Elmegreen2000c, WeiKro2004, WeiKro2006, OeyCla2005, SelMel2008, WeiKroBon2010}?  The result obtained here, from adding together the results from three 500-M$_\odot$ clouds, implies that the highest mass star is related to the total mass of the star-forming cloud (and its resulting cluster).  It must be noted, however, that most of the stars with masses greater than one solar mass are still accreting when the calculations are stopped and if they were evolved for longer the discrepancy may disappear.  Indeed, while there is a slight deficit of high-mass stars, there appears to be an excess of stars with $1.5-4$~M$_\odot$ that could accrete to erase the high-mass deficit if the calculations were followed further.  Furthermore, the four calculations are not strictly independent of each other since they all began with the same initial density and velocity structure.  It is likely that starting from clouds with equal masses but a range of initial conditions (e.g. different velocity fields) may result in a greater range of maximum stellar mass, and may smooth out the excess.

Below 0.03~M$_\odot$, the IMF is very poorly constrained observationally.  By combining the results of the three calculations, we find that the numerical IMF is reasonably well-described by a continuation of Chabrier's IMF.  The numbers from the combined sample fall slightly below those expected for the Chabrier IMF, but the deficit is not statistically significant.  As discussed above, there is no strong evidence for the abundance of the low-mass brown dwarfs depending on metallicity.   Although the lowest mass object produced by the solar-metallicity calculation of \cite{Bate2012} was 17.6 M$_{\rm J}$, the lowest mass object from all of the calculations (6.3 M$_{\rm J}$) was produced by the calculation with the highest metallicity, and the $Z=0.1$~Z$_\odot$ calculation produced four brown dwarfs with masses less than 20 M$_{\rm J}$ with a minimum mass of 12 M$_{\rm J}$.  The lowest mass object in each calculation had stopped accreting before the calculation was stopped.  As expected, the masses of the lowest-mass brown dwarfs are similar to the minimum masses predicted by models of the opacity limit for fragmentation \citep{LowLyn1976,Rees1976,Silk1977a,Silk1977b, BoyWhi2005}.  The lowest mass objects also have similar masses to the estimated masses of some of the lowest-mass objects observed in star-forming regions \citep[e.g.][]{ZapateroOsorioetal2000,Kirkpatricketal2001,ZapateroOsorioetal2002, Kirkpatricketal2006,Lodieuetal2008,Luhmanetal2008,Bihainetal2009,Weightsetal2009, Burgessetal2009,Luhmanetal2009a,Quanzetal2010, TodLuhMcL2010}.  Neither observations nor the simulations can currently determine an exact value for a cut-off mass, if one exists.

\begin{figure}
\centering
    \includegraphics[width=8.4cm]{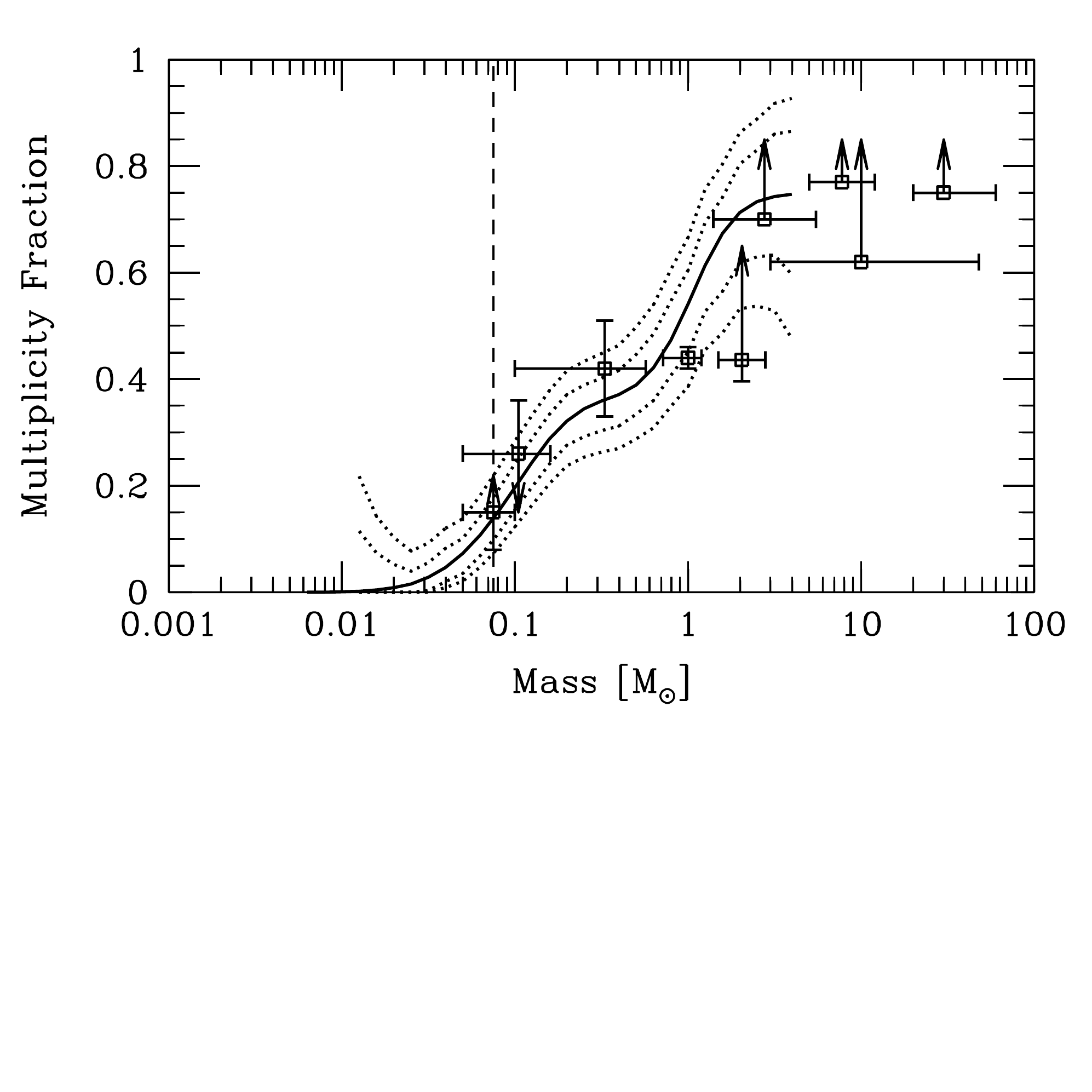} \vspace{-3.4cm}
\caption{Multiplicity fraction as a function of primary mass for the combined sample of the 535 stars and brown dwarfs from all the three calculations with the highest opacities ($Z\ge 0.1~{\rm Z}_\odot$).  The solid line gives the continuous multiplicity fraction from the calculations computed using a sliding log-normal average. The dotted lines give the approximate $1\sigma$ and $2 \sigma$ confidence intervals around the solid line.  The open black squares with error bars and/or upper/lower limits give the observed multiplicity fractions from the surveys of Close et al. (2003), Basri \& Reiners (2006), Fisher \& Marcy (1992), Raghavan et al. (2010), De Rosa et al. (2014), Kouwenhoven et al. (2007), Rizzuto et al. (2013), Preibisch et al. (1999) and Mason et al. (1998), from left to right.  The observed trend of increasing multiplicity with primary mass is well reproduced by the combined sample. }
\label{multiplicity_combined}
\end{figure}

\subsection{Multiplicity}

In Fig.~\ref{multiplicity_combined} we plot the multiplicity as a function of primary mass for the combined sample of 535 stars and brown dwarfs from the three calculations with the greatest opacities.  The solid line gives the continuous multiplicity fraction computed using a sliding log-normal average, while the dotted lines give the approximate $1\sigma$ and $2\sigma$ confidence intervals around the solid line due to the statistical uncertainties.  The numerical result is compared with the results of the observational surveys referred to in the figure caption.  The observed trend of increasing multiplicity with primary mass is well reproduced by the combined sample and the results for different ranges of primary mass are consistent with the multiplicities from current observational surveys.  

It should be noted that the statistical uncertainties from the combined numerical sample are smaller than the uncertainties from all of the observational surveys, with the exception of the \cite{Raghavanetal2010} survey.  Larger observational samples will be required in the future to avoid comparisons between numerical simulations and observations being limited by observational uncertainties.  In particular, the multiplicities of intermediate and massive stars are poorly constrained observationally, and there is little information on the multiplicity of low-mass brown dwarfs.  The multiplicity of VLM stars and brown dwarfs of $\approx 20$\% is typically derived from samples with primary masses close to 0.1~M$_\odot$, whereas the numerical simulations predict that the multiplicity for low-mass brown dwarfs (masses $<30$ Jupiter-masses) should be even lower ($<10$\%) \citep[see also][]{Bate2009a,Bate2012}.

Finally, we note that the surveys with which we are comparing the multiplicities are primarily of field stars rather than young stars.  This is necessary because surveys of young stars either do not sample a large range of separations and mass ratios, or the statistics are too poor.  However, in principle, there may be considerable evolution of the multiplicities between the age of the stars when the calculations were stopped ($\sim 10^5$ yrs) and a field population.  This issue was investigated by \cite{MoeBat2010} who took the end point of a hydrodynamical calculation from \cite{Bate2009a} and evolved it to an age of $10^7$~yrs using an N-body code under a variety of assumptions regarding the dispersal of the molecular cloud.  Importantly, \citeauthor{MoeBat2010} found that the multiplicity distribution evolved very little during dispersal of the molecular cloud and was surprisingly robust to different assumptions regarding gas dispersal.  They concluded that when star formation occurs in a clustered environment, the multiple systems that are produced are quite robust against dynamical disruption during continued evolution.  Therefore, we do not expect the multiplicities presented in Figs.~\ref{multiplicity} and \ref{multiplicity_combined} to evolve significantly as the stars evolve into a field population. We also note that \cite{ParReg2013} recently performed a series of N-body calculations to examine the dynamical evolution of binaries in young clusters.  They found that even when the separation distribution of binaries suffered evolution, the mass ratio distributions of binaries were unchanged.  

\begin{figure}
\centering
    \includegraphics[width=8.4cm]{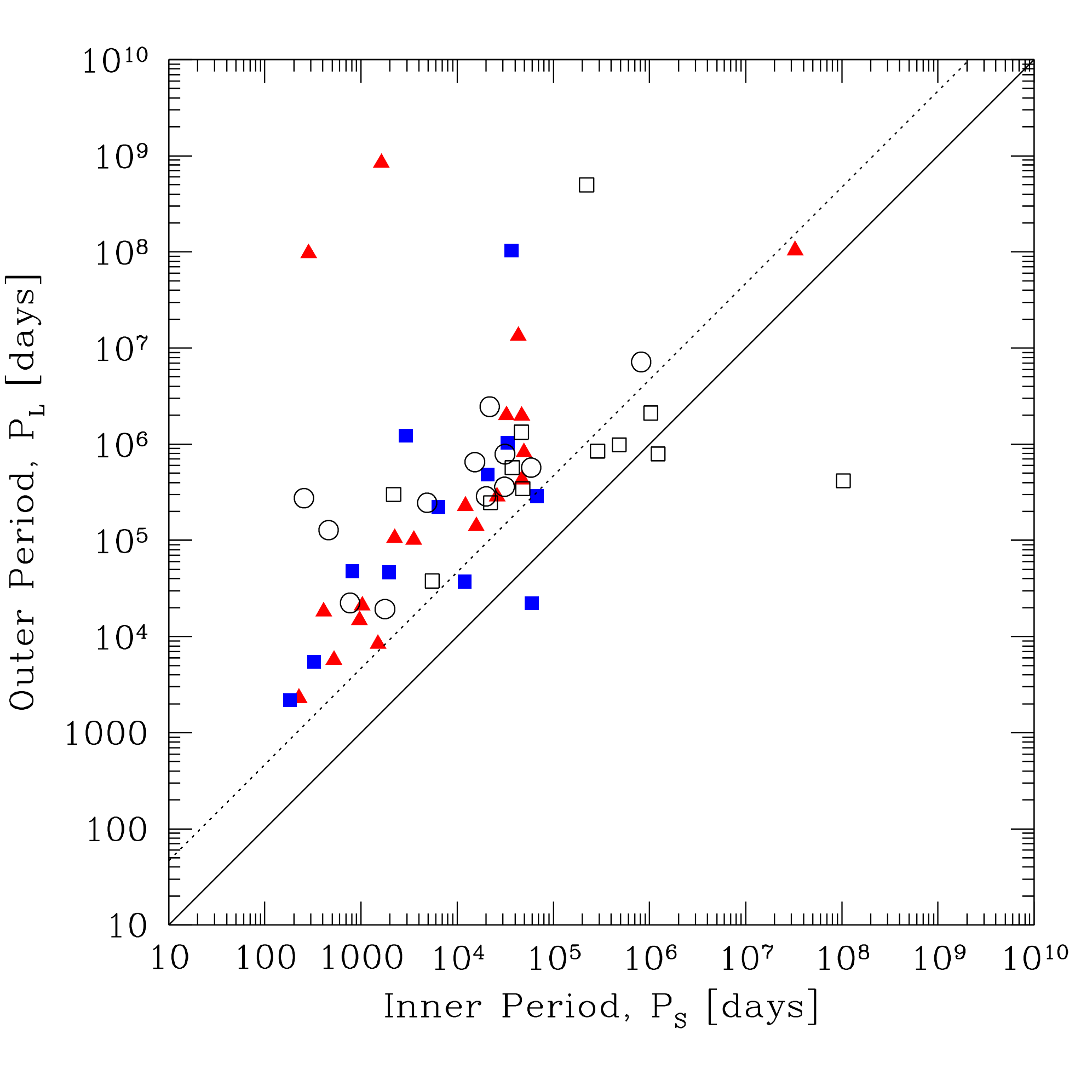}
\caption{The outer orbital period, $P_{\rm L}$ versus the inner orbital period, $P_{\rm S}$ for the 43 triple or quadruple systems  in the combined sample from the three radiation hydrodynamical calculations with the highest opacities ($Z\ge 0.1~{\rm Z}_\odot$).  Triples are plotted as red triangles, triples that are sub-components of quadruples are plotted as blue filled squares, quadruples that contain triples are plotted as black open squares, and quadruples that are composed of two pairs are plotted as black open circles. The solid and dotted lines denote equal periods and a period ratio $P_{\rm L}/P_{\rm S} = 4.7$, respectively.  Systems that lie below the dotted line are likely to be dynamically unstable and to undergo further evolution.}
\label{periodratios}
\end{figure}

\subsubsection{The frequencies of triple and quadruple systems}
\label{freq_high_order}

Consulting Table \ref{tablemult}, we find that the combined sample includes 252 single stars/brown dwarfs, 65 binaries, 19 triples and 24 quadruples.  This gives an overall frequency of triple and quadruple systems of $5.3_{-1.2}^{+1.4}$\% and $6.7_{-1.3}^{+1.6}$\%, respectively.  In Fig. \ref{periodratios}, we plot the outer orbital period, $P_{\rm L}$, versus the inner orbital period, $P_{\rm S}$,  for each of the triples and quadruples, including the triples that are sub-components of quadruples, and we distinguish between quadruples that are composed of two pairs, and those that are composed of a triple and a fourth object.  Note that there are three systems that are not hierarchical (lying below the solid line) that would certainly undergo further dynamical evolution.  There are several more for which $P_{\rm L}/P_{\rm S}<4.7$ that are also very likely to evolve (i.e. they lie below the dotted line), and in all there are 13 systems (3 triples and 10 quadruples, including 2 quadruples composed of pairs) for which $P_{\rm L}/P_{\rm S} < 4.7~(1-e_{\rm L})^{-1.8}(1+e_{\rm L})^{0.6}$, which is the stability criterion of \cite{MarAar2001}, where $e_{\rm L}$ is the outer orbital eccentricity and we have ignored the very weak dependence on the outer mass ratio of $(1+q_{\rm L})^{0.1}$.  Thus, further dynamical evolution would almost certainly decrease the above frequencies, particularly for the quadruple systems.

\cite{Bate2012} obtained $5.4_{-2.1}^{+3.1}$\% and $8.1_{-2.6}^{+3.5}$\%, respectively, while \cite{Bate2009a} found slightly lower values from the barotropic calculations, although the radiation hydrodynamic results agree within the uncertainties.  \cite{Bate2009a,Bate2012} found the frequencies of high-order multiples increased strongly with primary mass.  This also appears to be the case here.  For VLM primaries, there are no triples or quadruples out of 120 systems.  For M-dwarf primaries ($0.10-0.50$ M$_\odot$) the frequency of triples/quadruples is $10.4_{-2.4}^{+3.0}$\%, while for K-dwarfs to intermediate mass stars the frequency is $35 \pm 5$\%.

Comparing these results with observations,  \cite{FisMar1992} found 7 triples and 1 quadruple amongst 99 M-dwarf primaries giving a frequency of $8\pm 3$\%, in reasonable agreement.  However, the larger, more recent survey of M-dwarfs by \cite{Jansonetal2012} only found a frequency of high-order systems of $\approx 2$\%.  For solar-type primaries, the frequency of triple and higher-order multiple systems has been found to be $11\pm 1$\% by \citet{Raghavanetal2010} and $13 \pm 1$\% by \cite{Tokovinin2014}.  For primaries in the mass range $0.5-1.2$~M$_\odot$, the radiation hydrodynamical simulations give a frequency of $9/41 = 22\pm 7$\%.  But 5 of the 9 high-order systems break the stability criterion of \cite{MarAar2001}, so the eventual frequency may be similar to the observed fractions.  In summary, the frequencies of triples/quadruples obtained from the radiation hydrodynamical calculation tend to be larger than those that are observed, but further dynamical evolution is very likely to bring them into better agreement.

\cite{Bate2009a} found that quadruples composed of a triple and a fourth outer component (so-called 3+1 systems) out numbered quadruples composed of two inner binaries (2+2 systems) by 2:1 in a barotopic calculation.  The combined sample produces 12 of each, but 7 of the 3+1 systems are unstable while only two of the 2+2 systems are unstable so it is likely that 2+2 systems would significantly outnumber 3+1 systems if they were evolved for longer.  Observationally, \cite{Tokovinin2000a} found roughly equal numbers of such quadruples from a large but biased sample, while the very recent distance-limited sample of \cite{Tokovinin2014} contained nine 2+2 systems and only two 3+1 systems. The potential convergence of the relative numbers of 2+2 and 3+1 systems between the observed systems and numerical systems is pleasing, but fundamentally we need much larger samples in both cases. 

\begin{figure*}
\centering
    \includegraphics[width=5.5cm]{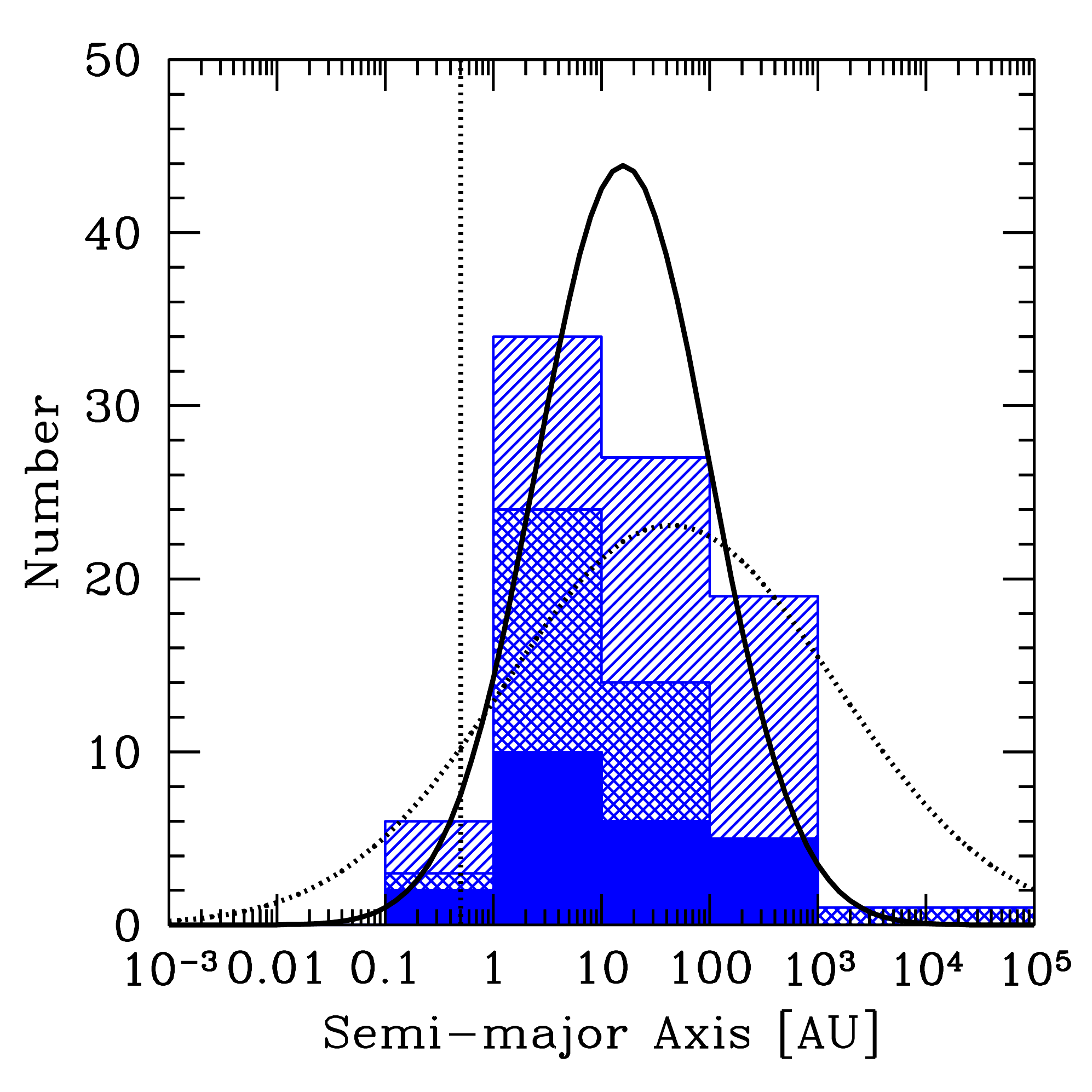}
    \includegraphics[width=5.5cm]{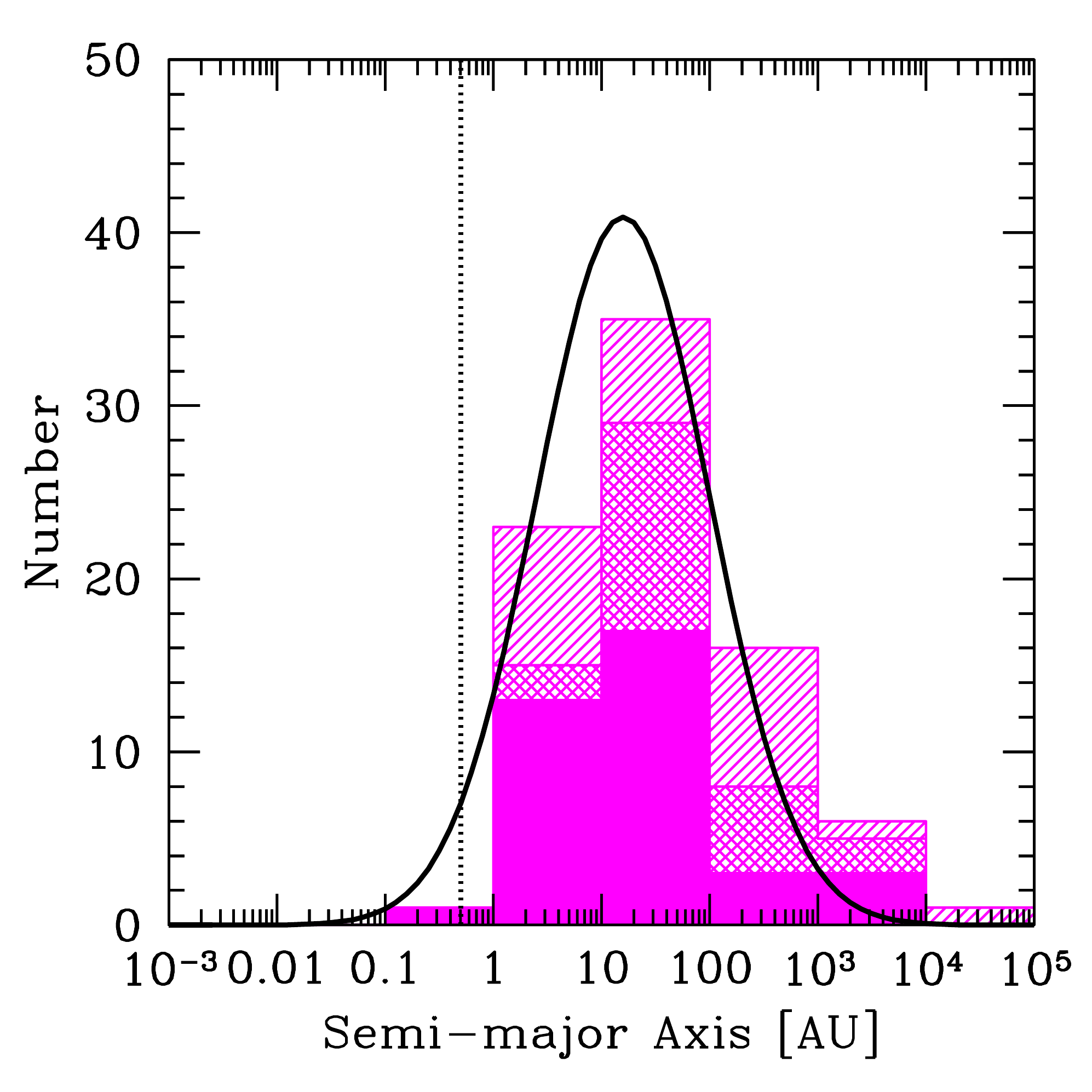}
    \includegraphics[width=5.5cm]{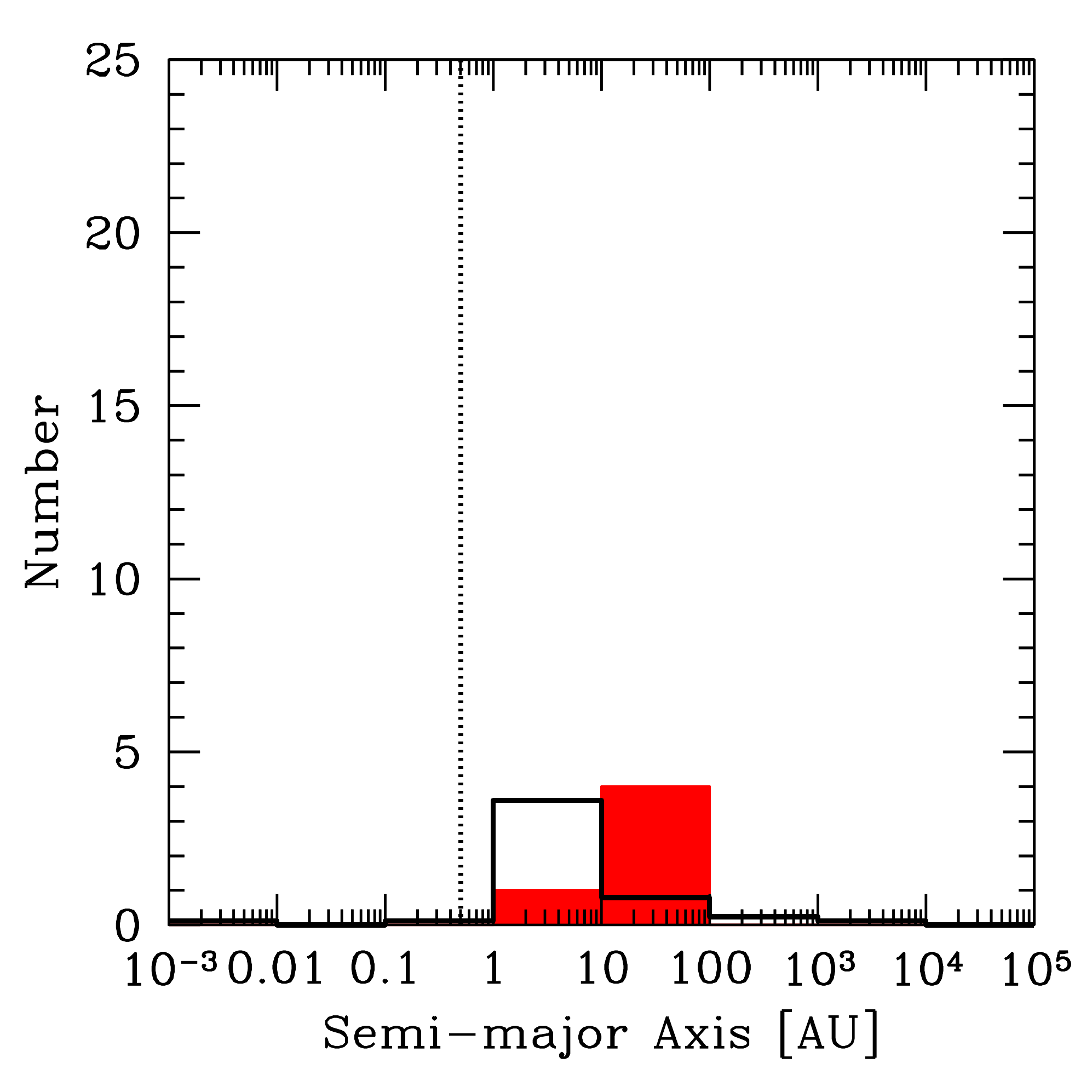}
\caption{The distributions of separations (semi-major axes) of multiple systems with solar and intermediate-mass primaries ($M_1>0.5$~M$_\odot$; left panel), M-dwarf primaries ($M_1 = 0.1 - 0.5$~M$_\odot$; centre panel)  and VLM primaries ($M_1<0.1$~M$_\odot$; right panel) for the combined sample from the three calculations with the highest opacities ($Z\ge 0.1~{\rm Z}_\odot$).  The solid, double hatched, and single hatched histograms give the orbital separations of binaries, triples, and quadruples, respectively (each triple contributes two separations, each quadruple contributes three separations).  The solid curve gives the M-dwarf separation distribution (scaled to match the area) from the M-dwarf survey of Janson et al.\ (2012), the dotted curve gives the separation distribution for solar-type primaries of Raghavan et al.\ (2010), and the open histogram in the VLM plot gives the separation distribution of the very-low-mass binary systems from the list at http://vlmbinaries.org/ . 
The vertical dotted line gives the resolution limit of the calculations as determined by the accretion radii of the sink particles.  }
\label{combined_separation_dist}
\end{figure*}

\subsection{Separation distributions of multiples}

%

As discussed in Section \ref{sec:separations}, the median separation for the multiple systems in the three calculations with the high opacities are all around 16~AU and there is no significant dependence on metallicity.  Combining the multiple systems from the three calculations (Fig.~\ref{combined_separation_dist}), the median of the 170 separations of the multiple systems with primaries with masses $>0.1$~M$_\odot$  is 17.3~AU and the standard deviation of the distribution is 0.94 dex.  Most of these systems have M-dwarf primaries.  Restricting the primaries to M-dwarfs (mass range $0.1-0.5$~M$_\odot$), the median of the 82 separations is 17.9~AU and the standard deviation is 0.91 dex.  \cite{FisMar1992} found that M-dwarf binaries have a median separation of $\approx 10$~AU.  Most recently, \cite{Jansonetal2012} found that the separation distribution of their M-dwarf sample was well fit by a log-normal distribution with a mean separation of 16 AU and a dispersion of 0.8 dex.  Thus, the separation distribution of the M-dwarf multiples is in very good agreement with the observed distributions, as can be seen in the centre panel of Fig.~\ref{combined_separation_dist}.  

On the other hand, both the median separation (40~AU) and the dispersion (1.52 dex) obtained by \cite{Raghavanetal2010} for solar-type stars are larger, whereas we obtain a very similar distribution for solar-type multiples (pimary mass range $0.8-1.2$~M$_\odot$) as for M-dwarf multiples with a median of 16~AU and a standard deviation of 0.72 dex.  We note that the number of close systems from the simulations is likely underestimated because of the lack of dissipation on small scales \citep[see][]{Bate2009a}.  Furthermore, the number of wide systems may be lower than found in the field because the stellar clusters that are formed are quite dense, meaning that it is difficult for wide binaries to exist within the clusters.  There appears to be a similar deficit of wide binaries in the Orion Nebula Cluster (\citealt*{BatClaMcC1998, ScaClaMcC1999}; \citealt{Reipurthetal2007}).  However, this does not necessarily mean that wide systems could not be produced as the stars joint the field population.  \cite{MoeBat2010} and \cite{Kouwenhovenetal2010} have shown that wide systems can be formed as a star cluster disperses \citep[see also][]{MoeCla2011}.

Among the combined sample, we have only 5 VLM binaries (right panel of Fig.~\ref{combined_separation_dist}).  All have separations less than 40 AU (their median is 26~AU, with a standard deviation of 0.42 dex).  This is at odds with observed VLM binaries because the vast majority have projected separations $\lsim 20$~AU.  However, \cite{Bate2009a}, who obtained 32 VLM multiples from a barotropic calculation, found that the median separation of VLM multiples decreased as the calculation was evolved from $\approx 30$~AU at $1.04~t_{\rm ff}$ to $\approx 10$~AU at $1.50~t_{\rm ff}$ because many were still accreting gas and interacting with other systems early on.  He concluded that VLM binaries may form with reasonably wide separations and evolve to smaller separations \citep[c.f.][]{BatBonBro2002b}.  Of the 5 VLM multiples in the combined sample, 3 were still accreting when the calculations were stopped, so it is likely that the separation distribution would continue to evolve if the calculations were followed further.  In addition to evolution of VLM binary separations during their formation, the observational studies of \cite{Closeetal2007} and \cite{Burgasseretal2007} suggest that young wide VLM binaries are disrupted, leading to the observed paucity of old wide VLM systems.  \citet{Closeetal2007} estimated that young VLM objects have a wide ($>100$ AU) binary frequency of $\sim 6$\%$\pm$3\% for ages less than 10 Myr, but only 0.3\%$\pm$0.1\% for field VLM objects.  Thus, it is at least plausible that the VLM binaries from the simulations are two wide because they neglect further evolution.

\begin{figure*}
\centering
    \includegraphics[width=5.5cm]{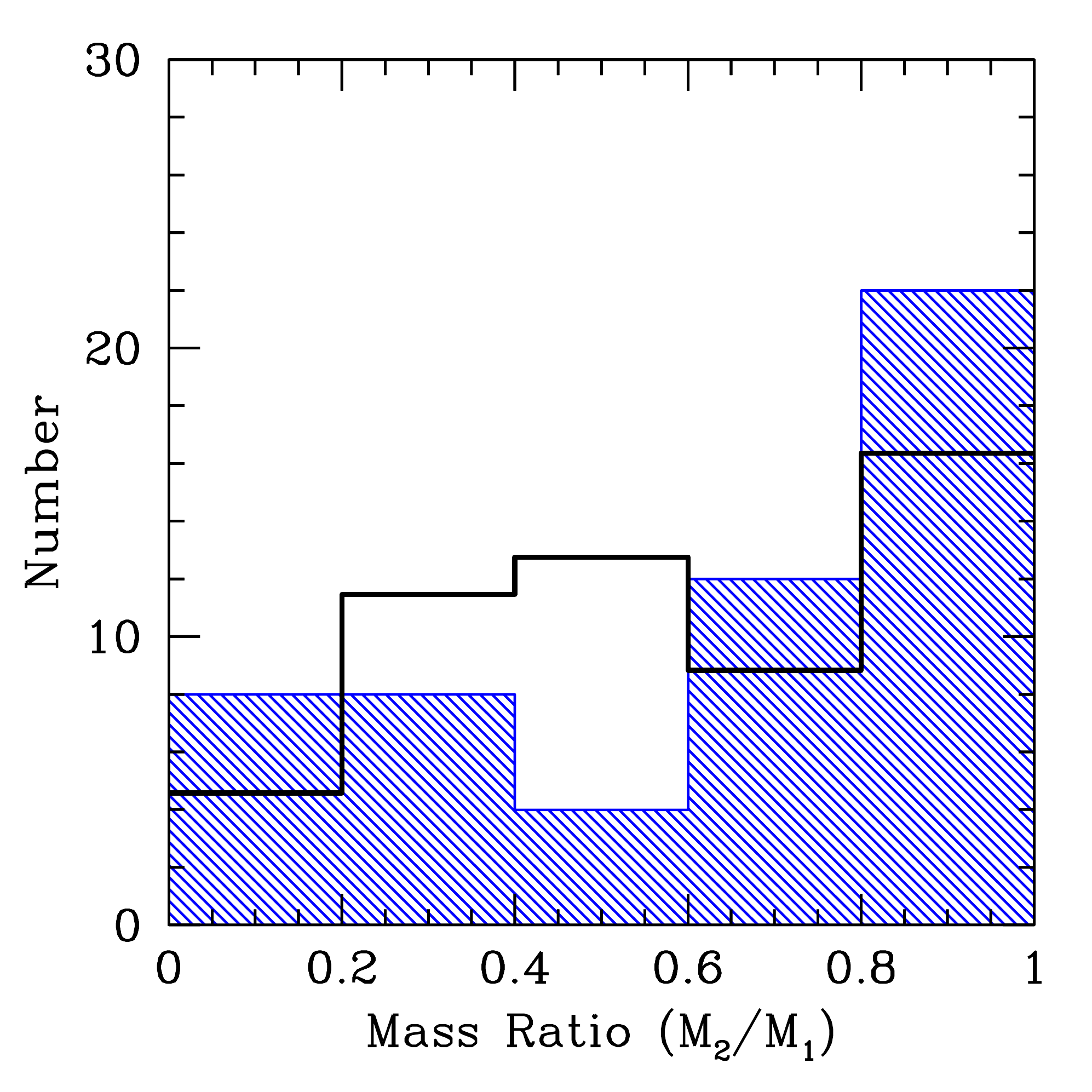}
    \includegraphics[width=5.5cm]{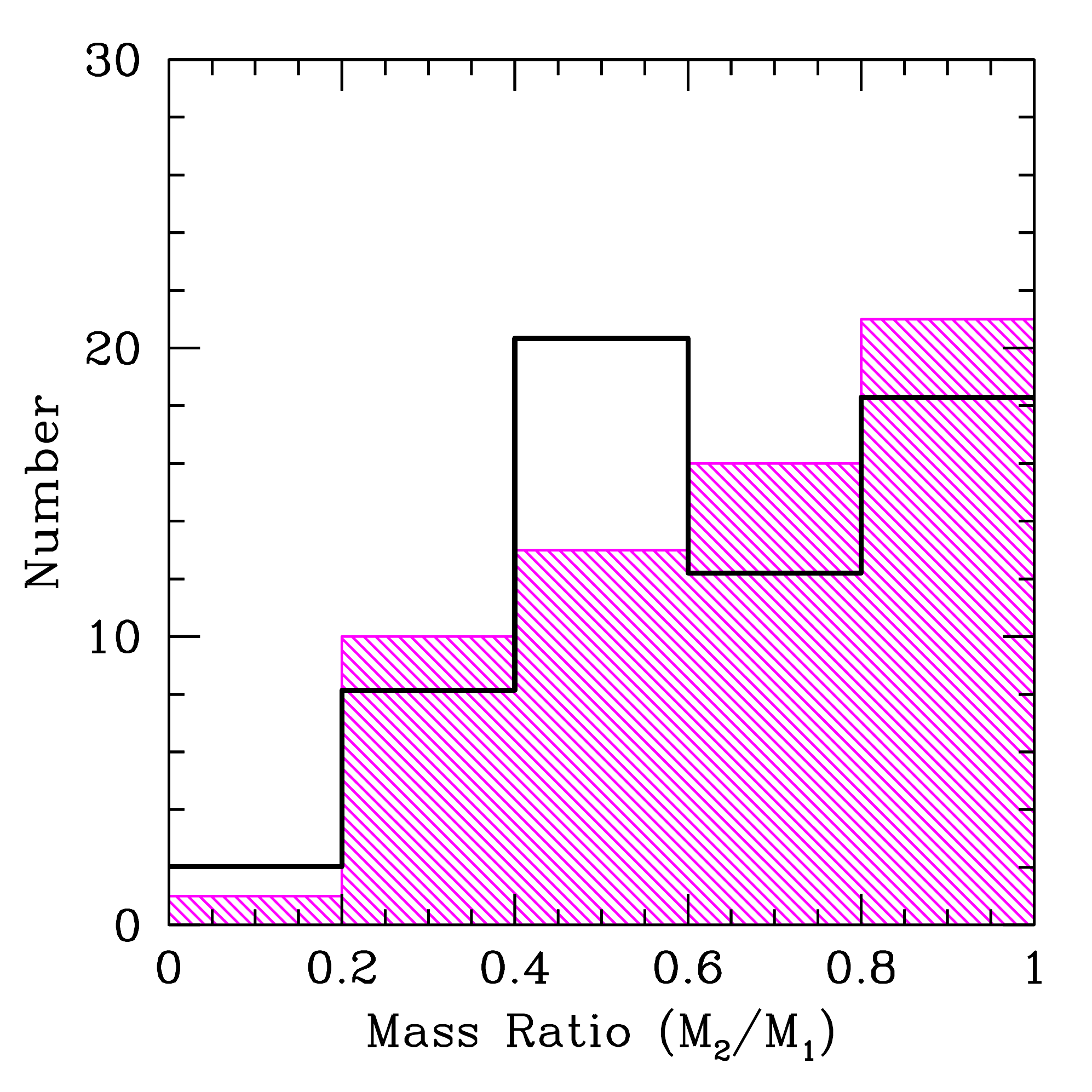} 
    \includegraphics[width=5.5cm]{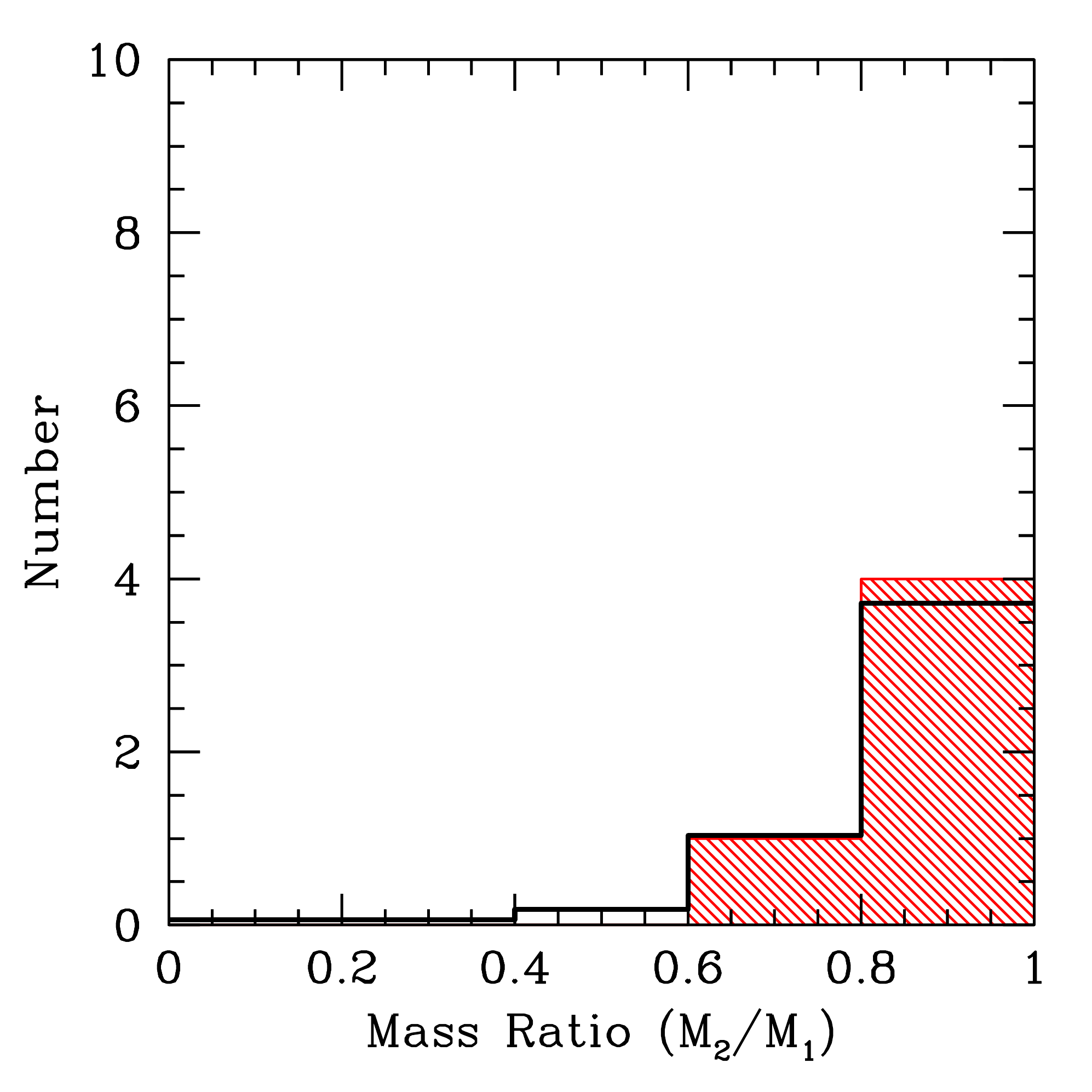}
    \caption{The mass ratio distributions of binary systems with stellar primaries in the mass ranges $M_1>0.5$ M$_\odot$ (left) and $M_1=0.1-0.5$ M$_\odot$ (M-dwarf; centre) and VLM primaries (right; $M_1<0.1$ M$_\odot$) produced by the combined sample from the three calculations with the highest opacities ($Z\ge 0.1~{\rm Z}_\odot$).  The solid black lines give the observed mass ratio distributions of \citet{Raghavanetal2010} for binaries with solar-type primaries (left), \citet{Jansonetal2012} for M-dwarfs (centre), and of the known very-low-mass binary systems from the list at http://vlmbinaries.org/ (right).  The observed mass ratio distributions have been scaled so that the areas under the distributions match those from the simulation results. The VLM binary mass ratio distribution and the M-dwarf distributions from the combined sample are in good agreement with the observed distributions, but among the more massive primaries there may be an excess of near equal-mass binaries. }
\label{combined_massratios}
\end{figure*}

\subsection{Mass ratio distributions of binaries}

As discussed in Section \ref{sec:massratios}, there is no evidence for the mass ratio distributions of the binaries produced by the radiation hydrodynamical simulations depending on the opacity.  In Fig.~\ref{combined_massratios} we plot the mass ratio distributions for the combined sample in three primary mass ranges:  the more massive binaries ($M_1>0.5$~M$_\odot$), M-dwarf binaries ($M_1=0.1-0.5$~M$_\odot$), and VLM binaries ($M_1<0.1$~M$_\odot$).  The numerical results are compared with the observed mass ratio distributions from \cite{Raghavanetal2010}, \cite{Jansonetal2012}, and the list of VLM binaries at http://vlmbinaries.org/, respectively.  The distributions include both binaries, and pairs that are components of higher-order systems.

For the most massive binaries, the numerical distribution is consistent with being flat with the exception of a large number of near equal-mass systems (`twins').  Forty percent of the systems have mass ratios $M_1/M_2>0.8$.  The observed distribution of \cite{Raghavanetal2010} also has a significant fraction of such systems (just over 1/3), though not quite as large a proportion as the combined numerical sample.  Performing a Kolmogorov-Smirnov test on the two distributions gives a 4\% probability of them being drawn from the same underlying distribution, so formally they are statistically indistinguishable.  The numerical and observed mass ratio distributions for both the M-dwarfs and the VLM binaries are clearly in very good agreement with each other.

Observationally, several older studies found evidence that the mass ratio distribution of binaries depends on primary mass.  \citet{DuqMay1991} found that the mass ratio distribution of solar-type binaries peaked at $M_2/M_1 \approx 0.2$, while \citet{FisMar1992} found a flat mass ratio distribution in the range $M_2/M_1 = 0.4-1.0$ for M-dwarf binaries, and VLM binaries have been found to have a strong preference for equal-mass systems \citep{Closeetal2003, Siegleretal2005,Reidetal2006}.

However, more recent studies have called this apparent trend into question.  \cite{Raghavanetal2010} found a flat mass ratio distribution for solar-type primaries in the range $M_2/M_1=0.2-0.95$, with a drop-off at lower mass ratios and a strong peak at nearly equal masses (so-called twins; \citealt{Tokovinin2000b}).  \citet{Jansonetal2012} also found that the M-dwarf mass ratio distribution is well fit by a uniform mass ratio distribution, but it can also be fit with a distribution that slowly rises towards equal masses.  In the Taurus-Auriga star-forming region, \cite{Krausetal2011} report a flat mass ratio distribution for primaries in the range $0.7-2.5$~M$_\odot$, but for primaries in the mass range $0.25-0.7$~M$_\odot$ they find a bias toward equal-mass systems.  Thus, apart from the apparent preference for equal-mass VLM binaries, mass ratio distributions seem to be more similar than indicated in the past.  In fact, \cite{RegMey2013} argue that the mass ratio distributions of M-dwarf and solar-type binaries are currently indistinguishable. 

\begin{figure}
\centering
\includegraphics[width=8cm]{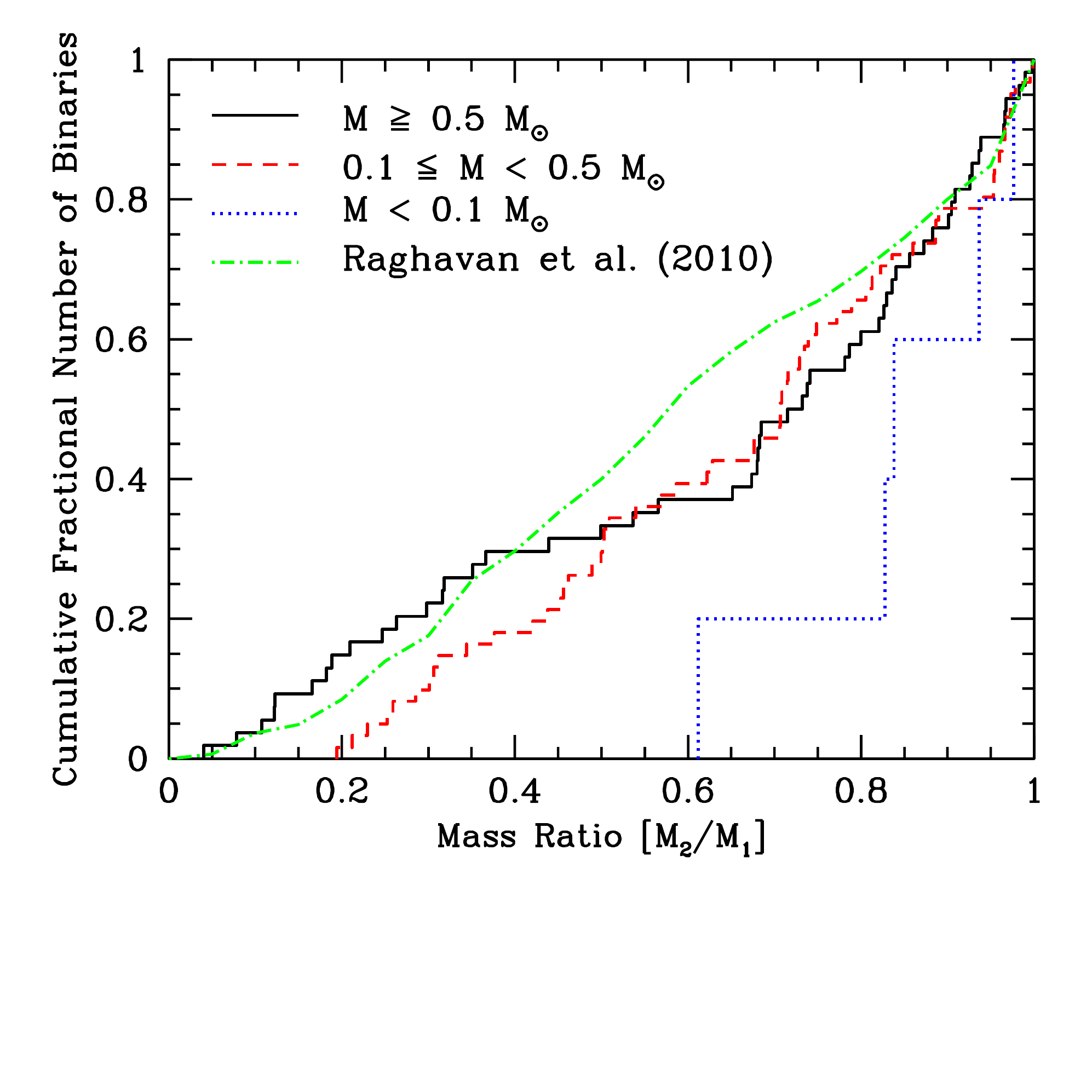}  \vspace{-1.5cm}
\caption{The cumulative mass ratio distributions of binary systems with stellar primaries in the mass ranges $M_1 > 0.5$ M$_\odot$ (black solid line) and $M_1=0.1-0.5$ M$_\odot$ (red dashed line) and VLM primaries (blue dotted line; $M_1<0.1$ M$_\odot$) produced by the combined sample from the three calculations with the highest opacities ($Z\ge 0.1~{\rm Z}_\odot$). There appears a preference for VLM binaries to have near equal masses.  However, a Kolmogorov-Smirnov test shows that all three distributions are consistent with being drawn randomly from  the same underlying distribution (there is a 50\% probability that the solar-type and M-dwarf distributions are drawn from the same underlying population, and a 13\% probability that the M-dwarf and VLM distributions are drawn from the same underlying population). We also plot the mass ratio distribution of binaries and pairs in higher-order systems for solar-type primaries as obtained by Raghavan et al. (2010) (green dot-dashed line).  A Kolmogorov-Smirnov test gives a 4\% probability of the $M_1 > 0.5$ M$_\odot$ and Raghavan et al. distributions being drawn from the same underlying population. }
\label{combined_cum_massratios}
\end{figure}

In Fig.~\ref{combined_cum_massratios}, we compare the cumulative mass ratio distributions of the VLM, M-dwarf, and more massive binaries obtained from the simulations with each other.  We also plot the distribution observed by \cite{Raghavanetal2010}.  Performing Kolmogorov-Smirnov tests on each pair of distributions, we find no evidence for a dependence of the mass ratio distribution on primary mass (see the figure caption for more details).

\subsubsection{Mass ratio versus separation}

\begin{figure}
\centering
    \includegraphics[width=8.4cm]{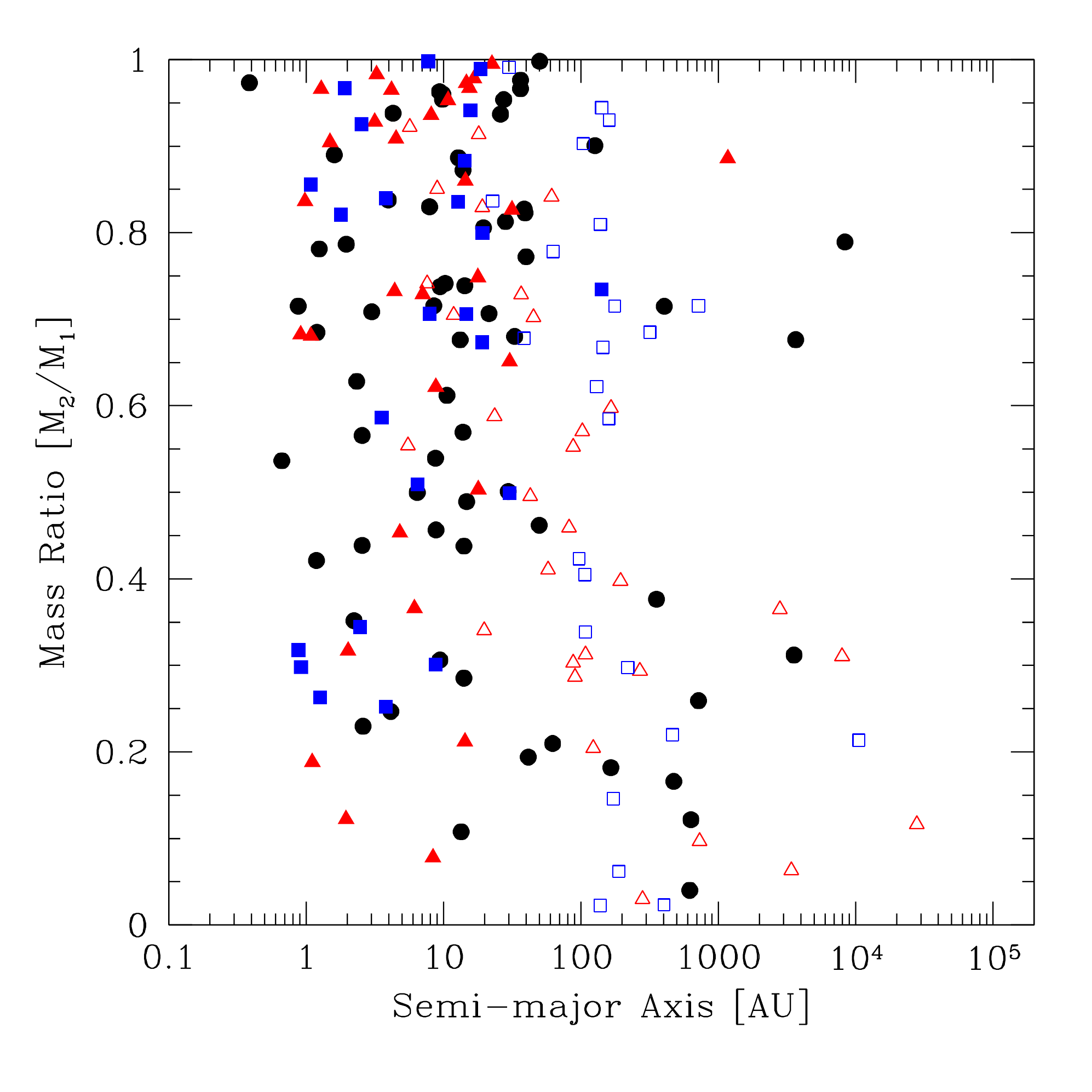}
\caption{The mass ratios of binaries (filled black circles), inner pairs in triples (filled red triangles), inner pairs in quadruples (filled blue squared),  the outer components of triples (open red triangles), and the widest components of quadruples (open blue squares) as a function of semi-major axis for the combined sample from the three radiation hydrodynamical calculations with the highest opacities ($Z\ge 0.1~{\rm Z}_\odot$).  For the outer components of triples, the outer mass ratio compares the mass of the outer component to the sum of the masses of the two inner components (the pair).  For quadruples involving a two binary components (inner pairs), the outer mass ratio is between the two pairs, and for quadruples involving a triple, the mass ratio is between the mass of the fourth component and the combined mass of the triple.  There is a clear relationship between mass ratio and separation with closer binaries having a greater fraction of near equal-mass systems.}
\label{a_q}
\end{figure}

In Fig.~\ref{a_q}, we plot mass ratios against separation (semi-major axis) for the binaries, triples, and quadruples from the combined sample.  The median mass ratios for {\it binary} separations in the ranges $1-10$, $10-100$, and $100-1000$ AU are $M_2/M_1=0.64, 0.67, 0.35$, respectively, so wide systems (separations $>100$~AU) appear to have smaller mass ratios than close systems. 
This is similar to the results obtained from earlier calculations. \cite{Bate2009a} found a clear relation between mass ratio and separation from barotropic calculations, with closer binaries having a preference for equal masses.  From the radiation hydrodynamical calculation with solar-metallicity, \cite{Bate2012} found weaker dependence of mass ratio on separation, but did note that there appeared to be a trend in which systems with separations $\lsim 100$~AU had more equal masses than for wider systems.  

In terms of the so-called twins, the combined sample contains 120 binaries (including pairs in triple and quadruple systems), of which there are 19 twins (pairs with mass ratios $M_2/M_1>0.95$) and all have semi-major axes less than $\approx 50$~AU.  This is in good agreement with observations that consistently find that closer binaries have a higher fraction of twins \citep{Soderhjelm1997, Tokovinin2000b, Halbwachsetal2003}.  \citet{Tokovinin2000b} found evidence for the frequency of twins falling off for orbital periods greater than 40 days, while \citet{Halbwachsetal2003} found that the fraction of near equal-mass systems ($M_2/M_1>0.8$) is always larger for shorter period binaries than longer period binaries regardless of the dividing value of the period (from just a few days up to 10 years).  The recent study of \cite{Raghavanetal2010} found the mass ratio distribution depends on period, with less than 1/4 of twins having periods longer than 200 years (separations $\approx 40$~AU) and no twins having periods greater than 1000 years (separations of $\approx 115$~AU).  However, the larger survey of \cite{Tokovinin2014} does not find much evidence for a dependence of mass ratio on period.  A trend of more equal-mass binaries with decreasing separation is expected from the evolution of protobinary systems accreting gas from an envelope \citep{BatBon1997, Bate2000}.  Furthermore, dynamical interactions between binaries and single stars tend to tighten binaries at the same time as increasing the binary mass ratio through exchange interactions.


\begin{figure*}
\centering
    \includegraphics[width=8.4cm]{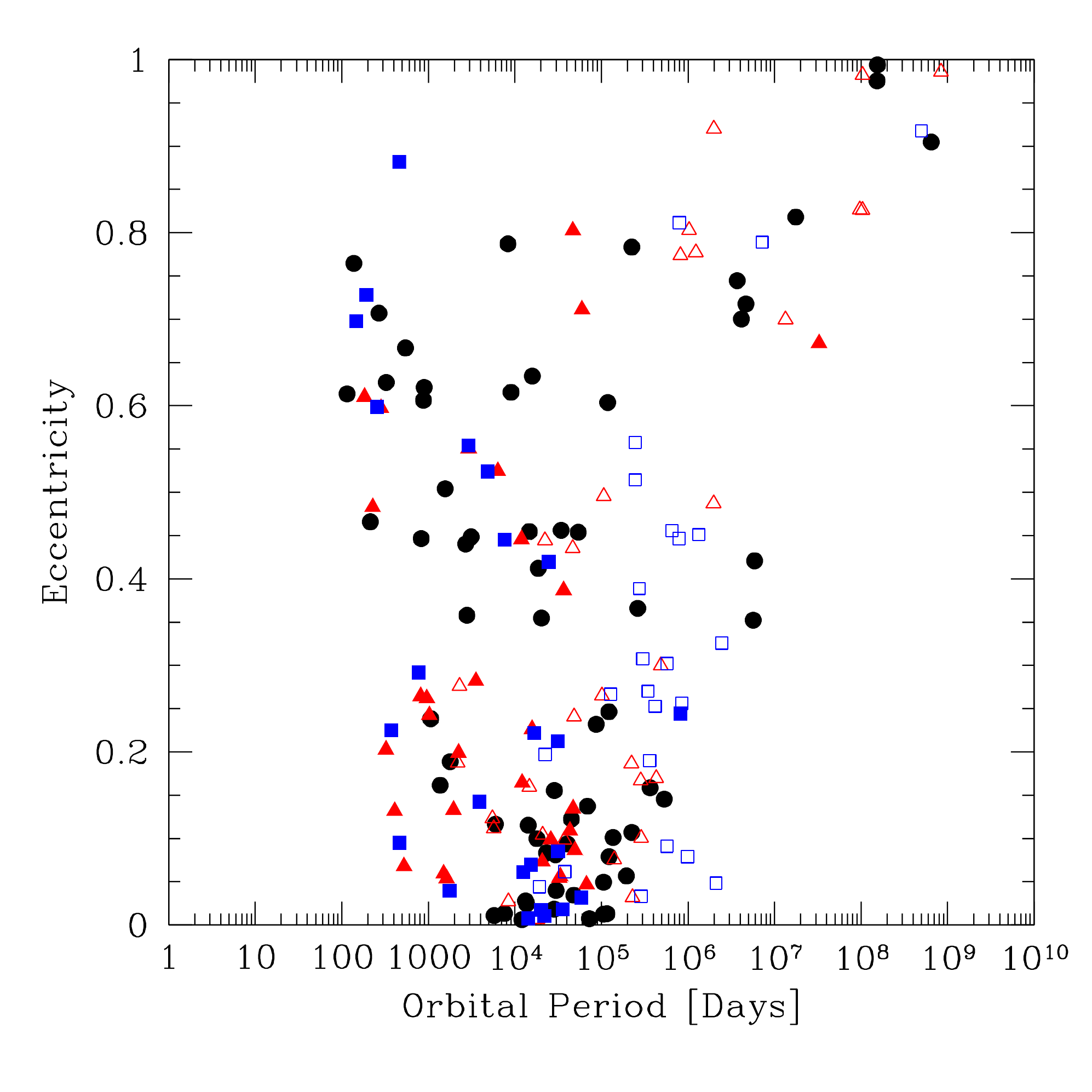}
    \includegraphics[width=8.4cm]{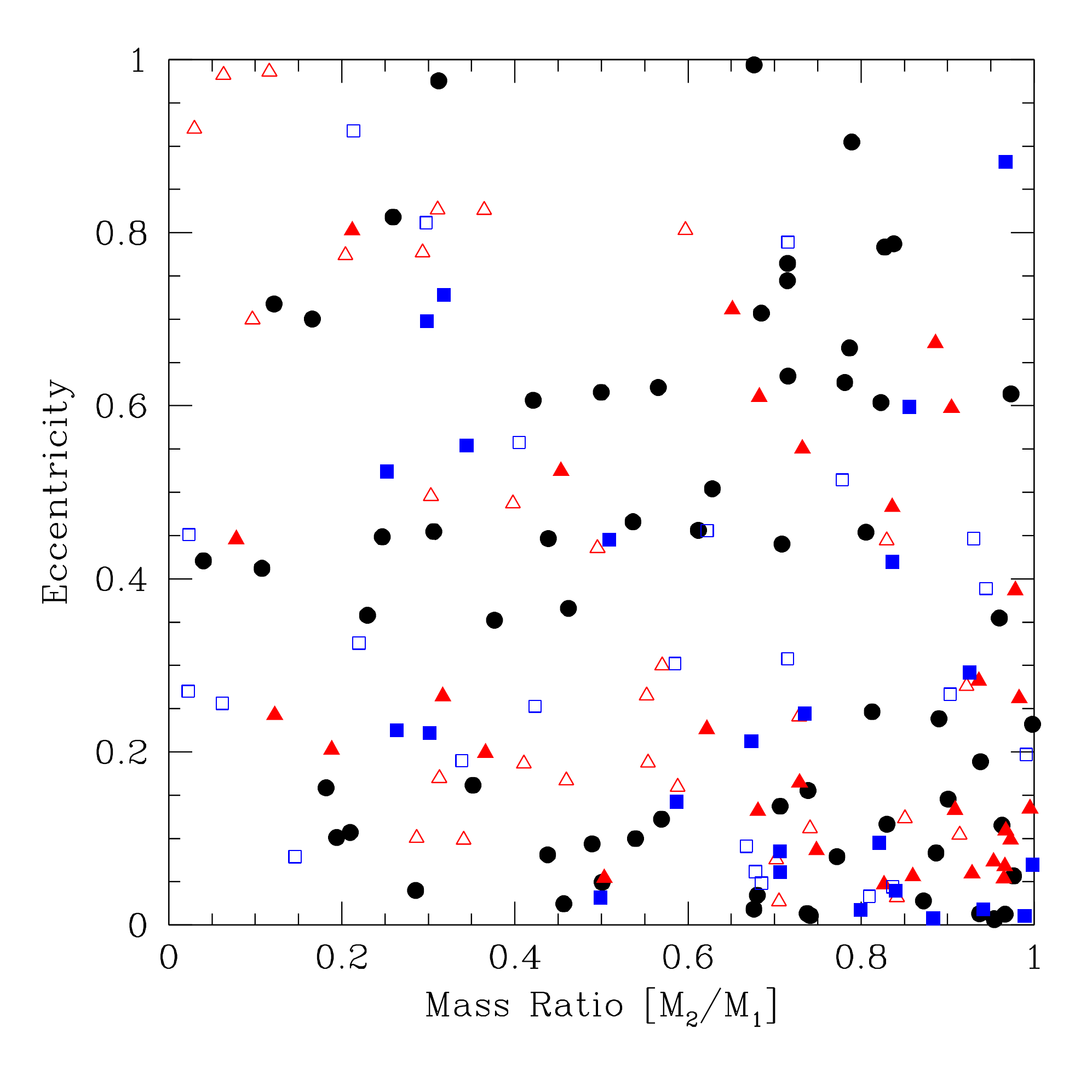}
\caption{The eccentricities of binaries (filled black circles), inner pairs in triples (filled red triangles), inner pairs in quadruples (filled blue squares), 
the outer components of triples (open red triangles), and the widest components of quadruples (open blue squares) as a function of orbital period (left) and mass ratio (right)  for the combined sample from the three radiation hydrodynamical calculations with the highest opacities ($Z\ge 0.1~{\rm Z}_\odot$). There are a significant number of binaries with eccentricities lower than 0.1 which are very rare observationally \citep{DuqMay1991, Halbwachsetal2003, Raghavanetal2010}.  However, it is also clear that most of these low-eccentricity binaries also have high mass ratios, similar to the observations of \citeauthor{Halbwachsetal2003} (\citeyear{Halbwachsetal2003}).}
\label{eccentricity}
\end{figure*}

\subsubsection{Mass ratios of triples and quadruples}

For stellar triple and quadruple systems, \citet{Tokovinin2008} reports that triples are observed to have a median outer mass ratio of 0.39 independent of the outer orbital period while quadruples involving two binary sub-components have a similar median outer mass ratio of $\approx 0.45$, but there appears to be a dependence on the outer orbital period with systems with shorter outer periods having higher mass ratios.  Of 31 triple systems, we obtain a median mass ratio of 0.51 (0.52 excluding the 12 triples which are members of quadruple systems).  

There are 12 quadruple systems consisting of two pairs, with outer mass ratios ranging from $M_3/(M_1+M_2)=0.22-0.99$ and outer periods $4.3 < \log_{10}(P_{\rm d})<6.9$ (measured in days).  \cite{Tokovinin2008} finds no outer mass ratios $<0.6$ for orbital periods $\log_{10}(P_{\rm d})<5.4$, but a wide range of outer mass ratios for longer periods.  In the combined sample, we have four quadruple systems consisting of two pairs and with outer orbital periods $\log_{10}(P_{\rm d})<5.4$ and all have outer mass ratios $>0.6$, while the other 8 quadruples consisting of two pairs have outer mass ratios ranging from 0.22--0.93 with a median outer mass ratio of 0.65 and three having outer mass ratios $<0.6$.  Thus, our quadruple systems are in agreement with the findings of \cite{Tokovinin2008}, although we note that the observed values are based on biased samples from catalogs and the observed systems tend to contain stars more massive than the Sun.

\subsection{Orbital eccentricities}

Observationally, there is an upper envelope to binary eccentricities at periods less than about one year, and binaries with periods less than 12 days are almost exclusively found to have circular orbits due to tidal circularisation \citep{DuqMay1991, Udryetal1998, Halbwachsetal2003, Raghavanetal2010}.  However, the radiation hydrodynamical calculations do not allow us to probe such small separations due to the absence of dissipation on scales $<0.5$~AU.  Observations also indicate that eccentricities $e<0.1$ are rare for periods greater than $\approx 100$ days (separations $\gsim 1$ AU).  \cite{Raghavanetal2010} finds no binaries with $e<0.1$ and orbital periods greater than 100 days, though they do find that the outer orbits of two triples and one quadruple have $e<0.1$.  \cite{DuqMay1991} and \cite{Raghavanetal2010} also find that the upper-eccentricity envelope is dominated by components of triple systems, possibly due to the action of the Kozai mechanism \citep{Kozai1962}.  Finally, \citet{Halbwachsetal2003} find that the eccentricities of binaries with mass ratios $M_2/M_1>0.8$ with periods greater than $\approx 10$ days (the tidal circularisation radius) are lower than for more unequal mass ratio systems.

In the left panel of Fig.~\ref{eccentricity} we plot the eccentricities versus orbital period for the binaries, triples and quadruples from the combined sample, including those that are sub-components of higher-order systems.  The mean eccentricity of all 175 orbits is $e=0.33 \pm 0.02$ with a standard deviation of 0.28.  The median is $e=0.25$. The mean eccentricity of binaries (including components of triples and quadruples) is $e=0.32 \pm 0.02$ with a standard deviation of 0.23.  The mean eccentricity of the triples and quadruples is $e=0.37 \pm 0.04$ with a standard deviation of 0.29.   The mean eccentricities obtained by \cite{Bate2009a} for the barotropic calculation with accretion radii of 0.5~AU and \cite{Bate2012} for the radiation hydrodynamic calculation with solar metallicity were $e=0.45$ and 0.35, respectively.  The median eccentricity from \cite{Raghavanetal2010} is about $e=0.4$, so there is reasonable agreement.

However, \cite{Raghavanetal2010} report a flat distribution of eccentricities for periods longer than 12 days out to $e=0.6$, whereas the combined sample produces approximately twice as many orbits with $e<0.2$ (77 orbits) compared to the the intervals $0.2 \leq e < 0.4$ (31 orbits) and $0.4 \leq e < 0.6$ (29 orbits).  In particular, there are 36 binaries with $e<0.1$, where as observed systems with $e<0.1$ are rare.  More than half of these (19) are components of triple or quadruple systems, which may be related to the finding of \cite{Raghavanetal2010} that components of higher-order multiple systems can have low eccentricities.  Furthermore, 21 of the 36 have mass ratios $M_2/M_1>0.8$ (right panel of Fig.~\ref{eccentricity}) which is in qualitative agreement with the finding of \citet{Halbwachsetal2003} that near-equal mass binaries have smaller eccentricities than more unequal mass ratio systems.  Both \cite{Bate2009a} and \cite{Bate2012} found evidence that near-equal mass binaries tend to have smaller eccentricities.  In the combined sample, the mean eccentricity of binaries with mass ratios $M_2/M_1<0.8$ is $e=0.37\pm 0.03$ (73 orbits) while for $M_2/M_1>0.8$ the median is $e=0.24\pm 0.04$ (47 orbits).  Thus, we also find evidence for a link between mass ratio and eccentricity such that near-equal mass systems have lower eccentricities, as is observed.  A possible explanation for this is that accretion, which drives binaries towards equal masses \citep{Artymowicz1983,Bate1997,BatBon1997,Bate2000}, may also provide dissipation which damps eccentricity.  However, given that there are so many systems with eccentricities less than $e=0.1$, it seems that the damping may be too effective in the numerical simulations.

\begin{figure*}
\centering
    \includegraphics[width=8.4cm]{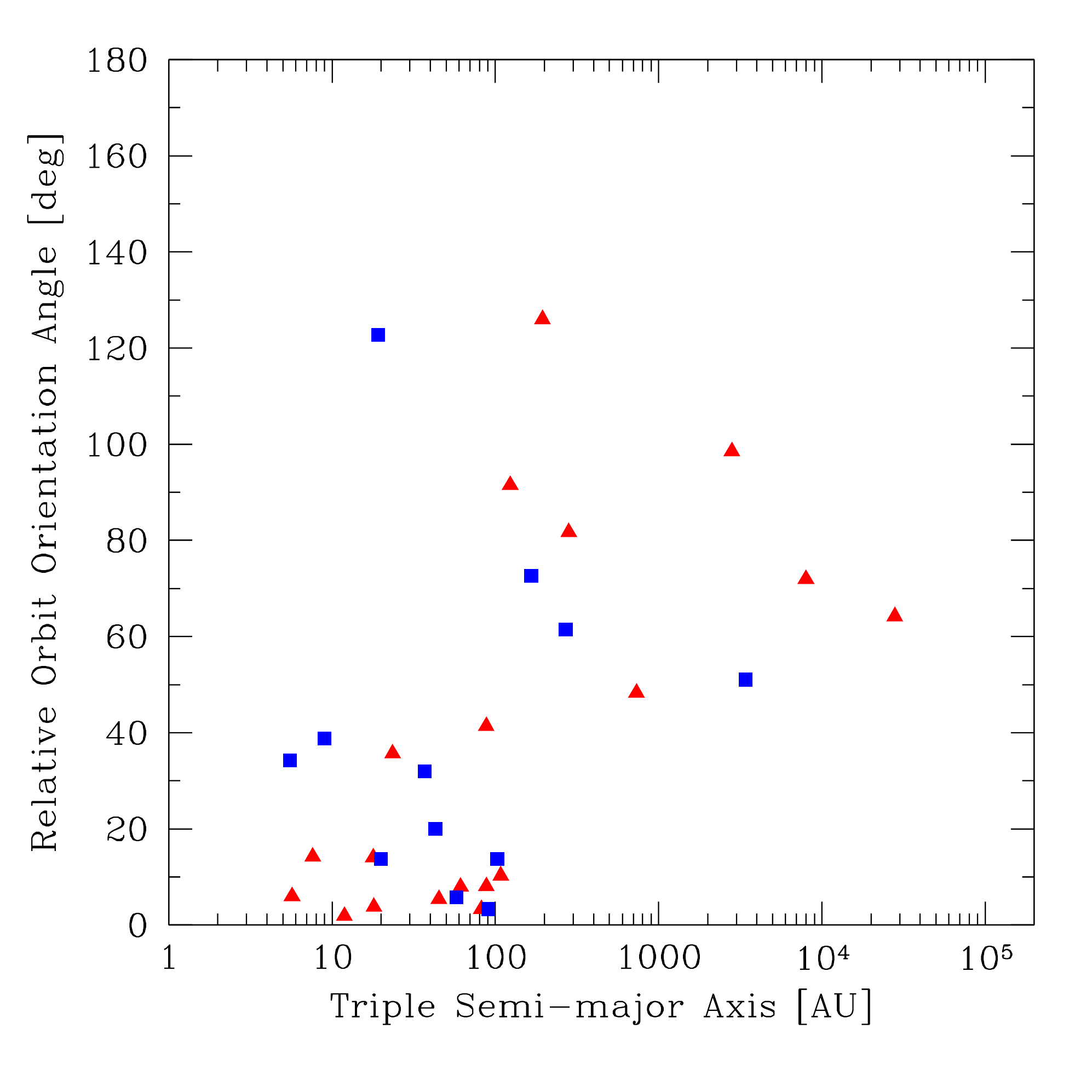}
    \includegraphics[width=8.4cm]{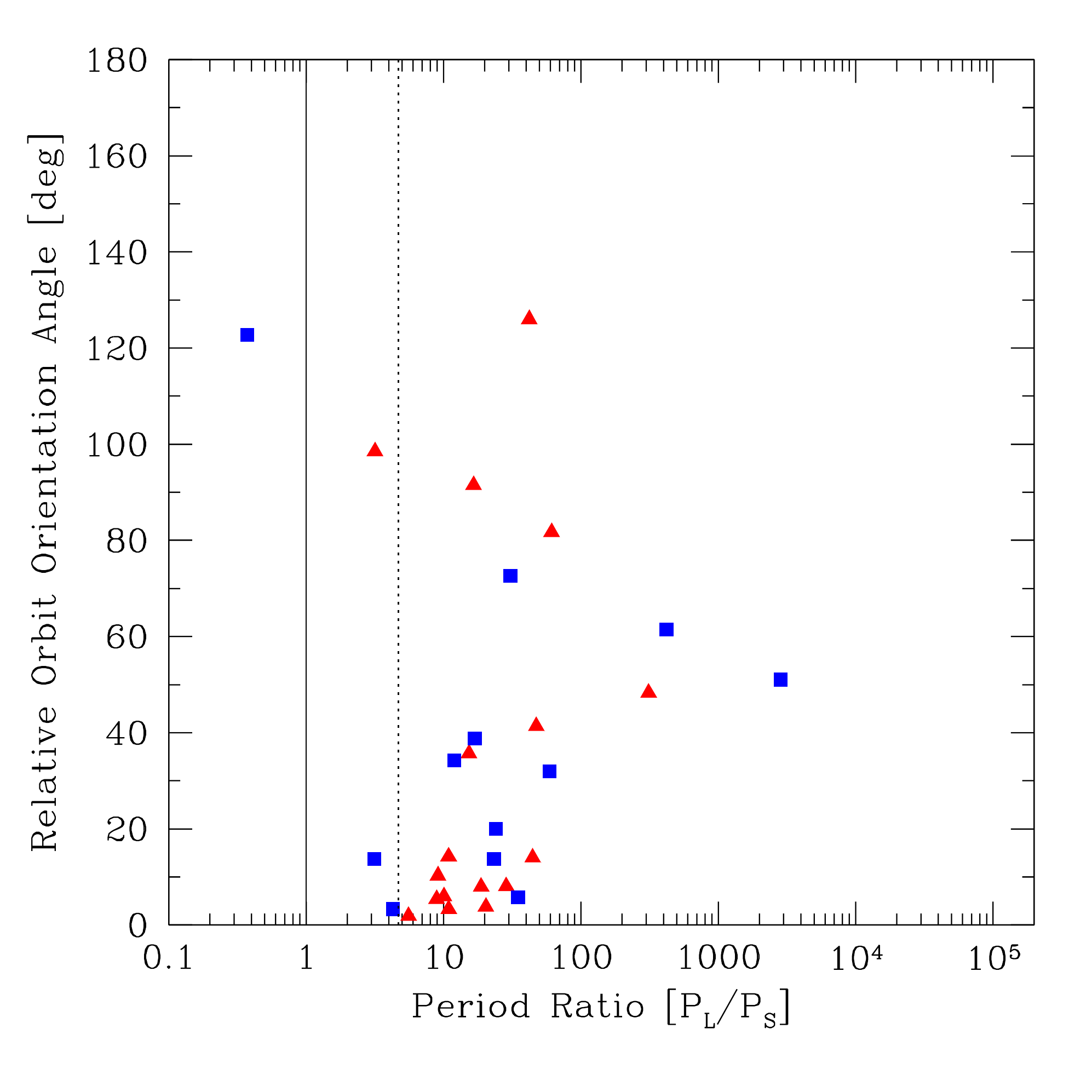}
\caption{The relative inclinations of the two orbital planes for the 31 triple systems included in the combined sample from the three radiation hydrodynamical calculations with the highest opacities ($Z\ge 0.1~{\rm Z}_\odot$).  Triples that are sub-components of quadruples are plotted as blue squares. We give plots of the relative orbital orientation angle versus the semi-major axis of the third component (left) and versus the period ratio of the long and short period orbits (right).  Wider triples and/or systems with larger period ratios tend to have larger relative orbital angles.  Note that the four systems with period ratios $P_{\rm L}/P_{\rm S} < 4.7$ (to the left of the dotted vertical line in the right panel) are likely to be dynamically unstable and to undergo further evolution.}
\label{triples:a_periodratio}
\end{figure*}

Finally, we note that VLM binaries are observed to have a preference for low eccentricities with a median value of 0.34 \citep{DupLiu2011}.  The barotropic calculation of \cite{Bate2009a} with small accretion radii also produced low-eccentricity VLM binaries \citep{Bate2010b}, with those VLM binaries with separations less than 30 AU having a mean eccentricity of 0.23.  Unfortunately, the combined sample only contains 5 VLM binaries, so there is little that can be said about their eccentricity distribution.  They have eccentricities ranging from 0.013 to 0.79.

\subsection{Relative alignment of orbital planes for triples}

For a hierarchical triple system there are two orbital planes, one corresponding to the short-period orbit and one to the long-period orbit.  Observationally, it is difficult to determine the relative orientation angle, $\Phi$, of the two orbits of a triple system due to the number of quantities that must be measured to fully characterise the orbits. However, the mean value of $\Phi$ can be measured simply from knowledge of the number of co-rotating and counter-rotating systems \citep{Worley1967, Tokovinin1993, SteTok2002}.

The first studies \citep{Worley1967,vanAlbada1968b} of the relative orbital orientations of triple systems found a small tendency towards alignment of the angular momentum vectors of the orbits.  Of 54 systems with known directions of the relative motions, 39 showed co-revolution and 15 counter-revolution resulting in a mean relative inclination angle of $\langle\Phi\rangle\approx 50^\circ$.  For 10 visual systems with known orbits, 5 systems were found to have $\Phi<90^\circ$, 2 had $\Phi>90^\circ$ and 3 were ambiguous.   \citet{Fekel1981} examined 20 systems with known orbits and periods of less than 100 years (for the wide orbit).  He found that 1/3 had non-coplanar orbits.  Finally, \citet{SteTok2002} performed the most detailed study to date.  From 135 visual triple systems for which the relative directions of the orbital motions are known they found $\langle\Phi\rangle=67^\circ\pm 9^\circ$ and this result was also consistent with 22 systems for which the orbits were known.  They also found a tendency for the mean relative orbital angular momentum angle to increase with increasing orbital period ratio (i.e.\ systems with more similar orbital periods tend to be more closely aligned).

The main barotropic calculation of \cite{Bate2009a} produced 40 triple systems (17 of which were sub-components of quadruple systems), with a mean relative orbital orientation angle of $\langle\Phi\rangle=65^\circ\pm 6^\circ$, in good agreement with the observed value.  

The combined sample provides 31 triple systems, 12 of which are components of quadruple systems.  The mean relative orbital orientation angle of the all these triple systems is $\langle\Phi\rangle=39^\circ\pm 7^\circ$, which is about 2.5$\sigma$ lower than the observed value.  For the 19 pure triples, $\langle\Phi\rangle=39^\circ\pm 9^\circ$, and further excluding the three unstable triples (see Section \ref{freq_high_order}) gives $\langle\Phi\rangle=29^\circ\pm 9^\circ$.  The relative angles are plotted in Fig.~\ref{triples:a_periodratio} as functions of semi-major axis and period ratio.  It can be seen that systems with larger period ratios tend to have larger relative orbital angles in qualitative agreement with observations.  This also applies to systems with wider third components in general.  We conclude that both the observed and simulated triple systems have a tendency towards orbital coplanarity, but as with the eccentricities it appears that the systems produced by the simulations may be overly dissipative.  Note also, that triples with $\Phi \sim 90^\circ$ will be subject to Kozai evolution and will merge, which is another way to reduce the frequency of high-order multiples (c.f. Section \ref{freq_high_order}).

\subsection{Relative alignment of discs and orbits}

Finally, we consider the relative alignment of the spins of the sink particles in binary systems.  Unfortunately there is not a direct analogy with real binary systems in this case because the sink particles are larger than stars and yet smaller than a typical disc.  The orientation of the sink particle spin thus represents the orientation of the total angular momentum of the star and the inner part of its surrounding disc.  This distinction is important because during the formation of an object the angular momentum usually varies with time as gas falls on to it from the turbulent cloud.  Thus, the orientation of the sink particle frequently differs substantially from the orientation of its resolved disc (if one exists) and, furthermore, the orientations of both the sink particles and their discs change with time while the object continues to accrete gas \citep[e.g.][]{BatLodPri2010}. The orientations may also evolve with time due to gravitational torques \citep{Bateetal2000}.

\begin{figure*}
\centering
    \includegraphics[width=4.7cm]{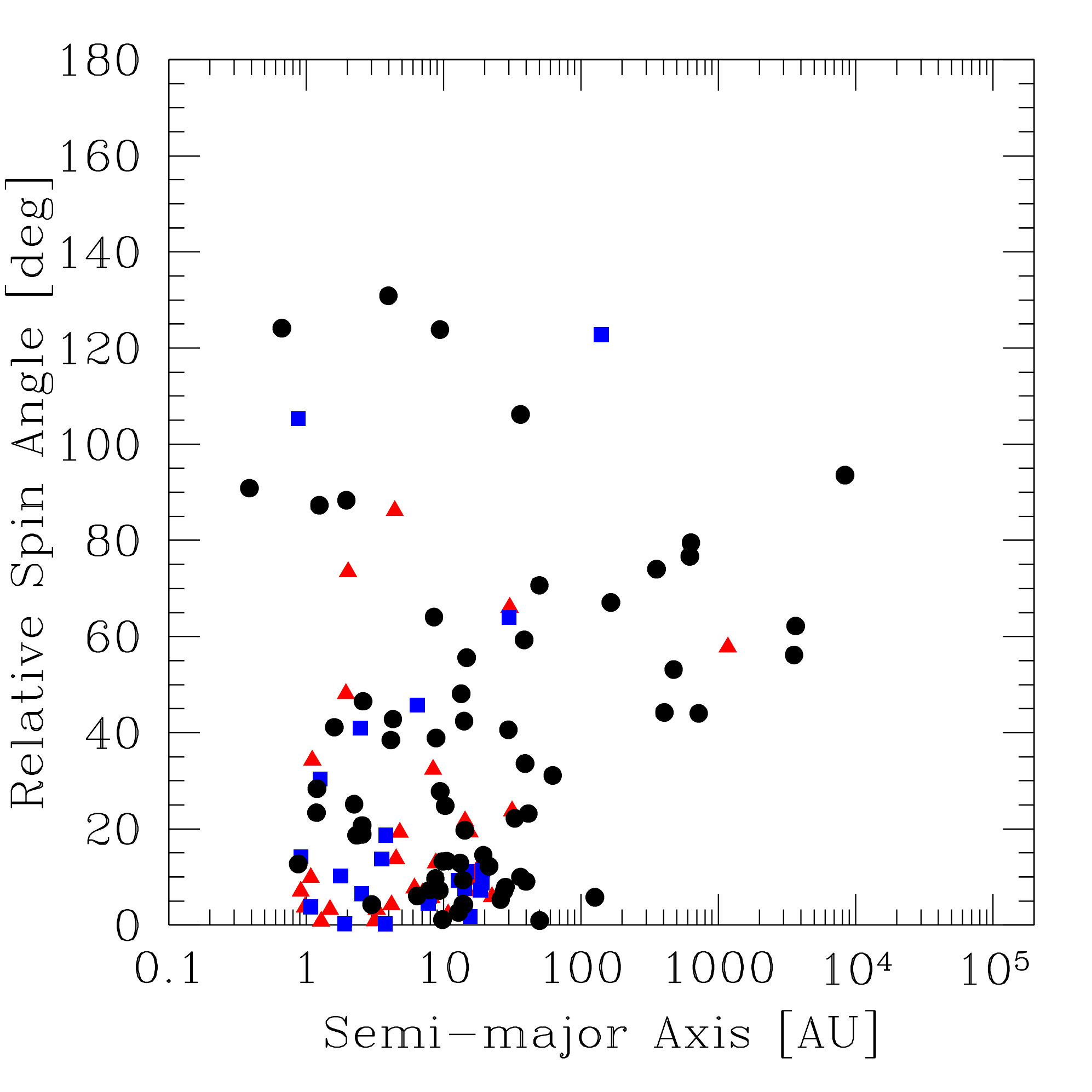} \hspace{-0.6cm}
    \includegraphics[width=4.7cm]{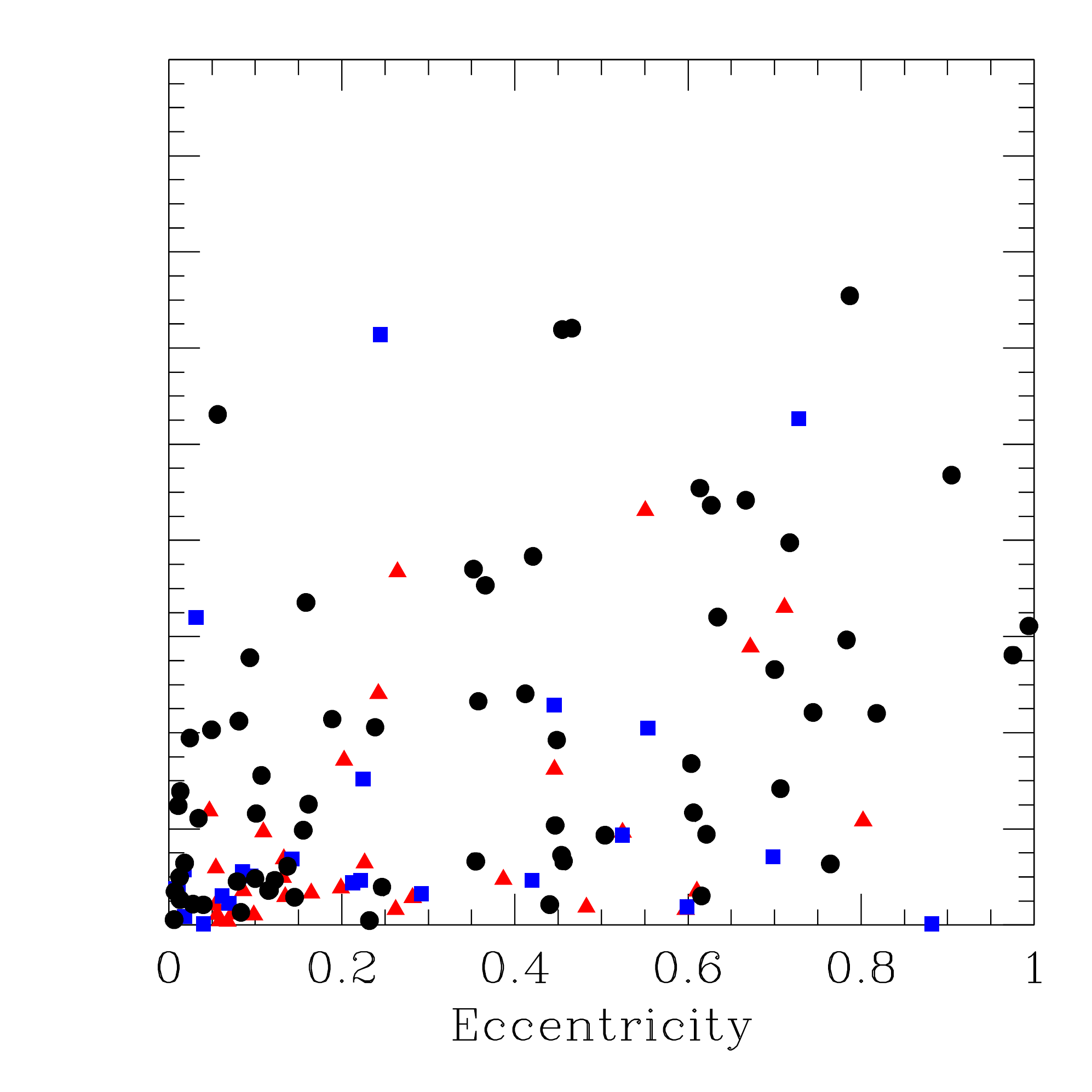} \hspace{-0.6cm}
    \includegraphics[width=4.7cm]{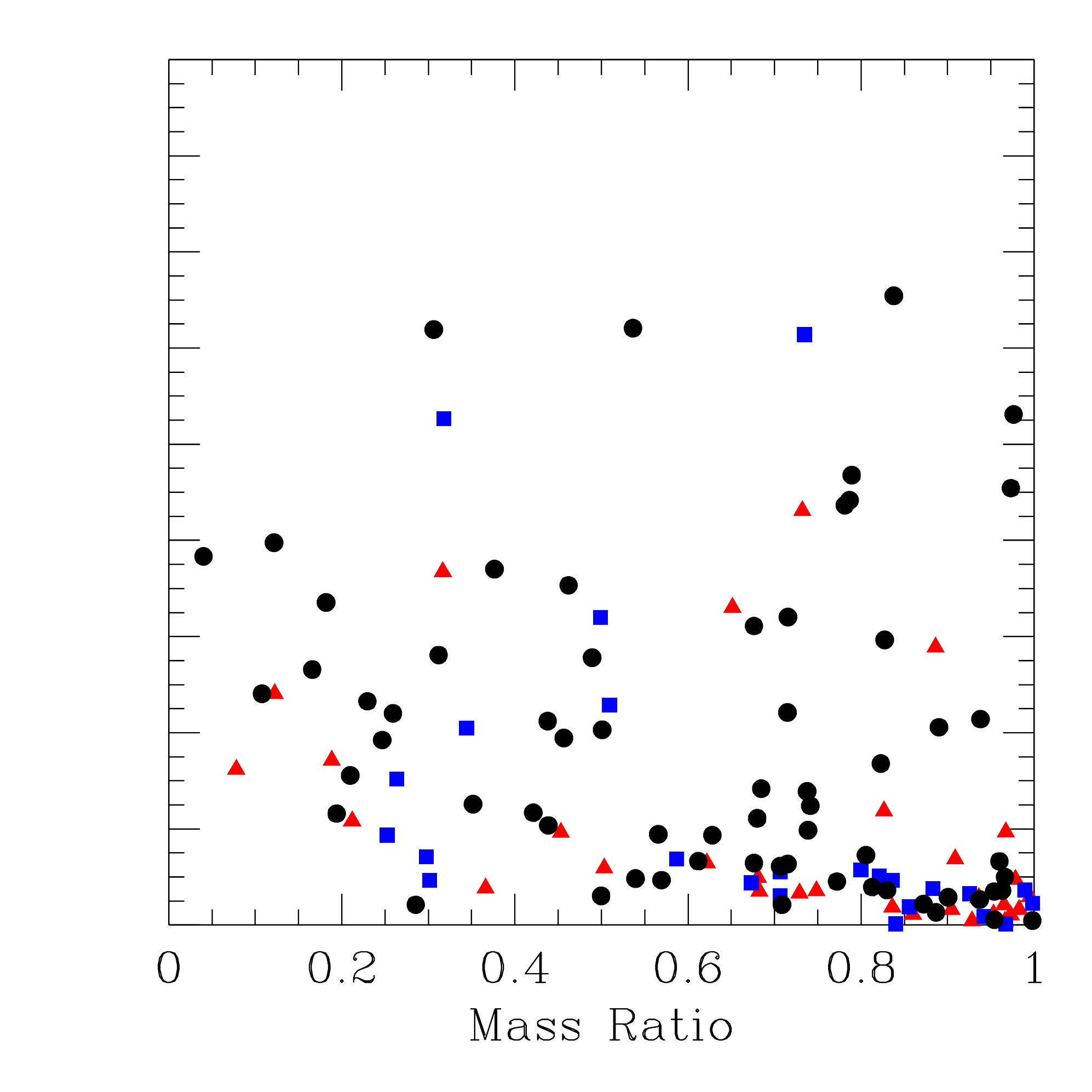} \hspace{-0.6cm}
    \includegraphics[width=4.7cm]{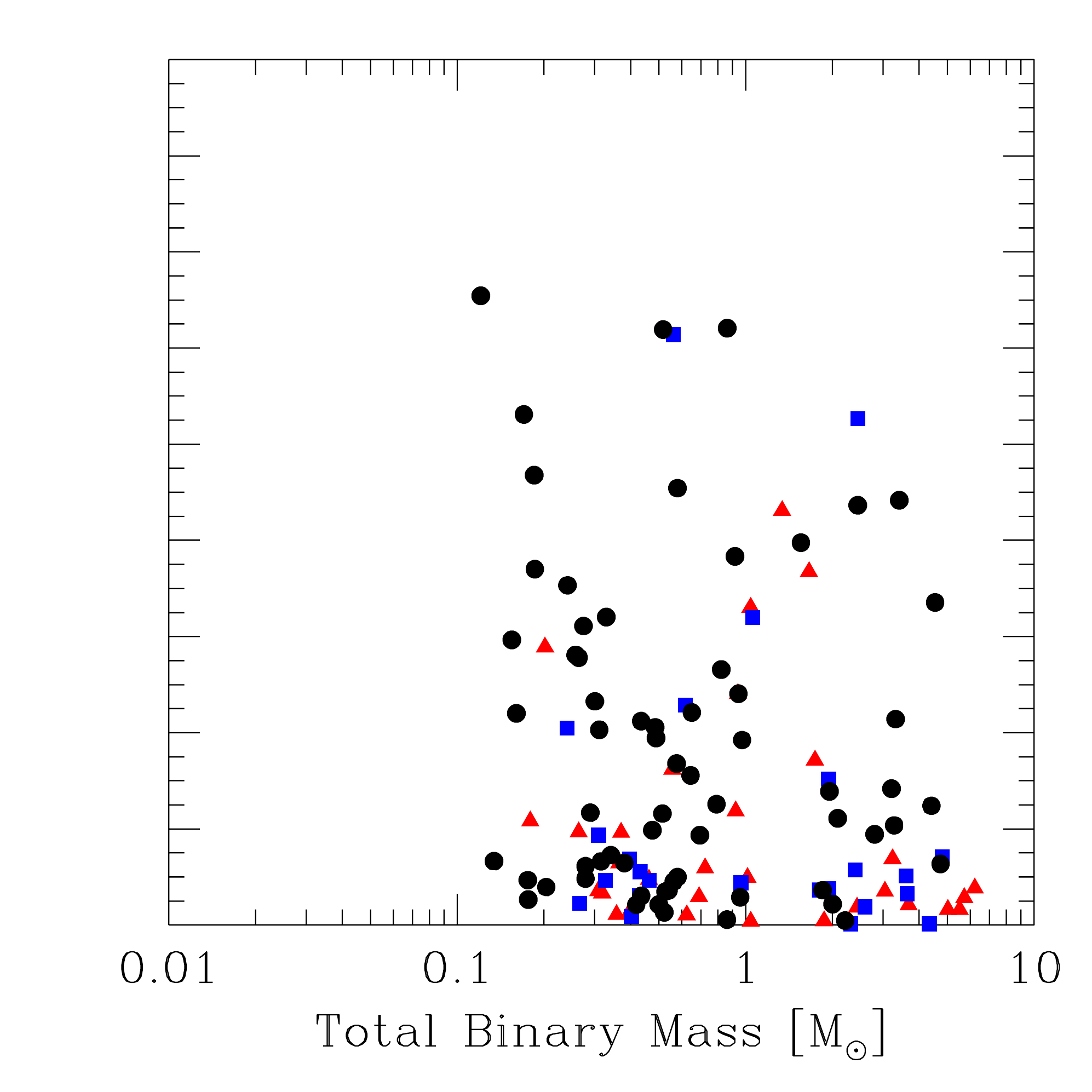}
\caption{The relative inclinations of the rotation axes of the sink particles (modelling stars and their inner discs) of the binaries included the combined sample from the three radiation hydrodynamical calculations with the highest opacities  ($Z\ge 0.1~{\rm Z}_\odot$) as functions of the binary's separation (left), eccentricity (centre-left), mass ratio (centre-right), and total mass (right).  We include binaries that are inner components of triples (red triangles) and quadruples (blue squares).  Most binaries for which the spins are closely aligned have semi-major axes $\lsim 30$~AU.  Binaries in which the spins are closely aligned also tend to have low eccentricities and high mass ratios, but the degree of alignment does not seem to depend on the binary's mass.
}
\label{spinspin}
\end{figure*}

\begin{figure*}
\centering
    \includegraphics[width=8.4cm]{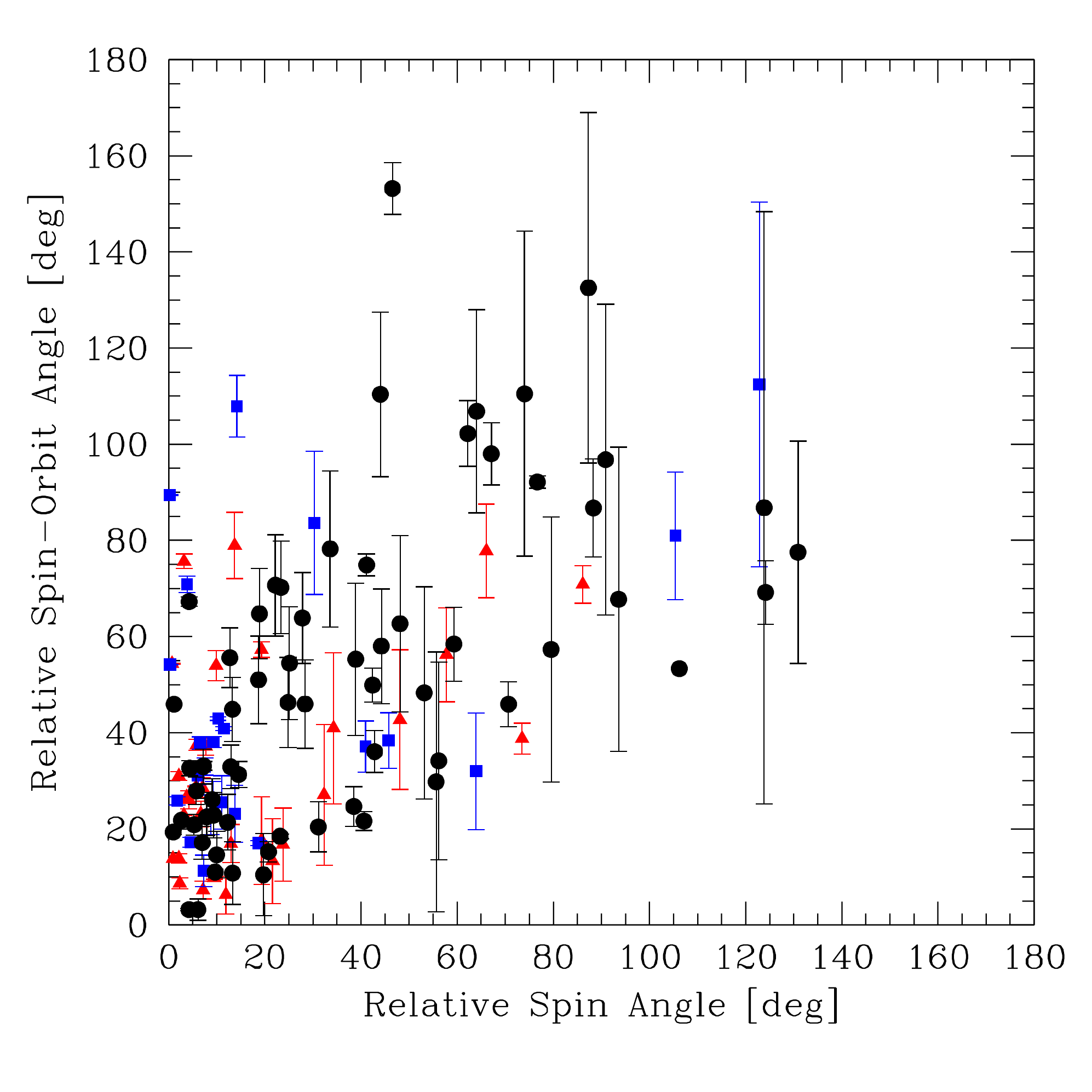}
    \includegraphics[width=8.4cm]{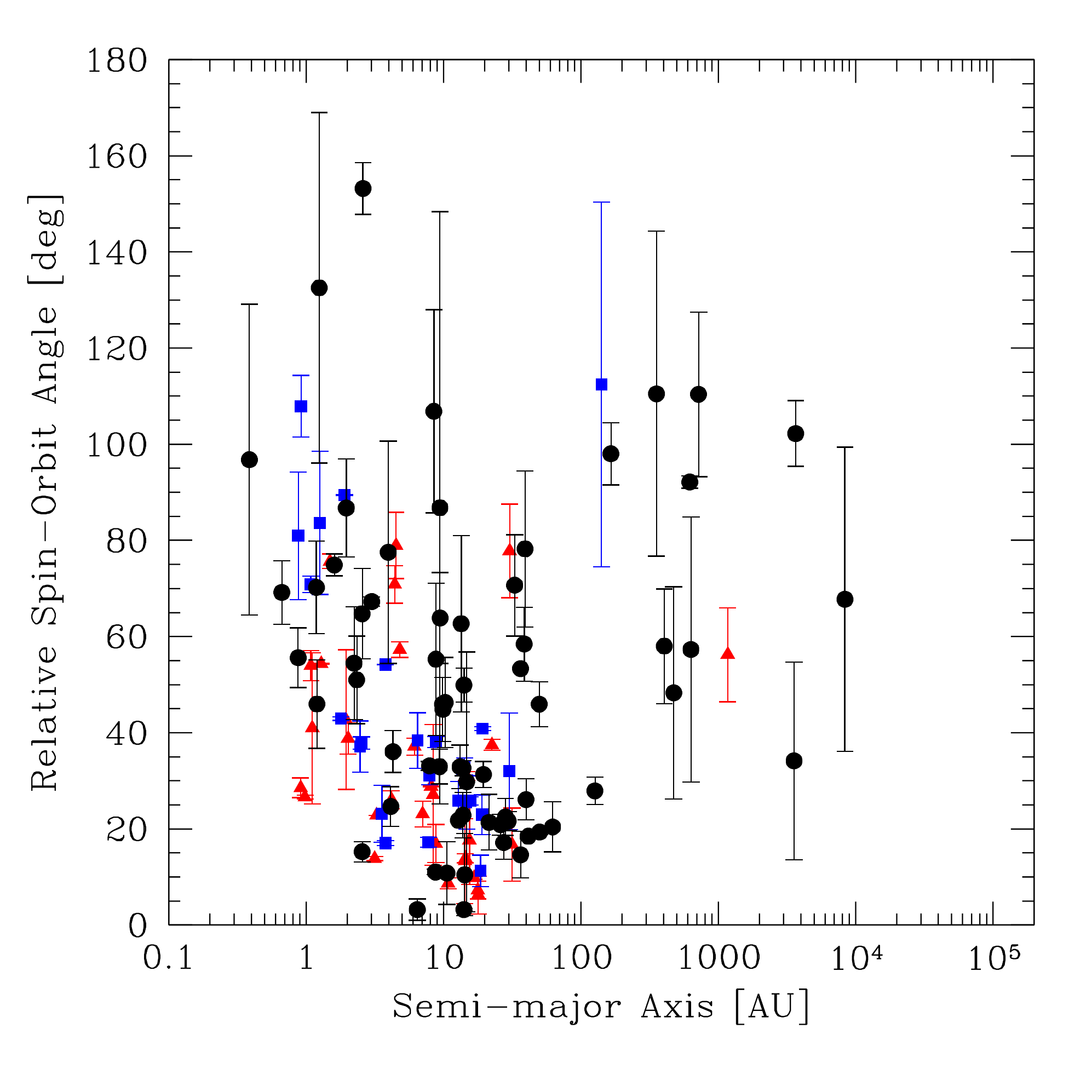}
\caption{The inclination angles of the rotation axes of the sink particles (modelling stars and their inner discs) of binaries relative to the inclination angle of the binary's orbital plane.  The binaries are taken from the combined sample from the three radiation hydrodynamical calculations with the highest opacities ($Z\ge 0.1~{\rm Z}_\odot$).  The relative spin-orbit angle is plotted as a function of the relative inclination angle of the two rotation axes of the sink particles (left) and the binary's separation (right).  For each binary we plot a point which gives the average of the angle between the primary's spin axis and that of the binary's orbit and the secondary's spin axis and binary's orbit, with the errorbars giving the actual values of these two angles.  We include binaries that are sub-components of triples (red triangles) and quadruples (blue squares).  Binaries for which the spins are closely aligned also tend to have spins that are closely aligned with the orbit.  Most binaries in which the orbit and spins are closely aligned have semi-major axes $\lsim 100$~AU.  }
\label{spinorbit}
\end{figure*}

Observationally, \citet{Weis1974} found a tendency for alignment between the stellar equatorial planes and orbital planes among primaries in F star binaries, but not A star binaries.  The orbital separations were mainly in the $10-100$ AU range.  Similarly, \citet{Guthrie1985} found no correlation for 23 A star binaries with separations 10-70 AU.   \citet{Hale1994} considered a larger sample of 73 binary and multiple systems containing solar-type stars and found evidence for approximate coplanarity between the orbital plane and the stellar equatorial planes for binary systems with separations less than $\approx 30$ AU and apparently uncorrelated stellar rotation and orbital axes for wider systems.  For higher-order multiple systems, however, non-coplanar systems were found to exist for both wide and close orbits.  Hale found no evidence to support a difference dependent on spectral type, eccentricity or age.  In terms of circumstellar discs, there is evidence for misaligned discs from observations of misaligned jets from protostellar objects \citep*{DavMunEis1994}, inferred jet precession \citep{Eisloffeletal1996,Davisetal1997}, and direct observations \citep{Koresko1998, Stapelfeldtetal1998}.  However, these are not statistically useful samples.  \citet*{MonMenDuc1998}, \citet*{DonJenMat1999}, \citet{Jensenetal2004}, \citet*{WolSteHen2001}, and \citet*{MonMenPer2006} used polarimetry to study the relative disc alignment in T Tauri wide binary and multiple systems and all found a preference for disc alignment in binaries. However, \citet{Jensenetal2004} also found that the wide components of triples and quadruples appear to have random orientations.  For more massive Herbig Ae/Be binaries, \cite{ Bainesetal2006} found that the circumprimary disc was preferentially aligned with the orbit and the larger study of \cite{Wheelwrightetal2011} also finds that the discs are preferentially aligned with the orbit.  

The barotropic calculations of \cite{Bate2009a} produced ambiguous results, with one calculation giving a strong tendency for alignment between sink particle spins, but another calculation with smaller accretion radii not showing any tendency for alignment \citep{Bate2011}.  However, the solar-metallicity radiation hydrodynamical calculation of \cite{Bate2012} produced strong tendencies for alignment between the spins of the components of binaries and for coplanarity of the orbital plane and the equatorial planes of the components for binaries.

In Fig.~\ref{spinspin} we plot the relative spin angles for the 120 binaries (including those that are components of triple and quadruple systems) as functions of semi-major axis and orbital eccentricity.  There are only 8 relative spin angles $>90^\circ$ and no relative spin angles greater than $140^\circ$, indicating a strong tendency for alignment.  The mean relative spin angle is $30^\circ\pm 3^\circ$, while the median angle is $17^\circ$. For the 65 pure binaries, the mean is $38^\circ\pm 4^\circ$ and the median is $28^\circ$, while for the binaries that are components of higher-order systems the mean is $21^\circ\pm 4^\circ$ and the median is $9^\circ$, so the spins of binaries that are the close components of higher-order multiples are more aligned.

Examining the left panel of Fig.~\ref{spinspin} it is clear that the tendency for alignment depends on separation:  of the 72 the binaries that have relative spin angles less than 25$^\circ$, all have separations less than $130$~AU (orbital periods $\approx 2000$~yrs) and 65 have separations less than 30~AU (orbital periods $\approx 200$~yrs).  Taking all binaries with semi-major axes less than 30~AU, the mean relative spin angle is $24^\circ\pm 3^\circ$, while those with longer periods have a mean of $52^\circ\pm 6^\circ$.  The centre-left panel of Fig.~\ref{spinspin} indicates that there may also be a relation between the relative spin angle and the eccentricity, with more circular systems having a stronger tendency for alignment.  Similarly, binaries with more equal masses tend to be more aligned (centre-right panel of Fig.~\ref{spinspin}).  Such relations may come about through accretion onto a binary system and/or gravitational torques between the stars and discs \cite[e.g.][]{Bateetal2000}, either of which would tend to align the components of the binary and may damp eccentricity, while accretion is expected to drive the system towards equal masses \citep{Bate2000}.  We note that the distribution of relative spin angles seems to be independent of the total mass of the binary (right panel of Fig.~\ref{spinspin}).

If the spins of the components of close binaries tend to be aligned with one another, one might also expect the spins to be aligned with the orbital plane of the binary.  Indeed, this is the case (left panel of Fig.~\ref{spinorbit}), though the alignment is not as strong as for the individual spins.  Taking the 72 binaries with relative spin angles less than 25$^\circ$, the mean spin-orbit angle is $32^\circ\pm 2^\circ$ with a standard deviation of $22^\circ$.  For the remaining 48 systems for which the spins are only weakly aligned, the mean spin-orbit angle is $66^\circ\pm 4^\circ$ and the standard deviation is also much larger ($36^\circ$).  It is also the case that the systems in which the spin axes are near to alignment with the orbit tend to be close systems with the vast majority having separations less than 100~AU (right panel of Fig.~\ref{spinorbit}).

In summary, for binaries with separations $\lsim 30-100$~AU, the radiation hydrodynamical calculations give a strong tendency for alignment between the spins of the components of binaries and for coplanarity of the orbital plane and the equatorial planes of the binary components.  These results are in good agreement with the observed coplanarity of binaries \citep{Hale1994} and in qualitative agreement with the many observational studies examining disc alignment mentioned above.

\section{Discussion}
\label{discussion}

\subsection{The dependence of stellar properties on metallicity}

From the results of four radiation hydrodynamical calculations that cover a range of opacities corresponding to  1/100 to 3 times solar metallicity (a factor of 300) that each produce at least 170 stars and brown dwarfs, we find no conclusive evidence for a dependence of the statistical properties of the stars and brown dwarfs on opacity.  Why is the opacity unimportant, and how do the results presented above compare to the results of observational surveys that investigate the dependence of stellar properties on metallicity?

\subsubsection{Comparison with previous theoretical results}

It is well established that the characteristic mass of stars produced by self-gravitating hydrodynamic calculations of star formation in molecular clouds using barotropic equations of state depend on the Jeans mass in the cloud \citep{BatBon2005, Jappsenetal2005, BonClaBat2006}.  This being the case, it is possible to propose that the characteristic stellar mass either increases or decreases with decreasing metallicity.  In the former case, dust cooling is less effective at lower metallicities in the low-density gas in the molecular cloud leading to an increase in the average temperature of the cloud.  This effect can be partially be seen in Fig.~\ref{global_temp} although, as discussed in Section \ref{initialconditions}, in reality the gas temperatures are expected to be even higher in the low metallicity cases because the calculations performed for this paper assume that the gas and dust are thermally well coupled which quickly breaks down at low-metallicity.  Therefore, if the characteristic stellar mass depends on the global Jeans mass, it should move to higher masses at lower metallicities.  Conversely, however, as gas collapses to form a protostar it will be able to cool more quickly at higher densities if the metallicity is lower because the optical depth will be reduced.  Thus, fragmentation (either in dense cores or protostellar discs) will potentially be able to occur at higher densities with lower metallicities, resulting in more fragmentation and a lower characteristic stellar mass.  Such changes in the effective equation of state of the gas due to different metallicity motivated the barotropic study of \cite{Bate2005} who found that, with the exception of the low-mass cut-off, the IMF was  insensitive to variations of the metallicity.

\begin{figure}
\centering \vspace{-0.5cm}
    \includegraphics[width=17.0cm]{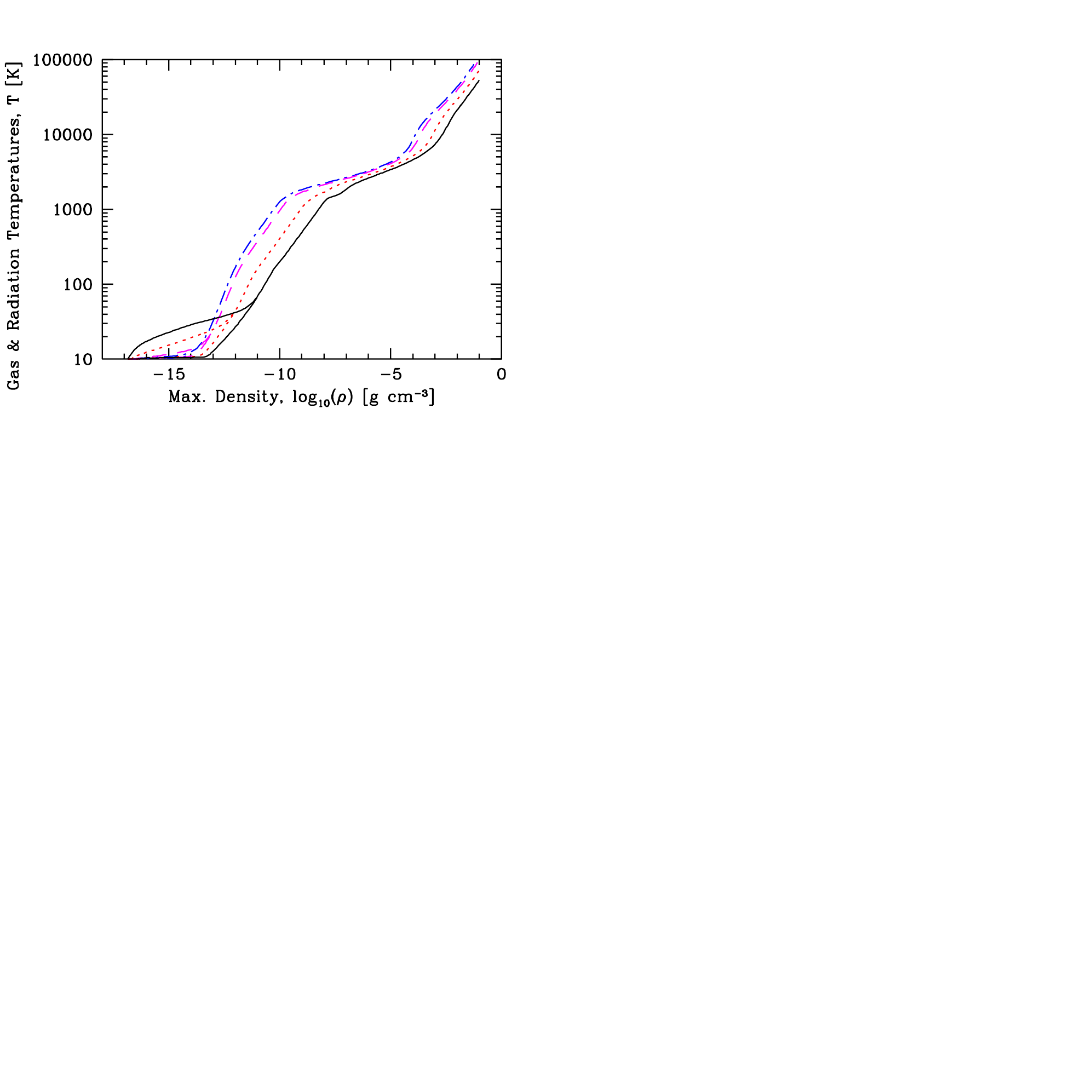}\vspace{-10.5cm}
\caption{The evolution of the maximum gas and radiation temperatures versus maximum density for radiation hydrodynamical calculations of the spherically-symmetric collapse of 1-M$_\odot$ molecular cloud cores with different metallicities: 1/100 Z$_\odot$ (solid black lines), 1/10 Z$_\odot$ (short-dashed red lines), Z$_\odot$ (long-dashed magneta lines), and 3~Z$_\odot$ (dot-dashed blue lines).  At low densities, two lines are visible for each of the different metallicities.  The upper line in each case is the gas temperature, while the lower line is the radiation temperature.  These are well coupled at high densities and/or metallicities, but are poorly coupled at low densities with low metallicity. }
\label{fig:maxdensitytemp}
\end{figure}

To illustrate this different thermal behaviour during the collapse of a molecular cloud core, we ran four calculations with varying metallicities of the collapse of initially uniform-density, stationary, spherical 1-M$_\odot$ molecular cloud cores with initial radii of $4\times 10^{16}$~cm embedded in an external radiation field with a temperature $T_{\rm r}=10$~K.  The evolution of the maximum gas and radiation temperatures with maximum density is shown for each calculation in Fig.~\ref{fig:maxdensitytemp}.  The differences are in good agreement with the results of \cite{OmuHosYos2010}, except that they do a much better job of modelling the thermodynamics at low-density. It is clear that at low densities, lower metallicities lead to higher gas temperatures (due to the less effective coupling of the gas and radiation), while at higher densities the temperatures are lower with lower metallicity (due to the increased cooling rate allowed by the lower optical depth).  The magnitudes of these effects are also in good agreement with those found in the similar recent calculations with $Z=0.1-1~{\rm Z}_\odot$ of \cite{Tomida2014}. Since the Jeans mass $M_{\rm Jeans} \propto T_{\rm g}^{3/2}/\rho$, at a given low-density the Jeans mass is higher with lower metallicity (reducing the fragmentation), while at higher densities the Jeans mass is lower (potentially promoting fragmentation).

The case for an increase in the characteristic stellar mass at lower metallicities relies on the link between the characteristic stellar mass and the global Jeans mass in the cloud.  However, \cite{Bate2009b} found that when hydrodynamical star formation calculations were performed using radiative transfer and a realistic equation of state the strong dependence of the characteristic stellar mass on the initial Jeans mass of the cloud was eliminated.  \citeauthor{Bate2009b} proposed a semi-analytic model for the characteristic mass of the IMF in which thermal heating of the molecular gas surrounding accreting protostars changed the effective Jeans mass.  Since the thermal heating depends on the luminosity of the protostars, which is unrelated to the initial global Jeans mass in the cloud, this removes the dependence of the characteristic stellar mass on the initial Jeans mass.  In the simplest case (taking the matter to be optically thin at all wavelengths with a grey opacity) the temperature distribution at radius $r$ from a protostar with luminosity $L_*$ is given by $T = (L_*/(4 \pi \sigma_{\rm B}))^{1/4} r^{-1/2}$, where $\sigma_{\rm B}$ is the Stefan-Boltzmann constant.  Assuming that the dust grains are in thermal equilibrium with this radiation field and that the gas is thermally coupled to the dust (as assumed in this paper), this may also be used to describe the gas temperature distribution.  However, it doesn't take into account the wavelength dependence of the dust absorption.  If the dust opacity depends on wavelength $\lambda$ as $\kappa \propto \lambda^{-\beta}$ then the temperature of the grains has a slightly different radial dependence $T \propto L_*^{1/4} r^{-2/(4+\beta)}$ \citep[e.g.][]{IveEli1997}, which reduces to the above formula when $\beta=0$, but $\beta \approx 2$ for interstellar grains.  In either of these equations, however, the radial temperature profile only depends on the properties of the dust grains, not the magnitude of the opacity, unless the clouds start to become optically thick.  For the calculations presented here, the opacities are at most $\approx 0.1$~cm$^2$~g$^{-1}$ at 20~K even with the highest metallicity (see Fig.~\ref{opacities}) and the typical column densities on the scale of the cloud range from $0.1-10$~g~cm$^{-2}$ (see Fig.~\ref{global_density}), so the clouds are marginally optically thin on large-scales even in the highest metallicity case (and very optically thin most cases).  Thus, for the calculations presented here, the temperature distribution at large distances from the accreting protostars is expected to be essentially independent of the metallicity.  Therefore, if the model of \cite{Bate2009b} is correct, the IMF should not vary with opacity, in agreement with the numerical results in this paper.

 \cite{Myersetal2011} performed radiation hydrodynamical simulations of star formation in which the opacity was varied by a factor of 20 and found no significant variation of the mass functions.  Their calculations had the same limitations as those we have presented here (i.e. they assumed that the gas and dust were thermally coupled, and neglected external sources of heating and gas emission line cooling), but they also proposed that the reason the IMF was independent of the metallicity was that the temperature distribution outside the photospheres of the protostellar cores was essentially independent of changes in the opacity due to metallicity.

\cite{Krumholz2011} took the arguments of \cite{Bate2009b} and \cite{Myersetal2011} one step further and tried to link the characteristic mass to fundamental physical constants.  In his simplest case, in which he assumes the temperature distribution varies with radius as $T_{\rm g} \propto r^{-1/2}$ as described above, the characteristic stellar mass does not depend on the magnitude of the opacity.  However, \cite{Krumholz2011} also develops a more complicated model of the thermal structure of protostars and predicts a dependence of the characteristic mass  on metallicity such that it increases by a factor of two for an order of magnitude decrease in opacity and a factor of 3--5 for two orders of magnitude decrease (i.e. approximately $M_{\rm ch} \propto Z^{-0.3}$).  Although this dependence is relatively weak, it is inconsistent with the results of the simulations presented in this paper -- we would easily have detected such changes in the characteristic stellar mass since our statistical uncertainties on the mean masses are no more than $\approx 0.05$~M$_\odot$ (c.f.\ Table 1 and Fig.~\ref{cumimf_comp}). 

Even if the thermal feedback from the protostars on cloud-scales is responsible for origin of the characteristic stellar mass, it is somewhat surprising that the different small-scale thermal behaviour of a collapsing molecular cloud core (e.g.\ Fig.~\ref{fig:maxdensitytemp}) does not result in differences in the small-scale fragmentation.  Naively, as mentioned above, the lower temperatures at a given density for lower opacity would be expected to increase fragmentation on small scales.  In fact, the weak variations in stellar properties noted in Section \ref{sec:results} may be due to this different small-scale thermal behaviour.  The lowest opacity calculation appeared to produce somewhat more brown dwarfs than the others, and the numbers of protostellar mergers increased strongly as the opacity was decreased below those expected for solar metallicities.  Although the statistical significance of these results is questionable, they are consistent with the idea that reducing the opacity promotes small-scale fragmentation.  Calculations of the fragmentation of individual pre-stellar cores to produce binary systems with different opacities have found similar effects, with the amount of fragmentation increasing and the typical binary separations decreasing as the opacity is decreased \citep{Machida2008, Machidaetal2009}. The way to be sure is to increase the statistical significance which will require large calculations in the future.

Finally, as noted in Sections \ref{introduction} and \ref{initialconditions}, while the calculations presented in this paper assume that the gas and dust are thermally well coupled and, thus, that the dominant cooling at low temperatures is dust cooling, this is not likely to be valid for metallicities much lower than the solar value.  Thermal decoupling of the gas and dust, cosmic ray and photoelectric heating of the gas, emission line cooling of the gas, and other effects all become much more important at low metallicities and must be included in future calculations to investigate the full dependence of stellar properties on metallicity. 

\subsubsection{Comparison with observations}

\cite{BasCovMey2010} recently reviewed the evidence for variations of the IMF, and concluded that there was no clear evidence of systematic variation with initial conditions, including metallicity variations, after the first few generations of stars.  For example, although studies of the thick disc population in our Galaxy obtain slopes that are somewhat shallower than found for the thin disc \citep{ReyRob2001,Vallenarietal2006}, the uncertainties are large.  Studies of OB associations in our Galaxy and stellar populations in nearby galaxies find that high-mass stars follow a standard Salpeter slope regardless of their metallicity (\citealt*{MasJohDeg1995}; \citealt{Masseyetal1995,Siriannietal2002,Sabbietal2008,Schmalzletal2008}), while some recent studies of nearby galaxies are able to begin probing below a solar mass but also fail to find a metallicity dependence \citep{Kaliraietal2013}.  Similarly, although studies of Galactic globular clusters tend to find somewhat higher characteristic stellar masses than found for young clusters, this is thought to be due to dynamical evolution rather than evidence of IMF variations \citep{PardeM2000, deMParPul2007}.   Thus, there seems to be little evidence that the IMF is sensitive to metallicity within the range $\approx 0.2-1$~Z$_\odot$.  Other evidence comes from examining the $\alpha$-element ratios (thought to be produced by core-collapse supernovae) which appear to be constant down to the lowest metallicities probed, suggesting that the high-mass end of the IMF has been invariant to redshifts of $z \sim 3-5$ \citep[][and references therein]{BasCovMey2010}.  Recently, there has been much interest in the possibility that the IMF may be bottom heavy in early-type galaxies \citep{Cenarroetal2003,vanCon2010,vanCon2011,Convan2012,Spinielloetal2012,Ferrerasetal2013}.  However, if this proves to be the case, it may be linked to the velocity dispersion in the galaxies rather than to metallicity \citep{laBarberaetal2013}.

Quite a few studies have investigated the question of whether the frequencies or properties of multiple stellar systems vary with metallicity.  However, the results have been mixed.  A number of studies of metal-deficient stars have indicated potentially lower multiplicities than solar-metallicity systems \citep{Carney1983, Strykeretal1985, AbtWil1987,AllPovHer2000,MarReb1992,Rastegaevetal2007, Rastegaevetal2008, RiaGizSam2008, LodZapMar2009, Jaoetal2009, Rastegaev2010}.  Others found similar multiplicities \citep{Ryan1992,Lathametal2002,ChaGou2004,ZapMar2004,ZinKohJah2004}, and a few studies have found higher multiplicities \citep{GreLin2007, Raghavanetal2010}.  The recent review of stellar multiplicity by \cite{DucKra2013} concludes that spectroscopic binaries are similar between Population II stars and their higher metallicity counterparts, but that the frequency of wide multiples with solar-mass primaries ($\gsim 10$~AU) is lower, resulting in slightly lower overall multiplicities.  \cite{Rastegaev2010} report a multiplicity of 33\% from 221 halo and thick-disc star primaries with metallicities less than 1/10 solar.  This compares to 44\% for nearby Sun-like stars \citep{Raghavanetal2010}.  For K and M-stars the multiplicities are around 26\% for metal-poor subdwarfs \citep{Jaoetal2009} versus approximately 37\% for dwarfs \citep{ReiGiz1997}.  However, for solar-mass primaries \cite{Rastegaev2010} also find that the distribution of orbital periods peaks at much lower periods ($\log P=2-3$ for Population II stars rather than $\log P\approx 5$ for Population I stars, with period $P$ measured in days), and a similar difference in the orbital separations has been suggested for M-star binaries \citep{RiaGizSam2008, LodZapMar2009}.  It is unclear whether these apparent differences in wide multiples are primordial or whether there has been significant evolution with time.

Overall, it seems from observations that metallicity does not play a large role in determining the properties of stellar systems.  This is consistent with our numerical results -- we find no firm evidence for a dependence of stellar properties on metallicity due to changes in the opacity.  However, it is interesting to note that the indications from the simulations that small-scale fragmentation may be slightly enhanced at low opacities, resulting in an increased frequency of protostellar mergers, is at least in the same sense as the observations that suggest metal-poor binaries may have closer separations than systems with solar metallicities.

\subsection{Potential for improving the agreement with observed stellar properties}

As discussed in Section \ref{sec:combined}, many of the statistical properties of the stellar systems produced by the calculations are in good agreement with observed systems.  However, there are some quantities where there is disagreement.  The first is that VLM multiple systems tend to have separations that are wider than is typically observed (Fig.~\ref{combined_separation_dist}).  Although wide VLM multiples are observed to be more common at young ages \citep{Closeetal2007}, and \cite{Bate2009a} found that the separation distribution of VLM multiples tended to move to smaller separations if the hydrodynamical calculations were evolved for longer, there is potential for improvement.  One area of concern is that the sink particle radius in the calculations is 0.5~AU which is uncomfortably close to the typical VLM binary separation of a few AU.  It is likely that interactions between the sink particles and their discs are not as dissipative as they should be and this may lead to larger orbital separations.  Therefore it would be desirable to increase the resolution by decreasing the sizes of the sink particles still further in future calculations.

Dissipation may also be the cause of the other areas of potential disagreement, but this time over dissipation rather than a lack of it.  As mentioned in Sections 4.5 and 4.6, there appears to be an excess of low eccentricity orbits and the orbital planes of triple systems may be more aligned than is observed.  This points to possible excess dissipation in modelling the orbits of multiple systems.  The potential excess of solar-type `twins' (binaries with near-equal masses) is also interesting in this regard.  Although the numerical mass ratio distribution is consistent with that found by \cite{Raghavanetal2010} (Section 4.4), one way to produce twins is via the accretion of gas with high angular momentum which tends to equalise the masses \citep{Bate2000} and if the effective viscosity is high it may also damp eccentricities and inclinations.  The numerical resolution in these calculations is only $7\times 10^4$ SPH particles per solar mass of gas which is much lower than would be typically used to study the evolution of individual protostars.  Artificial viscosity decreases with increased resolution, so again this is an area where increased resolution (this time in terms of increased particle number) may help.  Alternately, it may be possible to use some of the recently developed SPH viscosity switches \citep[e.g.][]{CulDeh2010, ReaHay2012} to reduce the effective viscosity.

\section{Conclusions}
\label{conclusions}

We have presented results from four radiation hydrodynamical simulations of star cluster formation that each resolve the opacity limit for fragmentation, protoplanetary discs (radii $\gsim 1$ AU), and multiple systems.  The calculations are identical except for the opacities that are used when modelling the radiative transfer.  We use opacities appropriate for metallicities ranging from 1/100 to three times the solar value (a factor of 300).  Each individual calculation produces at least 170 stars and brown dwarfs (modelled as sink particles with accretion radii of only 0.5~AU), sufficient to allow comparison of the statistical properties of the stars, brown dwarfs and multiple systems with the results of observational surveys.  Overall, the calculations display good agreement with a wide range of observed stellar properties, implying that the main physical processes involved in determining the properties stellar systems are gravity, gas dynamics (i.e.~dissipative $N$-body dynamics), and thermal feedback from protostars.  

However, there are a number of important caveats to this conclusion.  First, the star formation rate in the calculations is much higher than observed.  To solve this problem may require globally unbound molecular clouds and/or the inclusion of magnetic fields and kinetic feedback.  Second, we emphasise that changing the metallicity of a star-forming region affects more than just the opacity of the gas and dust.  In particular, the calculations performed for this paper have assumed that the gas and dust are thermally well coupled and, thus, that dust emission is the dominant gas coolant.  While this approximation is typical for calculations of star formation at solar metallicities and should also be a good at super-solar metallicites, it quickly breaks down as the metallicity is reduced.  Future calculations should examine both the effects of thermal feedback from protostars and the additional thermal effects at low metallicities in order to get a complete picture of how star formation depends on metallicity.
 
For the present study, our detailed conclusions for how star formation depends on opacity and how stellar properties obtained from the simulations compare with observations are as follows.

\begin{enumerate}
\item We find no statistically-significant dependence of the properties of the stars and brown dwarfs on opacity, despite varying the opacities by a factor of 300.  However, we do find that fragmentation of dense gas in protostellar cores and/or discs may increase with low opacities ($\lsim 1/10$ of the opacities corresponding to solar metallicities), potentially increasing the abundance of brown dwarfs. 
\item We find that protostellar collisions become significantly more frequent at lower opacities.  With opacities relevant for solar and super-solar metallicities, mergers are very rare.  But $\approx 10$\% of stars and brown dwarfs suffer a collision as they are forming with opacities of 1/100 the solar metallicity value.  This is also consistent with a higher degree of small-scale fragmentation at low metallicities.
\item All the calculations produce IMFs that are statistically indistinguishable from the parameterisation of the observed IMF by \cite{Chabrier2005}, and ratios of brown dwarfs to stars which are also in good agreement with observations.  Combining the results of the calculations with the three highest opacities ($Z \ge 0.1~{\rm Z}_\odot$ which produced 535 stars and brown dwarfs) also gives an IMF in good agreement with \citeauthor{Chabrier2005}'s parameterisation.
\item We find that multiplicity strongly increases with primary mass.  The results are in good agreement with the observed multiplicities of A--M stars and VLM objects.  For objects with primary masses in the range $0.03-0.20$ M$_\odot$ the multiplicity fraction is $0.14^{+0.07}_{-0.05}$ (95\% confidence interval).  But it is important to note that the multiplicity increases with primary mass steeply in this range and, thus, when comparing with observations it is important to take care to compare like with like.  We predict that the multiplicity continues to drop for lower-mass brown dwarfs. We also find that the frequency of high-order multiples (triples and quadruples) increases with primary mass.
\item We examine the separation distributions of binaries, triples and quadruples.  These are in good agreement with the observed distributions for M-stars, and reasonable agreement for higher-mass stars.  However, the VLM multiples tend to be wider than is typically observed.  This may be because VLM multiples evolve to closer separations with time \citep[as seen in the calculations of][]{Bate2009a} and the calculations are not followed long enough, or the numerical resolution may be insufficient.
\item The mass ratio distributions of solar-type, M-dwarf, and VLM binaries are consistent with observations.  Although the VLM binaries are consistent with a mass ratio distribution that is biased towards equal masses, we find that within the statistical uncertainties their mass ratio distribution is also consistent with those of higher-mass binaries.  We find that closer binaries tend to have a higher proportion of equal mass components in broad agreement with observed trends.
\item The distributions of eccentricities are broad, but there are too many low-eccentricity systems compared with observed systems.  This may be an indication that the hydrodynamical interactions between protostars are too dissipative.  There may also be a weak link between mass ratio and eccentricity such that `twins' have lower eccentricities, as is observed.
\item We investigate the relative orientation of the orbital planes of triple systems and the alignments between sink particle spins and the orbital planes in binaries.  We find a tendency for alignment between the orbital planes in triple systems, and we find that this tendency is stronger for closer systems, in qualitative agreement with observations.  However, the degree of alignment appears to be stronger than is observed.  Again, this may be an indication that the calculations are too dissipative.  Similarly, we find that in close binaries ($\lsim 100$~AU) the spins of sink particles tend to be aligned with each other, and the spins tend to be aligned with the orbital plane, in good agreement with existing observations.
\end{enumerate}

The dataset consisting of the output and analysis files from the calculations presented in this paper have been placed in the University of Exeter's Open Research Exeter (ORE) repository and can be accessed via the handle: http://hdl.handle.net/10871/14881

\section*{Acknowledgments}

MRB is grateful to the referee, P.\ Clark, for substantially improving the way the paper interfaces with past studies of the thermodynamic evolution of metal-poor clouds, and to A.\ Tokovinin for thoughtfully reading the manuscript and making many suggestions which helped improve the paper, particularly in the discussion of the properties of multiple stellar systems.
The calculations for this paper were performed on the University of Exeter Supercomputer, 
a DiRAC Facility jointly funded by STFC, the Large Facilities Capital Fund of BIS, and the University of Exeter.
This publication has made use of the Very-Low-Mass Binaries Archive housed at http://www.vlmbinaries.org and maintained by Nick Siegler, Chris Gelino, and Adam Burgasser.

\bibliography{/Users/mbate/Tex/mbate}

\end{document}